\documentclass[english]{article}
\usepackage[T1]{fontenc}
\usepackage[latin9]{inputenc}
\usepackage{geometry}
\usepackage{refstyle}
\usepackage{amsmath}
\usepackage{amssymb}
\usepackage{graphicx}
\usepackage{esint}

\makeatletter

\AtBeginDocument{\providecommand\eqref[1]{\ref{eq:#1}}}
\providecommand{\tabularnewline}{\\}
\RS@ifundefined{subref}
  {\def\RSsubtxt{section~}\newref{sub}{name = \RSsubtxt}}
  {}
\RS@ifundefined{thmref}
  {\def\RSthmtxt{theorem~}\newref{thm}{name = \RSthmtxt}}
  {}
\RS@ifundefined{lemref}
  {\def\RSlemtxt{lemma~}\newref{lem}{name = \RSlemtxt}}
  {}

\numberwithin{equation}{section}
\numberwithin{figure}{section}
\numberwithin{table}{section}

\newcommand{\blockZpos}[5]{
\setlength{\unitlength}{1cm}
\begin{picture}(3,1)
\linethickness{0.4mm}
\put(0.5,-0.5){\line(0,1){1}}
\put(2.5,-0.5){\line(0,1){1}}
\put(0.5,0){\line(1,0){2}}
\put(0.3,-0.8){$\phi_{#1}(\infty)$}
\put(2.3,-0.8){$\phi_{#2}(0)$}
\put(0.3,0.6){$\phi_{#3}(1)$}
\put(2.3,0.6){$\phi_{#4}(z)$}
\put(1.3,0.15){$\phi_{#5}$}
\end{picture}
}
\newcommand{\blockZ}[5]{
\setlength{\unitlength}{1cm}
\begin{picture}(3,1)
\linethickness{0.4mm}
\put(0.5,-0.5){\line(0,1){1}}
\put(2.5,-0.5){\line(0,1){1}}
\put(0.5,0){\line(1,0){2}}
\put(0.3,-0.8){$\phi_{#1}$}
\put(2.3,-0.8){$\phi_{#2}$}
\put(0.3,0.6){$\phi_{#3}$}
\put(2.3,0.6){$\phi_{#4}$}
\put(1.3,0.15){$\phi_{#5}$}
\end{picture}
}

\newcommand{\blockZgen}[5]{
\setlength{\unitlength}{1cm}
\begin{picture}(3,1)
\linethickness{0.4mm}
\put(0.5,-0.5){\line(0,1){1}}
\put(2.5,-0.5){\line(0,1){1}}
\put(0.5,0){\line(1,0){2}}
\put(0.4,-0.9){$#1$}
\put(2.4,-0.9){$#2$}
\put(0.4,0.7){$#3$}
\put(2.4,0.7){$#4$}
\put(1.4,0.1){$#5$}
\end{picture}
}

\newcommand{\blockOnemZpos}[5]{
\setlength{\unitlength}{1cm}
\begin{picture}(3,1.5)
\linethickness{0.4mm}
\put(1.5,-0.5){\line(0,1){1}}
\put(0.5,0.5){\line(1,0){2}}
\put(0.5,-0.5){\line(1,0){2}}
\put(0,-0.9){$\phi_{#1}(\infty)$}
\put(2.3,-0.9){$\phi_{#2}(0)$}
\put(0,0.7){$\phi_{#3}(1)$}
\put(2.3,0.7){$\phi_{#4}(z)$}
\put(1.6,0){$\phi_{#5}$}
\put(0.3,0.3){$ $}
\put(2.5,0.3){$ $}
\end{picture}
}
\newcommand{\blockOnemZ}[5]{
\setlength{\unitlength}{1cm}
\begin{picture}(3,1.5)
\linethickness{0.4mm}
\put(1.5,-0.5){\line(0,1){1}}
\put(0.5,0.5){\line(1,0){2}}
\put(0.5,-0.5){\line(1,0){2}}
\put(0,-0.9){$\phi_{#1}$}
\put(2.3,-0.9){$\phi_{#2}$}
\put(0,0.7){$\phi_{#3}$}
\put(2.3,0.7){$\phi_{#4}$}
\put(1.6,0){$\phi_{#5}$}
\put(0.3,0.3){$ $}
\put(2.5,0.3){$ $}
\end{picture}
}

\newcommand{\blockOnemZgen}[5]{
\setlength{\unitlength}{1cm}
\begin{picture}(3.2,1)
\linethickness{0.4mm}
\put(1.5,-0.5){\line(0,1){1}}
\put(0.5,0.5){\line(1,0){2}}
\put(0.5,-0.5){\line(1,0){2}}
\put(0.4,-0.9){$#1$}
\put(2.4,-0.9){$#2$}
\put(0.4,0.7){$#3$}
\put(2.4,0.7){$#4$}
\put(1.6,0){$#5$}
\put(0.3,0.3){$ $}
\put(2.5,0.3){$ $}
\end{picture}
}

\makeatother

\usepackage{babel}
\begin{document}

\title{Excited state TBA and renormalized TCSA in the scaling Potts model}

\author{M. Lencsés$^{1,2}$ and G. Takács$^{1,3}$\\
 ~\\
 $^{1}$MTA-BME \textquotedbl{}Momentum\textquotedbl{} Statistical
Field Theory Research Group\\
1111 Budapest, Budafoki út 8, Hungary\\
~\\
 $^{2}$Department of Theoretical Physics, Eötvös University\\
 1117 Budapest, Pázmány Péter sétány 1/A, Hungary\\
 ~\\
 $^{3}$Department of Theoretical Physics, \\
 Budapest University of Technology and Economics\\
1111 Budapest, Budafoki út 8, Hungary}

\date{14th July 2014}
\maketitle
\begin{abstract}
We consider the field theory describing the scaling limit of the Potts
quantum spin chain using a combination of two approaches. The first
is the renormalized truncated conformal space approach (TCSA), while
the second one is a new thermodynamic Bethe Ansatz (TBA) system for
the excited state spectrum in finite volume. For the TCSA we investigate
and clarify several aspects of the renormalization procedure and counter
term construction. The TBA system is first verified by comparing its
ultraviolet limit to conformal field theory and the infrared limit
to exact S-matrix predictions. We then show that the TBA and the renormalized
TCSA match each other to a very high precision for a large range of
the volume parameter, providing both a further verification of the
TBA system and a demonstration of the efficiency of the TCSA renormalization
procedure. We also discuss the lessons learned from our results concerning
recent developments regarding the low-energy scattering of quasi-particles
in the quantum Potts spin chain. 
\end{abstract}

\section{Introduction}

In this work we consider the field theory describing the vicinity
of the critical point of the three-state Potts quantum spin chain.
The model is defined on the Hilbert space 
\begin{equation}
\mathcal{H}_{\mathrm{chain}}=\bigotimes_{i}\left(\mathbb{C}^{3}\right)_{i}
\end{equation}
where $i$ labels the sites of the chain, and the quantum space $\mathbb{C}^{3}$
at site $i$ has the basis $|\alpha\rangle$, $\alpha=0,1,2$, corresponding
to the spin degrees of freedom. The dynamics is defined by the Hamiltonian
invariant under $\mathbb{S}_{3}$ permutation symmetry
\begin{equation}
H_{\mathrm{chain}}=-J\sum_{i}\sum_{\alpha=0}^{2}P_{i}^{\alpha}P_{i+1}^{\alpha}-Jg\sum_{i}\tilde{P}_{i}\label{eq:potts_chain_hamiltonian}
\end{equation}
where 
\begin{eqnarray}
P^{\alpha} & = & |\alpha\rangle\langle\alpha|-\frac{1}{3}\\
\tilde{P} & = & \frac{1}{3}\sum_{\alpha,\alpha'=0}^{2}\left(1-\delta_{\alpha\alpha'}\right)|\alpha\rangle\langle\alpha'|\nonumber 
\end{eqnarray}
The spin chain has a critical point at $g=1$ corresponding to a phase
transition between a paramagnetic $g>1$ and ferromagnetic $g<1$
case. The critical point  can be described with a conformal field
theory (CFT) with central charge $c=4/5$. The scaling limit of the
off-critical theory corresponds to a uniquely defined perturbation
of the fixed point CFT. This quantum field theory (QFT), called the
scaling Potts model is known to be integrable \cite{Zamolodchikov1988},
and its spectrum and scattering matrix was determined exactly \cite{Zamolodchikov1988,Koberle:1979sg,Chim1992}.
In section \ref{sec:Scaling-Potts-model-as-pCFT} we summarize the
necessary facts about perturbed CFT and its application to the scaling
Potts model; this also serves to specify our conventions and summarize
the most important known facts that are used later. 

In the main part of the paper we develop two methods to describe the
finite volume spectrum of the scaling QFT. The first of them is a
renormalized version of the truncated conformal space approach (TCSA).
The TCSA was introduced by Yurov and Zamolodchikov \cite{Yurov:1989yu},
and has been applied to numerous problems since then; among them we
mention a recent study of non-integrable perturbations of the Potts
conformal field theory \cite{Lepori:2009ip}. 

Recently, a renormalization group approach was proposed to treat the
cut-off dependence of TCSA, both for the case of boundary \cite{Feverati:2006ni,Watts:2011cr}
and bulk flows \cite{Konik:2007cb,Giokas:2011ix}. In the present
paper we mainly build on the results in the unpublished work by Giokas
and Watts \cite{Giokas:2011ix}, and work out the general theory of
counter terms in TCSA in section \ref{sec:Renormalization-in-TCSA}
together with its application to the scaling Potts model. We present
and explain the method in sufficient detail not only for the reproduction
of the results in this paper, but also to facilitate further applications.
A new aspect of our results is the construction of counter terms for
descendant states and the treatment of degenerate perturbation theory.

On the other hand, in section \ref{sec:Excited-state-TBA} we propose
TBA equations for the exact finite volume spectrum, in both the ferromagnetic
and paramagnetic phases of the Potts model. These are obtained by
starting from the ground state thermodynamical Bethe Ansatz (TBA)
equations \cite{Zamolodchikov:1989cf,Martins1991c,Fendley:1991xn}
and using simple arguments based on the analytical continuation approach
by Dorey and Tateo \cite{Dorey:1996re,Dorey:1997rb}. The resulting
excited TBA equations are first analyzed in the large volume (infrared,
IR) asymptotic regime, where they are demonstrated to match with the
exact S matrices of the scaling Potts model. In the small volume (ultraviolet,
UV) asymptotic regime, they are shown to agree with the spectrum of
conformal weights predicted by the fixed point CFT. 

In section \ref{sec:Numerical-comparison}, the renormalized TCSA
method is applied to obtain an accurate numerical finite volume spectrum
of the scaling Potts model. We compare the results to the predictions
of the TBA system and show that they match accurately and in detail.
This provides both a demonstration of the efficiency of the renormalized
TCSA, and also a detailed check of the correctness of the proposed
excited TBA equations. 

Finally, in section \ref{sec:Conclusions} we draw our conclusions.
The paper also contains three appendices: Appendix \ref{sec:CFT-data}
specifies the CFT structures which are used for the calculations in
the main text, Appendix \ref{sec:Derivation-of-the-UV-limit} contains
the general derivation of the UV limit of the excited TBA equations,
while Appendix \ref{sec:Comparison-tables} contains numerical tables
for the comparison between TBA and TCSA.

\section{Scaling Potts model as perturbed conformal field theory \label{sec:Scaling-Potts-model-as-pCFT}}

\subsection{The formalism of perturbed conformal field theory\label{sub:The-formalism-of}}

The idea of obtaining massive field theories as relevant perturbations
of their ultraviolet fixed points goes back to Zamolodchikov \cite{Zamolodchikov:1990jh}.
Here we summarize the necessary notations to set up the stage for
our calculations. Let us consider a theory defined on a Euclidean
space-time cylinder with spatial circumference $R$:
\begin{equation}
S=S_{CFT}+\mu\int_{-\infty}^{\infty}d\tau\int_{0}^{R}dx\Phi\left(\tau,x\right)\label{eq:pCFTaction}
\end{equation}
where $S_{CFT}$ is the action of a conformal field theory and the
perturbing operator $\Phi$ is a primary field with conformal dimensions
$h=\bar{h}$. Using complex coordinates $\zeta=\tau+ix$ 
\begin{equation}
S=S_{CFT}+\mu\int d^{2}\zeta\Phi\left(\zeta,\bar{\zeta}\right)
\end{equation}
and the corresponding Hamiltonian can be written as
\begin{equation}
H=H_{CFT}+\mu\int_{0}^{R}dx\Phi\left(0,x\right)\label{eq:pcft_hamiltonian}
\end{equation}
Under the exponential mapping to the conformal plane
\begin{equation}
z=e^{\frac{2\pi}{R}\zeta}
\end{equation}
the Hamiltonian can be expressed as 
\begin{equation}
H=H_{CFT}+\delta H
\end{equation}
where $H_{CFT}$ can be written in terms of Virasoro generators $L_{0}$
and $\bar{L}_{0}$ and the central charge $c$ of the CFT:

\begin{equation}
H_{CFT}=\frac{2\pi}{R}\left(L_{0}+\bar{L}_{0}-\frac{c}{12}\right)
\end{equation}
and the perturbing term is 
\begin{equation}
\delta H=\frac{2\pi}{R}\mu\frac{R^{2-2h}}{\left(2\pi\right)^{1-2h}}\Phi\left(1,1\right)
\end{equation}
Introducing a mass scale $m$ one can write $\mu=\kappa m^{2-2h}$
where $\kappa$ is a dimensionless constant, and define the dimensionless
volume $r=mR$ to obtain
\begin{equation}
\delta S=\kappa\frac{r^{2-2h}}{\left(2\pi\right)^{2-2h}}\int d^{2}z\left(z\bar{z}\right)^{h-1}\Phi\left(z,\bar{z}\right)
\end{equation}
The dimensionless Hamiltonian is defined as
\begin{eqnarray}
h\left(r\right) & = & \frac{H(R)}{m}=\frac{2\pi}{r}e\left(r\right)\nonumber \\
e\left(r\right) & = & L_{0}+\bar{L}_{0}-\frac{c}{12}+\lambda\Phi\left(z,\bar{z}\right)|_{z=\bar{z}=1}\nonumber \\
 &  & \lambda=\mu\frac{R^{2-2h}}{\left(2\pi\right)^{1-2h}}=\kappa\frac{r^{2-2h}}{\left(2\pi\right)^{1-2h}}\label{eq:dimless_pcft_hamiltonian}
\end{eqnarray}
where $e\left(r\right)$ is the so-called scaling function. In the
same units, the perturbing action takes the form
\begin{equation}
\delta S=\frac{\lambda}{2\pi}\int d^{2}z\left(z\bar{z}\right)^{h-1}\Phi\left(z,\bar{z}\right)
\end{equation}
For the scaling function of the vacuum, the perturbative expansion
is the following \cite{Zamolodchikov:1989cf}
\begin{eqnarray}
e_{0}(\lambda) & = & -\frac{c}{12}-\sum_{n=1}^{\infty}\frac{1}{n!}\left(-\frac{\lambda}{2\pi}\right)^{n}2\pi\left\{ \prod_{i=1}^{n-1}\int d^{2}z_{i}\left(z_{i}\bar{z}_{i}\right)^{h-1}\right\} \langle\Psi_{0}|\Phi\left(1,1\right)\prod_{i=1}^{n-1}\Phi(z_{i},\bar{z}_{i})|\Psi_{0}\rangle_{conn}\label{eq:cpt_scaling_function}\\
 & = & -\frac{c}{12}+\lambda\langle\Psi_{0}|\Phi\left(1,1\right)|\Psi_{0}\rangle-\frac{\lambda^{2}}{2\pi}\int_{\left|z\right|<1}d^{2}z(z\bar{z})^{h-1}\langle\Psi_{0}|\Phi\left(1,1\right)\Phi(z,\bar{z})|\Psi_{0}\rangle_{conn}+\mathcal{O}(\lambda^{3})\nonumber 
\end{eqnarray}
where the matrix elements are taken in the unperturbed CFT ($\lambda=0$),
the subscript $conn$ denotes the connected piece of the matrix element,
and radial ordering was taken into account by restricting $\left|z\right|<1$
and incorporating a factor of $2$ for the two identical contributions. 

Simple power counting in the integrals shows that for $h<1/2$ the
results are ultraviolet convergent, while for $h\geq1/2$ there are
ultraviolet divergences which are manifested in poles of gamma functions
resulting from the integration. However, due to the meromorphic dependence
of the perturbative coefficients on $h$, a finite result can be defined
by analytic continuation in $h$. In section \ref{sub:Cut-off-dependence-in}
we point out that the renormalization scheme defined by this procedure
is the preferred one when comparing to exact results from integrability.

\subsection{Scaling Potts model as a perturbed conformal field theory \label{sub:Scaling-Potts-model}}

The scaling limit of Potts model at the critical point is a minimal
conformal field theory with central charge 
\begin{equation}
c=\frac{4}{5}
\end{equation}
\cite{Belavin:1984vu,Dotsenko:1984if}. The spectrum of allowed primary
conformal weights is given by the Kac table
\begin{equation}
\left\{ h_{r,s}\right\} =\left(\begin{array}{ccccc}
0 & \frac{1}{8} & \frac{2}{3} & \frac{13}{8} & 3\\
\\
\frac{2}{5} & \frac{1}{40} & \frac{1}{15} & \frac{21}{40} & \frac{7}{5}\\
\\
\frac{7}{5} & \frac{21}{40} & \frac{1}{15} & \frac{1}{40} & \frac{2}{5}\\
\\
3 & \frac{13}{8} & \frac{2}{3} & \frac{1}{8} & 0
\end{array}\right)\qquad{r=1,\dots,4\atop s=1,\dots,5}
\end{equation}
The sectors of the Hilbert space are products of the irreducible representations
of the left and right moving Virasoro algebras which can be specified
by giving their left and right conformal weights as 
\begin{equation}
\mathcal{S}_{h,\bar{h}}=\mathcal{V}_{h}\otimes\mathcal{V}_{\bar{h}}
\end{equation}
There are two possible conformal field theory partition functions
for this value of the central charge \cite{Cappelli:1986hf}. The
one describing the Potts model is the $D_{4}$ modular invariant,
for which the complete Hilbert space is 
\begin{eqnarray}
\mathcal{H} & = & \mathcal{S}_{0,0}\oplus\mathcal{S}_{\frac{2}{5},\frac{2}{5}}\oplus\mathcal{S}_{\frac{7}{5},\frac{7}{5}}\oplus\mathcal{S}_{3,3}\nonumber \\
 &  & \oplus\mathcal{S}_{\frac{1}{15},\frac{1}{15}}^{+}\oplus\mathcal{S}_{\frac{1}{15},\frac{1}{15}}^{-}\oplus\mathcal{S}_{\frac{2}{3},\frac{2}{3}}^{+}\oplus\mathcal{S}_{\frac{2}{3},\frac{2}{3}}^{-}\nonumber \\
 &  & \oplus\mathcal{S}_{\frac{2}{5},\frac{7}{5}}\oplus\mathcal{S}_{\frac{7}{5},\frac{2}{5}}\oplus\mathcal{S}_{0,3}\oplus\mathcal{S}_{3,0}\label{eq:Dinvariantsectors}
\end{eqnarray}
The $D_{4}$ conformal field theory is invariant under the permutation
group $\mathbb{S}_{3}$ generated by two elements $\mathcal{Z}$ and
$\mathcal{C}$ with the relations
\begin{equation}
\mathcal{Z}^{3}=1\qquad\mathcal{C}^{2}=1\qquad\mathcal{CZC}=\mathcal{Z}^{-1}
\end{equation}
which have the signatures
\begin{equation}
\mbox{sign }\mathcal{Z}=+1\qquad\mbox{sign }\mathcal{C}=-1
\end{equation}
The sectors on the first line of (\ref{eq:Dinvariantsectors}) are
invariant under the action of the permutation group $\mathbb{S}_{3}$,
the two pairs on the second line each form the two-dimensional irreducible
representation, which is characterized by the following action of
the generators:
\begin{eqnarray}
\mathcal{C}|\pm\rangle & = & |\mp\rangle\nonumber \\
\mathcal{Z}|\pm\rangle & = & \mathrm{e}^{\pm\frac{2\pi i}{3}}|\pm\rangle
\end{eqnarray}
while the ones on the third line form the one-dimensional signature
representation where each element is represented by its signature.
These sectors are in one-to-one correspondence with the families of
conformal fields, and the primary field (the one with the lowest conformal
weight) in the family corresponding to $\mathcal{S}_{h,\bar{h}}$
has left and right conformal weights $h$ and $\bar{h}$; they are
denoted $\Phi_{h,\bar{h}}$ with an optional upper $\pm$ index for
fields forming a doublet of $\mathbb{S}_{3}$. In a family all other
fields have conformal weights that differ from those of the primary
by natural numbers. The conformal spin $s=h-\bar{h}$ gives the behaviour
under spatial translations; translational invariant fields must be
spinless i.e. $h=\bar{h}$.

The only $\mathbb{S}_{3}$-invariant spinless relevant field is
\begin{equation}
\Phi_{\frac{2}{5},\frac{2}{5}}
\end{equation}
which means that the Hamiltonian of the scaling limit of the off-critical
Potts model is uniquely determined \cite{Dotsenko:1984if}
\begin{equation}
H=H_{CFT}+\mu\int dx\Phi_{\frac{2}{5},\frac{2}{5}}\label{eq:scaling_potts_hamiltonian}
\end{equation}
where the dimensionful coupling $\mu$ is a scaled version of the
distance $g-1$ from the critical point of the spin chain (\ref{eq:potts_chain_hamiltonian}).
The sign of the coupling constant corresponds to the two phases: $\mu>0$
is the paramagnetic, while $\mu<0$ is the ferromagnetic phase. 

In the paramagnetic phase, the vacuum is non-degenerate and the spectrum
consists of a pair of particles $A$ and $\bar{A}$ of mass $m$ which
form a doublet under $\mathbb{S}_{3}$ \cite{Smirnov:1991uw}: 
\begin{eqnarray}
\mathcal{C}|A(\theta)\rangle=|\bar{A}(\theta)\rangle & \qquad & \mathcal{Z}|A(\theta)\rangle=\mathrm{e}^{\frac{2\pi i}{3}}|A(\theta)\rangle\nonumber \\
\mathcal{C}|\bar{A}(\theta)\rangle=|A(\theta)\rangle & \qquad & \mathcal{Z}|\bar{A}(\theta)\rangle=\mathrm{e}^{-\frac{2\pi i}{3}}|\bar{A}(\theta)\rangle\label{eq:s3action-on-aabar}
\end{eqnarray}
The mass $m$ is expressed with the coupling $\mu$ via the relation
\cite{Fateev:1993av}
\begin{eqnarray}
\mu & = & \kappa m^{6/5}\nonumber \\
 &  & \kappa=\frac{\Gamma\left(\frac{3}{10}\right)\left[\Gamma\left(\frac{2}{3}\right)\Gamma\left(\frac{5}{6}\right)\right]^{6/5}}{4\times2^{1/5}\pi^{8/5}\Gamma\left(\frac{7}{10}\right)}\sqrt{\frac{\Gamma\left(-\frac{1}{5}\right)\Gamma\left(\frac{7}{5}\right)}{\Gamma\left(-\frac{2}{5}\right)\Gamma\left(\frac{6}{5}\right)}}=0.1643033\dots\label{eq:massgap_relation}
\end{eqnarray}
The generator $\mathcal{C}$ is identical to charge conjugation ($\bar{A}$
is the antiparticle of $A$). Choosing units in which $\hbar=c=1$,
two-dimensional Lorentz invariance implies that the energy and momentum
of the particles can be parametrized by the rapidity $\theta$:
\begin{equation}
E=m\cosh\theta\quad p=m\sinh\theta
\end{equation}
The two-particle scattering amplitudes are
\begin{eqnarray}
S_{AA}(\theta_{12}) & = & S_{\bar{A}\bar{A}}(\theta_{12})=\frac{\sinh\left(\frac{\theta_{12}}{2}+\frac{\pi i}{3}\right)}{\sinh\left(\frac{\theta_{12}}{2}-\frac{\pi i}{3}\right)}\nonumber \\
S_{A\bar{A}}(\theta_{12}) & = & S_{\bar{A}A}(\theta_{12})=-\frac{\sinh\left(\frac{\theta_{12}}{2}+\frac{\pi i}{6}\right)}{\sinh\left(\frac{\theta_{12}}{2}-\frac{\pi i}{6}\right)}\label{eq:highT_Smatrix}
\end{eqnarray}
where $\theta_{12}=\theta_{1}-\theta_{2}$ is the rapidity difference
of the incoming particles. This $S$ matrix was confirmed by thermodynamic
Bethe Ansatz \cite{Zamolodchikov:1989cf}. We remark that the pole
in the $S_{AA}=S_{\bar{A}\bar{A}}$ amplitudes at 
\begin{equation}
\theta_{12}=\frac{2\pi i}{3}
\end{equation}
corresponds to the interpretation of particle $\bar{A}$ as a bound
state of two particles $A$ and similarly $A$ as a bound state of
two $\bar{A}$s, under the bootstrap principle (a.k.a. ``nuclear
democracy''). Accordingly, the above amplitudes satisfy the bootstrap
relations
\begin{eqnarray}
S_{A\bar{A}}(\theta) & = & S_{AA}(\theta+\pi i/3)S_{AA}(\theta-\pi i/3)\nonumber \\
S_{AA}(\theta) & = & S_{A\bar{A}}(\theta+\pi i/3)S_{A\bar{A}}(\theta-\pi i/3)
\end{eqnarray}
The pole in $S_{A\bar{A}}=S_{\bar{A}A}$ amplitudes at
\begin{equation}
\theta_{12}=\frac{\pi i}{3}
\end{equation}
has the same interpretation, but in the crossed channel.

The excitations in the ferromagnetic phase are topologically charged
\cite{Chim1992}. The vacuum is three-fold degenerate
\begin{equation}
|0\rangle_{a}\qquad a=-1,0,1
\end{equation}
where the action of $\mathbb{S}_{3}$ is 
\begin{equation}
\mathcal{Z}|0\rangle_{a}=|0\rangle_{a+1\bmod3}\qquad\mathcal{C}|0\rangle_{a}=|0\rangle_{-a}
\end{equation}
and the excitations are kinks of mass $m$ interpolating between adjacent
vacua. The kink of rapidity $\theta$, interpolating from $a$ to
$b$ is denoted by 
\begin{equation}
K_{ab}(\theta)\qquad a-b=\pm1\bmod3
\end{equation}
 and can be interpreted as a spin flip up/down (depending on the sign).
The scattering processes of the kinks are of the form 
\begin{equation}
K_{ab}(\theta_{1})+K_{bc}(\theta_{2})\rightarrow K_{ad}(\theta_{1})+K_{dc}(\theta_{2})
\end{equation}
with the scattering amplitudes equal to
\begin{equation}
S\left(a{d\atop b}c\right)(\theta_{12})=\begin{cases}
S_{AA}(\theta_{12})\quad\mbox{if}\quad b & =d\\
S_{A\bar{A}}(\theta_{12})\quad\mbox{if}\quad a & =c
\end{cases}\label{eq:lowT_Smatrix}
\end{equation}
This essentially means that apart from the restriction of kink succession
dictated by the vacuum indices (adjacency rules) the following identifications
can be made 
\begin{eqnarray}
K_{ab}(\theta) & \equiv & A(\theta)\qquad a-b=+1\bmod3\nonumber \\
K_{ab}(\theta) & \equiv & \bar{A}(\theta)\qquad a-b=-1\bmod3\label{eq:para_ferro_ident}
\end{eqnarray}
in all other relevant physical aspects (such as the bound state interpretation
given above). 

By looking at the conformal fusion rules implied by the three-point
couplings \cite{Fuchs:1989kz,Petkova:1988yf,Petkova:1994zs}, it turns
out that the perturbing operator acts separately in the following
four sectors:
\begin{eqnarray}
\mathcal{H}_{0} & = & \mathcal{S}_{0,0}\oplus\mathcal{S}_{\frac{2}{5},\frac{2}{5}}\oplus\mathcal{S}_{\frac{7}{5},\frac{7}{5}}\oplus\mathcal{S}_{3,3}\nonumber \\
\mathcal{H}_{\pm} & = & \mathcal{S}_{\frac{1}{15},\frac{1}{15}}^{\pm}\oplus\mathcal{S}_{\frac{2}{3},\frac{2}{3}}^{\pm}\nonumber \\
\mathcal{H}_{1} & = & \mathcal{S}_{\frac{2}{5},\frac{7}{5}}\oplus\mathcal{S}_{\frac{7}{5},\frac{2}{5}}\oplus\mathcal{S}_{0,3}\oplus\mathcal{S}_{3,0}\label{eq:Hilbert_space_sectors}
\end{eqnarray}
so the Hamiltonian can be diagonalized separately in each of them.
It is also the case that the Hamiltonian is exactly identical in the
sectors $\mathcal{H}_{+}$ and $\mathcal{H}_{-}$. The reason for
this is that the charge conjugation symmetry $\mathcal{C}$ acts on
the sectors $\mathcal{H}_{+}$ and $\mathcal{H}_{-}$ by swapping
them. The subspaces $\mathcal{H}_{0}$ and $\mathcal{H}_{1}$ contain
neutral states (i.e. invariant under $\mathcal{Z}$), which are even/odd
under the action of $\mathcal{C}$, respectively.

The spectrum is invariant under $\mu\rightarrow-\mu$ in sectors $\mathcal{H}_{0}$
and \emph{$\mathcal{H}_{1}$}. The latter fact is the consequence
of a $\mathbb{Z}_{2}$ symmetry in these sectors under which the parities
in $\mathcal{H}_{0}$ are 
\begin{eqnarray}
\mbox{even} & : & \mathcal{S}_{0,0}\quad\mathcal{S}_{\frac{7}{5},\frac{7}{5}}\nonumber \\
\mbox{odd} & : & \mathcal{S}_{\frac{2}{5},\frac{2}{5}}\quad\mathcal{S}_{3,3}
\end{eqnarray}
and so the perturbing operator $\Phi_{\frac{2}{5},\frac{2}{5}}$ is
odd. In $\mathcal{H}_{1}$ this $\mathbb{Z}_{2}$ acts by swapping
\begin{equation}
\mathcal{S}_{0,3}\leftrightarrow\mathcal{S}_{3,0}\qquad\mathcal{S}_{\frac{2}{5},\frac{7}{5}}\leftrightarrow\mathcal{S}_{\frac{7}{5},\frac{2}{5}}
\end{equation}
This symmetry leaves the fixed point Hamiltonian $H_{*}$ and the
conformal operator product expansion (OPE)%
\footnote{Cf. Appendix \ref{sub:Structure-constants}.%
} in these sectors invariant%
\footnote{The conformal fusion rules do not allow the extension of this symmetry
to the $\mathcal{H}_{\pm}$. %
}; away from the critical point, it can be interpreted as the realization
of the well-known low/high-temperature (Kramers-Wannier) duality at
the level of the scaling field theory.

\section{Renormalization in TCSA \label{sec:Renormalization-in-TCSA}}

\subsection{General theory of cut-off dependence \label{sub:General-theory-of-cut-off-dependence}}

First we recall the derivation of (\ref{eq:cpt_scaling_function})
using standard time-independent perturbation theory. Taking a Hamiltonian
of the form 
\begin{equation}
H=H_{0}+\lambda V
\end{equation}
where $H_{0}$ has the following spectrum
\begin{equation}
H_{0}|n\rangle=E_{n}^{(0)}|n\rangle
\end{equation}
and supposing that the ground state is non-degenerate, the ground
state energy in the perturbed model can be written as
\begin{equation}
E_{0}(\lambda)=E_{0}^{(0)}+\lambda\langle0|V|0\rangle+\lambda^{2}\sum_{k\neq0}\frac{\langle0|V|k\rangle\langle k|V|0\rangle}{E_{0}^{(0)}-E_{k}^{(0)}}+O(\lambda^{3})
\end{equation}
Using the Schwinger representation for the energy denominators, the
second order can be written as 
\begin{eqnarray}
\sum_{k\neq0}\frac{\langle0|V|k\rangle\langle k|V|0\rangle}{E_{0}^{(0)}-E_{k}^{(0)}} & = & -\sum_{k\neq0}\langle0|V|k\rangle\int_{0}^{\infty}d\tau e^{-(E_{k}^{(0)}-E_{0}^{(0)})\tau}\langle k|V|0\rangle=-\int_{0}^{\infty}d\tau\sum_{k\neq n}\langle n|e^{\tau H_{0}}Ve^{-\tau H_{0}}|k\rangle\langle k|V|n\rangle\nonumber \\
 & = & -\int_{0}^{\infty}d\tau\langle0|V(\tau)V(0)|0\rangle_{conn}=-\frac{1}{2}\int_{-\infty}^{\infty}d\tau\langle n|\mathcal{T}V(\tau)V(0)|n\rangle_{conn}
\end{eqnarray}
where the integral representation is valid due to $E_{k}^{(0)}>E_{0}^{(0)}$;
this is where the restriction to ground state appears. We obtain
\begin{equation}
E_{0}(\lambda)=E_{0}^{(0)}+\lambda\langle0|V|0\rangle-\frac{\lambda^{2}}{2}\int_{-\infty}^{\infty}d\tau\langle n|\mathcal{T}V(\tau)V(0)|n\rangle_{conn}+O(\lambda^{3})
\end{equation}
We note that the all order expansion is 
\begin{equation}
E_{0}=E_{0}^{(0)}-\sum_{n=1}^{\infty}\frac{(-\lambda)^{n}}{n!}\int d\tau_{1}\dots\int d\tau_{n-1}\langle0|\mathcal{T}V(\tau_{1})\dots V(\tau_{n-1})V(0)|0\rangle_{conn}\label{eq:allorder_pt}
\end{equation}
which can be proven by writing the ground state energy as the limit
\cite{Klassen:1990dx}
\begin{equation}
E_{0}=E_{0}^{(0)}+\lim_{T\rightarrow\infty}\left\{ -\frac{1}{T}\log\langle0|\mathcal{T}\exp\left(-\lambda\int_{-T/2}^{T/2}d\tau V(\tau)\right)|0\rangle\right\} 
\end{equation}
which is a standard trick using that asymptotically long Euclidean
time evolution projects onto the exact ground state, this time employed
in the interaction picture where the Hamiltonian is $V(\tau)$. Expanding
the exponential in terms of time-ordered multi-point functions, the
logarithm replaces them by their connected parts, and finally one
of the time integrals cancels with the prefactor $1/T$ due to time
translation invariance, resulting in (\ref{eq:allorder_pt}). Substituting
\begin{equation}
\lambda V(\tau)=\mu\int_{0}^{R}dx\Phi(\tau,x)
\end{equation}
and using space translation invariance leads to (\ref{eq:cpt_scaling_function})\@.

Let us now introduce a projection operator $P_{\Lambda}$, which projects
to states with unperturbed energy $E_{n}^{(0)}\leq\Lambda$. We can
split the Hamiltonian into a low and a high energy part writing
\begin{eqnarray}
H & = & H_{\Lambda}+\lambda\Delta V\nonumber \\
 &  & H_{\Lambda}=H_{0}+\lambda P_{\Lambda}VP_{\Lambda}\nonumber \\
 &  & \Delta V=V-P_{\Lambda}VP_{\Lambda}=P_{\Lambda}V\bar{P}_{\Lambda}+\bar{P}_{\Lambda}VP_{\Lambda}+\bar{P}_{\Lambda}V\bar{P}_{\Lambda}
\end{eqnarray}
where $\bar{P}_{\Lambda}=1-P_{\Lambda}$ is the projector to states
above the cut-off. Now suppose $E_{n}^{(0)}<\Lambda$ and write 
\begin{eqnarray}
E_{n} & = & E_{n}^{(0)}+\lambda V_{nn}+\lambda^{2}\sum_{k\neq n}\frac{V_{nk}V_{kn}}{E_{n}^{(0)}-E_{k}^{(0)}}+O(\lambda^{3})
\end{eqnarray}
Summing up all terms in the perturbation series with intermediate
states below $\Lambda$ produces the eigenvalue $E_{n}(\Lambda)$
of $H_{\Lambda}$. For the contribution of higher order states we
keep only the second order corrections:
\begin{equation}
E_{n}=E_{n}(\Lambda)+\lambda^{2}\sum_{E_{k}^{(0)}>\Lambda}\frac{V_{nk}V_{kn}}{E_{n}^{(0)}-E_{k}^{(0)}}+O(\lambda^{3})
\end{equation}
Since $E_{n}^{(0)}\leq\Lambda<E_{k}^{(0)}$, we can use the Schwinger
proper time representation to obtain
\begin{eqnarray}
E_{n} & = & E_{n}(\Lambda)-\lambda^{2}\int_{0}^{\infty}d\tau\sum_{E_{k}^{(0)}>\Lambda}\langle n|V(\tau)|k\rangle\langle k|V(0)|n\rangle+O(\lambda^{3})\nonumber \\
 & = & E_{n}(\Lambda)-\lambda^{2}\int_{0}^{\infty}d\tau\langle n|V(\tau)\bar{P}_{\Lambda}V(0)|n\rangle+O(\lambda^{3})
\end{eqnarray}
which describes the cut-off dependence to second order in $\lambda$.
It is obvious that this derivation can be systematically extended
to higher orders as well. It is also possible to write down the difference
between the energy levels computed with cut-offs $\Lambda$ and $\Lambda+\Delta\Lambda$
in the form
\begin{equation}
E_{n}(\Lambda+\Delta\Lambda)-E_{n}(\Lambda)=-\lambda^{2}\int_{0}^{\infty}d\tau\langle n|V(\tau)\tilde{P}_{\Lambda,\Delta\Lambda}V(0)|n\rangle+O(\lambda^{3})
\end{equation}
where 
\begin{equation}
\tilde{P}_{\Lambda,\Delta\Lambda}=\bar{P}_{\Lambda}-\bar{P}_{\Lambda+\Delta\Lambda}=P_{\Lambda+\Delta\Lambda}-P_{\Lambda}
\end{equation}
is the projector to states with 
\begin{equation}
\Lambda<E_{n}^{(0)}<\Lambda+\Delta\Lambda
\end{equation}

\subsection{Cut-off dependence in TCSA\label{sub:Cut-off-dependence-in}}

\subsubsection{The truncated conformal space approach\label{sub:The-truncated-conformal}}

In conformal field theory with periodic boundary conditions the Hilbert
space can be written as
\begin{equation}
\mathcal{H}=\bigoplus_{k}\mathcal{V}_{h_{k}}\otimes\mathcal{V}_{\bar{h}_{k}}
\end{equation}
where the $\mathcal{V}_{h}$ are irreducible representations of the
Virasoro algebra. The basis of a conformal module $\mathcal{V}_{h_{k}}\otimes\mathcal{V}_{\bar{h}_{k}}$
is spanned by vectors $\left|k,\{n,\bar{n}\},\alpha\right\rangle $
which satisfy
\begin{eqnarray*}
L_{0}\left|k,\{n,\bar{n}\},\alpha\right\rangle  & = & (h_{k}+n)\left|k,\{n,\bar{n}\},\alpha\right\rangle \\
\bar{L}_{0}\left|k,\{n,\bar{n}\},\alpha\right\rangle  & = & (\bar{h}_{k}+\bar{n})\left|k,\{n,\bar{n}\},\alpha\right\rangle 
\end{eqnarray*}
so that $n$ and $\bar{n}$ denote the left and right descendant numbers,
and $\alpha$ indexes the independent vectors at the same level $(n,\bar{n})$.
The momentum operator is
\begin{equation}
P=\frac{2\pi}{R}(L_{0}-\bar{L}_{0})
\end{equation}
and the eigenvalue of $L_{0}-\bar{L}_{0}$ is the conformal spin.

The basic idea behind TCSA is to truncate the conformal Hilbert space
at some level $n$, which plays the role of the ultraviolet cut-off
parameter $\Lambda$. Using the machinery of conformal field theory
one then computes explicitly the matrix elements of the dimensionless
Hamiltonian (\ref{eq:dimless_pcft_hamiltonian}) restricted to the
truncated conformal space:
\begin{eqnarray}
h_{n}(r) & = & \frac{2\pi}{r}e_{n}(r)\nonumber \\
e_{n}(r) & = & L_{0}+\bar{L}_{0}-\frac{c}{12}+\lambda\, P_{n}\mathcal{V}P_{n}\label{eq:TCSA_hamiltonian}
\end{eqnarray}
and compute its spectrum by numerical diagonalization. The eigenvalue
$e_{\Psi,n}(r)$ of the operator $e_{n}(r)$ corresponding to a given
energy level $\Psi$ is the scaling function of the corresponding
state with truncation $n$, and the truncation is represented explicitly
by $P_{n}$, which is the projector to the subspace with states having
descendant level less or equal than $n$. 

To obtain the interaction matrix $\mathcal{V}$, it is necessary to
take into account that the natural bases of conformal modules are
not orthonormal. Denoting the metric on the conformal Hilbert space
by 
\begin{equation}
G_{ij}=\langle i|j\rangle
\end{equation}
the interaction matrix elements are
\begin{eqnarray}
\mathcal{V}_{ij} & = & \sum_{k}\left(G^{-1}\right)_{ik}B_{kj}\label{eq:conformal_metric_in_perturbation}\\
B_{ij} & = & \langle i|\Phi(z,\bar{z})|j\rangle|_{z=\bar{z}=1}\delta_{s_{i}s_{j}}\nonumber 
\end{eqnarray}
where $s_{i,j}$ are the conformal spins of the states $|i\rangle$
and $|j\rangle$, the selection rule resulting from the integration
of the perturbing field over the volume. The matrix elements of $B$
between primary states are given in (\ref{eq:opesect1},\ref{eq:opesect2},\ref{eq:opesect4});
for descendant states they can be constructed from the primary ones
by a recursive application of the conformal Ward identities (\ref{eq:conformal_ward_identities}).
To describe the dependence on the cut-off, we implement the procedure
introduced in subsection \ref{sub:General-theory-of-cut-off-dependence}.
The method we follow is the eventual basis of the TCSA renormalization
group method introduced in \cite{Feverati:2006ni} (see also \cite{Konik:2007cb}),
which was applied to theories on the cylinder by Giokas and Watts
\cite{Giokas:2011ix} for the case when the operator product expansion
of $\Phi(z)\Phi(0)$ contains only the identity and $\Phi$ together
with their descendants. Since the scaling Potts model is not in this
class, and also for the sake of later applications, we give a review
of the formalism below.

\subsubsection{General theory of counter terms in TCSA\label{sub:General-theory-of-counter-terms}}

The TCSA provides a non-perturbative tool to handle perturbed conformal
field theories, and the aim of the TCSA renormalization procedure
is to speed up the convergence of the method and also to deal with
ultraviolet divergences when necessary. Since the perturbation is
supposed to be relevant, the running coupling flows to zero at high
energies. As a result, the influence of the high energy degrees of
freedom can be treated perturbatively. Suppose we consider a quantity
$Q$, for which TCSA with a cut-off at level $n$ gives $Q_{TCSA}\left(n\right)$
and let us write the exact value as follows:
\begin{equation}
Q=Q_{TCSA}\left(n\right)+\delta Q\left(n\right)
\end{equation}
where $\delta Q\left(n\right)$ is a counter term which can either
go to zero (in the convergent case) or even be divergent when $n$
increases. The counter term can be constructed by computing the contribution
$Q_{l}$ of the $l$th level 
\begin{eqnarray}
Q & = & Q_{TCSA}\left(n\right)+\sum_{l=n+1}^{\infty}Q_{l}\nonumber \\
 & = & Q_{TCSA}\left(n\right)+\sum_{l=1}^{\infty}Q_{l}-\sum_{l=1}^{n}Q_{l}\label{eq:gen_CT_derivation}
\end{eqnarray}
therefore the counter term can be written as
\begin{equation}
\delta Q\left(n\right)=\sum_{l=1}^{\infty}Q_{l}-\sum_{l=1}^{n}Q_{l}\label{eq:gen_CT}
\end{equation}
Depending on the weight $h$ of the perturbation, the first sum on
the second line can be either convergent or divergent. In the divergent
case it is necessary to use an appropriate regularization method and
renormalization scheme. For example, in integrable field theories
we usually compare our results to predictions of the exact $S$-matrix
or form factors resulting from the bootstrap, or to scaling functions
predicted by the thermodynamic Bethe Ansatz. In many cases, the model
we investigate is just a member of a family of perturbed CFTs, where
$h$ varies across the range of possible theories, and the exact predictions
depend analytically on $h$. Therefore the relevant scheme is provided
by analytically continuation from the range of parameter space where
the theory is ultraviolet finite ($h<1/2$).

\subsubsection{Counter terms for scaling functions\label{sub:Counter-terms-for-scaling-functions}}

If we choose our quantity $Q$ as a finite volume energy level $E_{\Psi,n}(R)$
and substitute
\begin{equation}
\lambda V(\tau)=\mu\int_{0}^{R}dx\Phi(\tau,x)
\end{equation}
we obtain 
\begin{eqnarray}
E_{\Psi,n}(R)-E_{\Psi,n-1}(R) & = & -\mu^{2}\int_{0}^{R}dx\int_{0}^{R}dx'\int_{0}^{\infty}d\tau\langle\Psi|\Phi(\tau,x)\tilde{P}_{n}\Phi(0,x')|\Psi\rangle_{CFT}+O(\mu^{3})\nonumber \\
 & = & -\mu^{2}R\int_{0}^{R}dx\int_{0}^{\infty}d\tau\langle\Psi|\Phi(\tau,x)\tilde{P}_{n}(\Psi)\Phi(0,0)|\Psi\rangle_{CFT}+O(\mu^{3})
\end{eqnarray}
where $\tilde{P}_{n}$ is the projector on states at level $n$, and
we used translation invariance to eliminate one spatial integral,
which in turn restricts the intermediate states to ones which have
the same Lorentz momentum (or conformal spin) as $\Psi$; the corresponding
restricted projector is denoted by $\tilde{P}_{n}(\Psi)$. Passing
to the scaling function we obtain
\begin{equation}
e_{\Psi,n}(r)-e_{\Psi,n-1}(r)=-\frac{\mu^{2}}{2\pi}\int_{0}^{R}dx\int_{0}^{\infty}d\tau\langle\Psi|\Phi(\tau,x)\tilde{P}_{n}(\Psi)\Phi(0,0)|\Psi\rangle_{CFT}+O(\mu^{3})
\end{equation}
Mapping this expression on the conformal plane we finally obtain
\begin{equation}
e_{\Psi,n}(r)-e_{\Psi,n-1}(r)=-\frac{\lambda^{2}}{2\pi}\int_{\left|z\right|<1}d^{2}z\left(z\bar{z}\right)^{h-1}\langle\Psi|\Phi(1,1)\tilde{P}_{n}(\Psi)\Phi(z,\bar{z})|\Psi\rangle_{CFT}+\mathcal{O}\left(\lambda^{3}\right)\label{eq:second_order_level_contrib}
\end{equation}
It is clear that in order to evaluate the counter term it is necessary
to construct the contribution of a given level $n$ to the conformal
correlators.

\subsubsection{Evaluating the level contributions\label{sub:Evaluating-the-level-contributions}}

As pointed out in \cite{Beria2013}, the most systematic way to obtain
it is by considering the Kallen-Lehmann spectral representation. In
general the unperturbed state $|\Psi\rangle$ can be written as a
linear combination of conformal states; this is necessary to allow
for degenerate perturbation theory, which is relevant due to the high
degeneracy in the conformal Hilbert space. Therefore we consider the
two-point function of the perturbation between two conformal states
$|i,\{n_{i},\bar{n}_{i}\},\alpha_{i}\rangle$ and $|j,\{n_{j},\bar{n}_{j}\},\alpha_{j}\rangle$,
which are from conformal modules with conformal weights $(h_{i},\bar{h}_{i})$
and $(h_{j},\bar{h}_{j})$ and have descendant levels $(n_{i},\bar{n}_{i})$
and $(n_{j},\bar{n}_{j})$; the $\alpha_{i,j}$ index a basis in the
conformal modules at the given level. Inserting a complete set of
states we obtain
\begin{eqnarray}
 &  & \langle i,\{n_{i},\bar{n}_{i}\},\alpha_{i}|\Phi\left(0,0\right)\Phi\left(\tau,x\right)|j,\{n_{j},\bar{n}_{j}\},\alpha_{j}\rangle\label{eq:spectral_expansion}\\
 &  & =\sum_{k,n,\alpha}\langle i,\{n_{i},\bar{n}_{i}\},\alpha_{i}|\Phi\left(0,0\right)|k,\{n,\bar{n}\},\alpha\rangle\langle k,\{n,\bar{n}\},\alpha|\Phi\left(\tau,x\right)|j,\{n_{j},\bar{n}_{j}\},\alpha_{j}\rangle\nonumber 
\end{eqnarray}
where the states $\left|k,\{n,\bar{n}\},\alpha\right\rangle $ form
an orthonormal basis of the conformal module with conformal weight
$\left(h_{k},\bar{h}_{k}\right)$ at descendant level $\left(n,\bar{n}\right)$.
Note that translational invariance (via the spatial integrals) enforces
\begin{equation}
h_{i}-\bar{h}_{i}+n_{i}-\bar{n}_{i}=h_{k}-\bar{h}_{k}+n-\bar{n}=h_{j}-\bar{h}_{j}+n_{j}-\bar{n}_{j}
\end{equation}
in the matrix elements that contribute to (\ref{eq:second_order_level_contrib}).
Using the space-time translation operator $e^{-H\tau-iPx}$ we can
write:
\begin{eqnarray}
 &  & \langle i,\{n_{i},\bar{n}_{i}\},\alpha_{i}|\Phi\left(0,0\right)\Phi\left(\tau,x\right)|j,\{n_{j},\bar{n}_{j}\},\alpha_{j}\rangle\nonumber \\
 &  & =\sum_{k,n,\alpha}\langle i,\{n_{i},\bar{n}_{i}\},\alpha_{i}|\Phi\left(0,0\right)|k,\{n,\bar{n}\},\alpha\rangle\langle k,\{n,\bar{n}\},\alpha|e^{H\tau+iPx}\Phi\left(0,0\right)e^{-H\tau-iPx}|j,\{n_{j},\bar{n}_{j}\},\alpha_{j}\rangle
\end{eqnarray}
Mapping to the complex plane using $z=e^{\frac{2\pi}{R}\left(\tau+ix\right)}$
one obtains
\begin{eqnarray}
\left(\frac{2\pi}{R}\right)^{2h}\left(z\bar{z}\right)^{h}\Phi\left(z,\bar{z}\right) & = & \left(\frac{2\pi}{R}\right)^{2h}e^{\frac{2\pi}{R}\left(L_{0}+\bar{L}_{0}-\frac{c}{12}\right)\tau+i\frac{2\pi}{R}\left(L_{0}-\bar{L}_{0}\right)x}\nonumber \\
 &  & \times\Phi\left(1,1\right)e^{-\frac{2\pi}{R}\left(L_{0}+\bar{L}_{0}-\frac{c}{12}\right)\tau-i\frac{2\pi}{R}\left(L_{0}-\bar{L}_{0}\right)x}
\end{eqnarray}
which gives 
\begin{equation}
\Phi\left(z,\bar{z}\right)=\left(z\bar{z}\right)^{-h}z^{L_{0}}\bar{z}^{\bar{L}_{0}}\Phi\left(1,1\right)z^{-L_{0}}\bar{z}^{-\bar{L}_{0}}
\end{equation}
Inserting this expression into (\ref{eq:spectral_expansion}) we obtain
\begin{eqnarray}
 &  & \langle i,\{n_{i},\bar{n}_{i}\},\alpha_{i}|\Phi\left(1,1\right)\Phi\left(z,\bar{z}\right)|j,\{n_{j},\bar{n}_{j}\},\alpha_{j}\rangle\nonumber \\
 &  & =\sum_{k,n,\alpha}\langle i,\{n_{i},\bar{n}_{i}\},\alpha_{i}|\Phi\left(1,1\right)|k,\{n,\bar{n}\},\alpha\rangle\langle k,\{n,\bar{n}\},\alpha|\Phi\left(1,1\right)|j,\{n_{j},\bar{n}_{j}\},\alpha_{j}\rangle\nonumber \\
 &  & \qquad\quad\times z^{h_{k}+n-h_{j}-n_{j}-h}\bar{z}^{\bar{h}_{k}+\bar{n}-\bar{h}_{j}-\bar{n}_{j}-h}
\end{eqnarray}
so the contribution of level $(n,\bar{n})$ from a given primary field
with conformal dimensions $h_{k},\bar{h}_{k}$ can be found by first
splitting the matrix element into conformal blocks
\begin{eqnarray}
 &  & \langle i,\{n_{i},\bar{n}_{i}\},\alpha_{i}|\Phi\left(1,1\right)\Phi\left(z,\bar{z}\right)|j,\{n_{j},\bar{n}_{j}\},\alpha_{j}\rangle\nonumber \\
 &  & =\sum_{k}\langle i,\{n_{i},\bar{n}_{i}\},\alpha_{i}|\Phi\left(1,1\right)\mathcal{P}_{k}\Phi\left(z,\bar{z}\right)|j,\{n_{j},\bar{n}_{j}\},\alpha_{j}\rangle
\end{eqnarray}
with $\mathcal{P}_{k}$ being the projector onto the conformal module
$\mathcal{V}_{h_{k}}\otimes\mathcal{V}_{\bar{h}_{k}}$, and then considering
the coefficient of the term $z^{n}\bar{z}^{\bar{n}}$ in the Taylor
expansion of the function 
\begin{equation}
z^{-(h_{k}-h_{j}-n_{j}-h)}\bar{z}^{-(\bar{h}_{k}-\bar{h}_{j}-\bar{n}_{j}-h)}\langle i,\{n_{i},\bar{n}_{i}\},\alpha_{i}|\Phi\left(1,1\right)\mathcal{P}_{k}\Phi\left(z,\bar{z}\right)|j,\{n_{j},\bar{n}_{j}\},\alpha_{j}\rangle
\end{equation}

\subsection{Constructing counter-terms to scaling functions}

\subsubsection{\label{sub:The-ground-state}The ground state scaling function}

For the ground state, the computations are simpler, since from (\eqref{cpt_scaling_function})
one can write an explicit formula for the contributions up to level
$n$ in the form
\begin{equation}
e_{0,n}\left(\lambda\right)=-\frac{c}{12}-\frac{\lambda^{2}}{2\pi}\int_{\left|z\right|<1}d^{2}z\left(z\bar{z}\right)^{h-1}\langle0|\Phi\left(1,1\right)P_{n}\Phi\left(z,\bar{z}\right)|0\rangle+\mathcal{O}\left(\lambda^{3}\right)
\end{equation}
Expanding the conformal two-point function into a binomial series
one obtains 
\begin{equation}
\langle0|\Phi\left(1,1\right)\Phi\left(z,\bar{z}\right)|0\rangle=\frac{1}{\left(1-z\right)^{2h}\left(1-\bar{z}\right)^{2\bar{h}}}=\sum_{m=0}^{\infty}\sum_{\bar{m}=0}^{\infty}\frac{\Gamma\left(2h+m\right)}{\Gamma\left(2h\right)\Gamma\left(m+1\right)}\frac{\Gamma\left(2h+\bar{m}\right)}{\Gamma\left(2h\right)\Gamma\left(\bar{m}+1\right)}z^{m}\bar{z}^{\bar{m}}
\end{equation}
Performing the angular integral selects the terms with $m=\bar{m}$
and in addition gives a factor of $2\pi$. Now using the spectral
expansion argument with $h_{i}=\bar{h}_{i}=h_{j}=\bar{h}_{j}=0$,
$h_{k}=\bar{h}_{k}=h$ and $n_{i}=\bar{n}_{i}=n_{j}=\bar{n}_{j}=0$,
the second order TCSA contribution to the ground state scaling function
coming from the $m$-th level is the following

\begin{eqnarray}
\tilde{e}_{0,m} & = & -\int_{0}^{1}dr\, r^{2h-1+2m}\left(\frac{\Gamma\left(2h+m\right)}{\Gamma\left(2h\right)\Gamma\left(m+1\right)}\right)^{2}\nonumber \\
 & = & -\frac{1}{2\left(h+m\right)}\left(\frac{\Gamma\left(2h+m\right)}{\Gamma\left(2h\right)\Gamma\left(m+1\right)}\right)^{2}\label{eq:ground_state_level_contrib}
\end{eqnarray}
With this the scaling function from TCSA truncated to level $n$ is
given by 
\begin{eqnarray}
e_{0,n}\left(\lambda\right) & = & -\frac{c}{12}-\sum_{m=1}^{n}\tilde{e}_{0,m}\lambda^{2}+\mathcal{O}\left(\lambda^{3}\right)\label{eq:vacuum_level_contrib}
\end{eqnarray}
The level $m$ contribution (\ref{eq:ground_state_level_contrib})
will be tested against TCSA in subsection \ref{sub:Level-contributions}.

\subsubsection{Determining the counter term\label{sub:Determining-the-counter-term}}

For large $m$ one can expand
\begin{eqnarray}
\tilde{e}_{0,m} & = & -\frac{1}{2\left(h+m\right)}\left(\frac{\Gamma\left(2h+m\right)}{\Gamma\left(2h\right)\Gamma\left(m+1\right)}\right)^{2}\nonumber \\
 & =- & \frac{1}{2\Gamma(2h)^{2}}m^{4h-3}-\frac{4h^{2}-3h}{2\Gamma(2h)^{2}}m^{4h-4}-\frac{24h^{4}-44h^{3}+21h^{2}-h}{6\Gamma(2h)^{2}}m^{4h-5}\nonumber \\
 &  & -\frac{32h^{6}-104h^{5}+116h^{4}-49h^{3}+5h^{2}}{6\Gamma(2h)^{2}}m^{4h-6}+\mathcal{O}\left(m^{4h-7}\right)
\end{eqnarray}
The summation up to the TCSA cut-off level $n$ can be performed using
\begin{equation}
\sum_{m=1}^{n}m^{\gamma}=H_{n,-\gamma}
\end{equation}
where $H_{n,-\gamma}$ is the so-called generalized harmonic number.
For large $n$ it has the expansion
\begin{equation}
H_{n,-\gamma}=\zeta\left(-\gamma\right)+\frac{n^{\gamma+1}}{\gamma+1}+\frac{n^{\gamma}}{2}+\frac{\gamma n^{\gamma-1}}{12}+\frac{\left(-\gamma^{3}+3\gamma^{2}-2\gamma\right)n^{\gamma-3}}{720}+\ldots
\end{equation}
Now considering the construction of the counter term as described
in (\ref{eq:gen_CT}), the first term $\zeta\left(-\gamma\right)$
cancels with the corresponding infinite sum term. So the counter term
for the ground state scaling function at level $n$ is given by 
\begin{eqnarray}
e_{0}(r) & = & e_{0,n}\left(r\right)+\delta e_{0,n}\left(r\right)+\mathcal{O}\left(\lambda^{3}\right)\label{eq:ground_state_counterterm}\\
\delta e_{0,n}\left(r\right) & = & \lambda^{2}n^{4h-2}\frac{1}{4\left(2h-1\right)\Gamma\left(2h\right)^{2}}+\lambda^{2}n^{4h-3}\left(\frac{1+2h}{4\Gamma\left(2h\right)^{2}}\right)+\lambda^{2}n^{4h-4}\left(\frac{24h^{3}+4h^{2}-13h-4}{24\Gamma(2h)^{2}}\right)+\dots\nonumber 
\end{eqnarray}
If the perturbing operator has dimension $h>1/2$ the first correction
to the counter term is divergent; more terms (involving also ones
which are of higher order in $\lambda$) become divergent as the weight
of the perturbation increases. In this case the above formula gives
a prescription to renormalize the divergent TCSA result.

\subsection{Excited states}

\subsubsection{Construction of counter terms in general\label{sub:Construction-of-counter-for-excited}}

The construction of the counter term to the scaling function of excited
states requires the level contributions of the correlator $\left\langle i\left|\Phi\left(1,1\right)\Phi\left(z,\bar{z}\right)\right|i\right\rangle $.
For simplicity let us first suppose that the state $\left|i\right\rangle $
is a highest weight vector; then the correlator can be written in
terms of left and right chiral conformal blocks:
\begin{equation}
\langle i|\Phi\left(1,1\right)\Phi\left(z,\bar{z}\right)|i\rangle=\sum_{j}\left(C_{i\Phi}^{j}\right)^{2}\mathcal{F}_{ii}^{\phi\phi}\left(j|z\right)\bar{\mathcal{F}}_{ii}^{\bar{\phi}\bar{\phi}}\left(j|\bar{z}\right)
\end{equation}
where 
\begin{equation}
\mathcal{F}_{ii}^{\phi\phi}\left(j|z\right)\bar{\mathcal{F}}_{ii}^{\phi\phi}\left(j|\bar{z}\right)=\sum_{|k\rangle\in\mathcal{V}_{h_{j}}\otimes\mathcal{V}_{\bar{h}_{j}}}\langle i|\Phi\left(1,1\right)|k\rangle\langle k|\Phi\left(z,\bar{z}\right)|i\rangle
\end{equation}
the small $\phi$ refers to the chiral component (the perturbation
has the same left and right moving weight, therefore they are identical),
and 
\begin{equation}
C_{i\Phi}^{j}=\langle j|\Phi\left(1,1\right)|i\rangle
\end{equation}
is the CFT structure constant. From \ref{sub:Evaluating-the-level-contributions},
the contribution of level $\left(n,\bar{n}\right)$ comes from the
coefficient of 
\begin{equation}
z^{h_{j}+n-h_{i}-n_{i}-h}\bar{z}^{\bar{h}_{j}+\bar{n}-\bar{h}_{i}-\bar{n}_{i}-h}
\end{equation}
In principle, this coefficient can be evaluated using the Virasoro
symmetry for $\mathcal{F}_{ii}^{\Phi\Phi}\left(j|z\right)$: the lowest
order coefficient is by convention normalized to one, and coefficients
of subsequent powers can be computed using the conformal Ward identities
(\ref{eq:conformal_ward_identities}) to evaluate descendant matrix
elements in terms of primary ones. However, this gives a recursive
method from which it is very hard to extract the large $n$ behaviour
of the coefficients, which is necessary for the explicit construction
of the counter terms. 

An alternative method that leads to a systematic large $n$ expansion
of the required coefficients is the following \cite{Giokas:2011ix}.
First we expand the conformal blocks in the dual channel (i.e. in
terms of $1-z$) using the duality relations
\begin{equation}
\mathcal{F}_{ij}^{kl}\left(p|z\right)=\sum_{q}F_{pq}\left[\begin{array}{cc}
k & l\\
i & j
\end{array}\right]\mathcal{F}_{ij}^{kl}\left(q|1-z\right)
\end{equation}
where the $F$ are the so-called fusion coefficients. With the following
pictorial notation
\begin{eqnarray}
\mathcal{F}_{ij}^{kl}\left(p|z\right)=\blockZgen{i}{j}{k}{l}{p} & \qquad & \mathcal{F}_{ij}^{kl}\left(q|1-z\right)=\blockOnemZgen{i}{j}{k}{l}{q}\nonumber \\
\end{eqnarray}
one can write

\begin{eqnarray}
\blockZgen{i}{i}{\phi}{\phi}{j} & = & \sum_{k}F_{jk}\left[\begin{array}{cc}
\phi & \phi\\
i & i
\end{array}\right]\blockOnemZgen{i}{i}{\phi}{\phi}{k}\nonumber \\
\label{eq:duality_maintex}
\end{eqnarray}
The chiral conformal blocks in the dual channel can be expanded as
\begin{eqnarray}
\blockOnemZgen{i}{j}{k}{l}{p} & = & \left(1-z\right)^{-h_{k}-h_{l}+h_{p}}\sum_{r=0}^{\infty}B_{r}\left[\begin{array}{cc}
k & l\\
i & j
\end{array}p\right]\left(1-z\right)^{r}\label{eq:dual_channel_block_expansion}\\
\nonumber 
\end{eqnarray}
with $B_{0}\left[\begin{array}{cc}
k & l\\
i & j
\end{array}p\right]=1$, and the rest of the coefficients $B_{r}$ determined by Virasoro
symmetry via the Ward identities (\ref{eq:conformal_ward_identities}).
Introducing the shorthands $\mathcal{F}_{jk}(i)=F_{jk}\left[\begin{array}{cc}
\phi & \phi\\
i & i
\end{array}\right]$ and $B_{r}(i,k)=B_{r}\left[\begin{array}{cc}
\phi & \phi\\
i & i
\end{array}k\right]$ we can write
\begin{eqnarray}
\left\langle i\left|\Phi\left(1,1\right)\Phi\left(z,\bar{z}\right)\right|i\right\rangle  & = & \sum_{j}\left(C_{i\Phi}^{j}\right)^{2}\left(\blockZgen{i}{i}{\phi}{\phi}{j}\right)\left(\blockZgen{\bar{i}}{\bar{i}}{\phi}{\phi}{\bar{j}}\right)\nonumber \\
\label{eq:getlevelcontrib_expansion}\\
 & = & \sum_{j}\left(C_{i\Phi}^{j}\right)^{2}\sum_{k}\mathcal{F}_{jk}(i)\left(\blockOnemZgen{i}{i}{\phi}{\phi}{k}\right)\sum_{k'}\mathcal{F}_{\bar{j}k'}(\bar{i})\left(\blockOnemZgen{\bar{i}}{\bar{i}}{\phi}{\phi}{k'}\right)\nonumber \\
\nonumber \\
 & = & \sum_{j,k,k'}\left(C_{i\Phi}^{j}\right)^{2}\mathcal{F}_{jk}(i)\mathcal{F}_{\bar{j}k'}(\bar{i})\sum_{r,\bar{r}=0}^{\infty}B_{r}(i,k)B_{\bar{r}}(\bar{i},k')\left(1-z\right)^{-2h+h_{k}+r}\left(1-\bar{z}\right)^{-2\bar{h}+\bar{h}_{k'}+\bar{r}}\nonumber 
\end{eqnarray}
Reading off the coefficient of a required power of the form $z^{n+\gamma}$,
where $n$ is the descendant level we are interested in, is then possible
using the following consideration%
\footnote{We are very grateful to G. Watts for the idea underlying this consideration.%
}. Suppose we have a function $f(z)$ that has singular points at $0$,
$1$ and $\infty$, and the following expansions around $z=0$ and
$z=1$:
\begin{equation}
f(z)=\sum_{n=0}^{\infty}C_{n}z^{n+\gamma}=\sum_{i=0}^{\infty}A_{i}(1-z)^{-\alpha_{i}}
\end{equation}
where the exponents $\alpha_{i}$ decrease with $i$. Note that these
properties are satisfied by the conformal blocks appearing in (\ref{eq:getlevelcontrib_expansion}).
Then 
\begin{equation}
C_{n}=\oint_{\mathcal{C}_{0}}\frac{dz}{2\pi i}z^{-n-\gamma-1}f(z)=\oint_{\mathcal{C}_{1}}\frac{dz}{2\pi i}z^{-n-\gamma-1}\sum_{i=0}^{\infty}A_{i}(1-z)^{-\alpha_{i}}
\end{equation}
where we deformed the contour $\mathcal{C}_{0}$ encircling $z=0$
to $\mathcal{C}_{1}$ enclosing the real line segment between $z=1$
to $z=\infty$. Exchanging the sum with the integration and substituting
the discontinuity of the $(1-z)^{-\alpha_{i}}$ terms gives 
\begin{equation}
C_{n}=\sum_{i=0}^{\infty}A_{i}\frac{\sin\pi\alpha_{i}}{\pi}\int_{1}^{\infty}dt\, t^{-n-\gamma-1}\left(t-1\right)^{-\alpha_{i}}=\sum_{i=0}^{\infty}\frac{\Gamma\left(\alpha_{i}+n+\gamma\right)}{\Gamma\left(\alpha_{i}\right)\Gamma\left(1+n+\gamma\right)}A_{i}\label{eq:integral_for_extraction}
\end{equation}
which provides the required coefficient as a series summed over $i$.
However, given that we aim at constructing the counter term to finite
order in $1/n$, and in view of the behaviour
\begin{equation}
\frac{\Gamma\left(\alpha+n+\gamma\right)}{\Gamma\left(\alpha\right)\Gamma\left(1+n+\gamma\right)}=\frac{1}{\Gamma(\alpha)}\left(\frac{1}{n+\gamma}\right)^{-\alpha+1}\left(1+\frac{\alpha(\alpha-1)}{2(n+\gamma)}+O\left(\frac{1}{(n+\gamma)^{2}}\right)\right)
\end{equation}
we only need to keep a finite number of terms from the $i$ sum. The
subsequent steps are the same as in subsection \ref{sub:The-ground-state}.
Once the level $n$ contribution to the matrix element has been extracted,
the level $n$ contribution to the scaling function of state $i$
is given by 
\begin{eqnarray}
e_{i,n}(r)-e_{i,n-1}(r) & = & -\frac{\lambda^{2}}{2\pi}\tilde{e}_{i,n}+\mathcal{O}\left(\lambda^{3}\right)\label{eq:excited_state_level_contrib}\\
 &  & \tilde{e}_{i,n}=\int_{\left|z\right|<1}d^{2}z\left(z\bar{z}\right)^{h-1}\langle i|\Phi\left(1,1\right)\tilde{P}_{n}\Phi\left(z,\bar{z}\right)|i\rangle\nonumber 
\end{eqnarray}
where the integral can be performed the same way as in (\ref{eq:ground_state_level_contrib}).

We remark that the step of exchanging the sum with the integral is
only valid for terms in which $n+\gamma+\alpha_{i}>0$. Since the
$\alpha_{i}$ in general decrease without lower bound, for any finite
$n$ this only holds for finitely many terms in the sum. As a result,
the $1/n$ expansion gives an asymptotic series, as discussed later
in subsection \ref{sub:Stationary--pair}.

To construct the counter term for the scaling function of descendant
states, some modifications are needed. First of all, the descendant
level of the state shifts the exponent of the wanted power of $z$
and $\bar{z}$, resulting in a shift in the dependence on the truncation
level, as observed previously in \cite{Giokas:2011ix}. In addition,
the conformal blocks for the descendant states must be constructed
from the primary ones, which can be accomplished using the Ward identities
(\ref{eq:conformal_ward_identities}). We now proceed to present two
examples, the first of which is a simple application of the method,
while the second demonstrates both the treatment of degeneracies in
the conformal Hilbert space and the procedure for descendant states.

\subsubsection{The first $A\bar{A}$ two-particle state in the Potts model \label{sub:The-first-neutral-two-particle-state}}

In the scaling three-state Potts model the first excited state in
sector $\mathcal{H}_{0}$ a two-particle state which in the scattering
picture consists of two stationary particles, one of which is of species
$A$ and the other is $\bar{A}$. The UV limit of this excited state
level corresponds to the highest weight vector in the conformal module
\begin{equation}
S_{\frac{2}{5},\frac{2}{5}}
\end{equation}
this can be seen either from TCSA or using the excited TBA equation
introduced later. Therefore the excited state scaling function has
the limiting value (\ref{eq:confblocklevelcoeffs})
\begin{equation}
e_{1}(0)=-\frac{1}{12}\cdot\frac{4}{5}+2\cdot\frac{2}{5}=\frac{11}{15}
\end{equation}
All fields that occur in the calculation below have identical left
and right conformal weights, so it is useful to introduce the shorter
notation
\begin{equation}
\Phi_{r,s}=\Phi_{h_{r,s},h_{r,s}}
\end{equation}
The conformal state is created by the primary field $\Phi_{2,1}$:
\begin{equation}
|\Phi_{2,1}\rangle=|2/5,2/5\rangle=\Phi_{2,1}(0,0)|0\rangle
\end{equation}
Using the conformal fusion rules
\begin{equation}
\Phi_{2,1}\times\Phi_{2,1}=\mathbb{I}+\Phi_{3,1}
\end{equation}
the relevant two-point function can be expanded into conformal blocks
as follows:

\begin{eqnarray}
\left\langle \Phi_{2,1}\left|\Phi_{2,1}\left(1,1\right)\Phi_{2,1}\left(z,\bar{z}\right)\right|\Phi_{2,1}\right\rangle  & = & \left(C_{\Phi_{2,1}\Phi_{2,1}}^{\mathbb{I}}\right)^{2}\left|\blockZ{2,1}{2,1}{2,1}{2,1}{1,1}\right|^{2}+\left(C_{\Phi_{2,1}\Phi_{2,1}}^{\Phi_{3,1}}\right)^{2}\left|\blockZ{2,1}{2,1}{2,1}{2,1}{3,1}\right|^{2}\nonumber \\
\nonumber \\
\label{eq:twoptconfblocks1}
\end{eqnarray}
where the operation of taking the modulus squared corresponds to the
product of holomorphic ($z$-dependent) and antiholomorphic ($\bar{z}$-dependent)
factors. 

The level $n$ contribution can be constructed from the coefficients
of $(z\bar{z})^{n-2h_{2,1}}$ in the first term and of $(z\bar{z})^{h_{3,1}+n-2h_{2,1}}$
in the second term of the correlator, respectively. Using the duality
relations (\ref{eq:duality_maintex}) we can rewrite the two terms
as 
\begin{eqnarray}
\blockZ{2,1}{2,1}{2,1}{2,1}{1,1} & = & \mathcal{F}_{\mathbb{I}\mathbb{I}}[\phi_{2,1}]\blockOnemZ{2,1}{2,1}{2,1}{2,1}{1,1}+\mathcal{F}_{\mathbb{I}\phi_{3,1}}[\phi_{2,1}]\blockOnemZ{2,1}{2,1}{2,1}{2,1}{3,1}\nonumber \\
\nonumber \\
\blockZ{2,1}{2,1}{2,1}{2,1}{3,1} & = & \mathcal{F}_{\phi_{3,1}\mathbb{I}}[\phi_{2,1}]\blockOnemZ{2,1}{2,1}{2,1}{2,1}{1,1}+\mathcal{F}_{\phi_{3,1}\phi_{3,1}}[\phi_{2,1}]\blockOnemZ{2,1}{2,1}{2,1}{2,1}{3,1}\nonumber \\
\end{eqnarray}
From (\ref{eq:confblocklevelcoeffs}), the series expansions of the
dual channel conformal blocks are the following
\begin{eqnarray}
\blockOnemZ{2,1}{2,1}{2,1}{2,1}{1,1} & = & \left(1-z\right)^{-2h_{2,1}}\left(1+\frac{2h_{2,1}^{2}}{c}\left(1-z\right)^{2}+\mathcal{O}\left(\left(1-z\right)^{3}\right)\right)\nonumber \\
\nonumber \\
\blockOnemZ{2,1}{2,1}{2,1}{2,1}{3,1} & = & \left(1-z\right)^{-2h_{2,1}+h_{3,1}}\left(1+\frac{h_{3,1}}{2}\left(1-z\right)+\mathcal{O}\left(\left(1-z\right)^{2}\right)\right)\nonumber \\
\nonumber \\
\label{eq:dual_expansion1}
\end{eqnarray}
Keeping only the leading terms and applying (\ref{eq:integral_for_extraction}),
we can rewrite the conformal blocks in (\ref{eq:twoptconfblocks1})
in the following way: 
\begin{eqnarray}
\blockZ{2,1}{2,1}{2,1}{2,1}{1,1} & = & \sum_{n=0}^{\infty}\left\{ \mathcal{F}_{\mathbb{I}\mathbb{I}}[\phi_{2,1}]\left(\frac{\Gamma\left(n\right)}{\Gamma\left(2h_{2,1}\right)\Gamma\left(n+1-2h_{2,1}\right)}+\dots\right)\right.\nonumber \\
\\
 &  & +\left.\mathcal{F}_{\mathbb{I}\phi_{3,1}}[\phi_{2,1}]\left(\frac{\Gamma\left(n-h_{3,1}\right)}{\Gamma\left(2h_{2,1}-h_{3,1}\right)\Gamma\left(n+1-2h_{2,1}\right)}+\dots\right)\right\} z^{n-2h_{2,1}}\nonumber 
\end{eqnarray}
\begin{eqnarray}
\blockZ{2,1}{2,1}{2,1}{2,1}{3,1} & = & \sum_{n=0}^{\infty}\left\{ \mathcal{F}_{\phi_{3,1}\mathbb{I}}[\phi_{2,1}]\left(\frac{\Gamma\left(n+h_{3,1}\right)}{\Gamma\left(2h_{2,1}\right)\Gamma\left(n+1+h_{3,1}-2h_{2,1}\right)}+\dots\right)\right.\nonumber \\
\\
 &  & +\left.\mathcal{F}_{\phi_{3,1}\phi_{3,1}}[\phi_{2,1}]\left(\frac{\Gamma\left(n\right)}{\Gamma\left(2h_{2,1}-h_{3,1}\right)\Gamma\left(n+1+h_{3,1}-2h_{2,1}\right)}+\dots\right)\right\} z^{h_{3,1}+n-2h_{2,1}}\nonumber 
\end{eqnarray}
where the ellipsis indicate terms which are subleading for large $n$,
resulting from the subleading terms in (\ref{eq:dual_expansion1}).
Putting together the left and right moving parts, and performing the
integral (\ref{eq:excited_state_level_contrib}), the level $n$ contribution
to the coefficient of $\lambda^{2}$ can be expressed as
\begin{eqnarray}
\tilde{e}_{1,n} & = & -\frac{\left(C_{\Phi_{2,1}\Phi_{2,1}}^{\mathbb{I}}\right)^{2}}{2h_{2,1}+2\left(n-2h_{2,1}\right)}\Bigg(\mathcal{F}_{\mathbb{I}\mathbb{I}}[\phi_{2,1}]\frac{\Gamma\left(n\right)}{\Gamma\left(2h_{2,1}\right)\Gamma\left(n+1-2h_{2,1}\right)}+\dots\nonumber \\
 &  & +\mathcal{F}_{\mathbb{I}\phi_{3,1}}[\phi_{2,1}]\frac{\Gamma\left(n-h_{3,1}\right)}{\Gamma\left(2h_{2,1}-h_{3,1}\right)\Gamma\left(n+1-2h_{2,1}\right)}+\dots\Bigg)^{2}\nonumber \\
 &  & -\frac{\left(C_{\Phi_{2,1}\Phi_{2,1}}^{\Phi_{3,1}}\right)^{2}}{2h_{2,1}+2\left(n-2h_{2,1}+h_{3,1}\right)}\Bigg(\mathcal{F}_{\phi_{3,1}\mathbb{I}}[\phi_{2,1}]\frac{\Gamma\left(n+h_{3,1}\right)}{\Gamma\left(2h_{2,1}\right)\Gamma\left(n+1-2h_{2,1}+h_{3,1}\right)}+\dots\nonumber \\
 &  & +\mathcal{F}_{\phi_{3,1}\phi_{3,1}}[\phi_{2,1}]\frac{\Gamma\left(n\right)}{\Gamma\left(2h_{2,1}-h_{3,1}\right)\Gamma\left(n+1-2h_{2,1}+h_{3,1}\right)}+\dots\Bigg)^{2}
\end{eqnarray}
From this expression one can construct the large $n$ counter term
as for the ground state case in \ref{sub:The-ground-state}. The leading
behaviour of the counter term is 
\begin{equation}
\delta e_{1,n}\left(r\right)=\lambda^{2}n^{4h-2}\frac{1}{4\left(2h-1\right)\Gamma\left(2h\right)^{2}}+\dots\label{eq:aabar_level_contrib}
\end{equation}
($h=h_{2,1}$) which is the same as for the ground state. The reason
is that this comes from the identity operator in the operator product
expansion of the perturbing operator with itself, and the matrix elements
of this term are independent of the state considered, so this term
is universal.

\subsubsection{The second $A\bar{A}$ two-particle state and the first $AAA$ three-particle
state\label{sub:The-second-AA-and-AAA}}

Both from TCSA and excited states TBA, the ultraviolet limit of the
scaling function of the second $A\bar{A}$ state is
\begin{equation}
e_{2}(0)=-\frac{1}{12}\frac{4}{5}+2\cdot\frac{2}{5}+2=\frac{41}{15}
\end{equation}
The zero-momentum part of the Hilbert space of the $\mathcal{M}_{5,6}$
minimal model at this level is doubly degenerate: it is spanned by
$L_{-1}\bar{L}_{-1}\left|\frac{2}{5},\frac{2}{5}\right\rangle $ and
$\left|\frac{7}{5},\frac{7}{5}\right\rangle $. So one has to use
degenerate perturbation theory and diagonalize the perturbing operator
in this subspace, which leads to the two eigenstates
\begin{equation}
\left|\pm\right\rangle =\frac{1}{\sqrt{2}}\left(\left|\frac{7}{5},\frac{7}{5}\right\rangle \pm\frac{1}{2h_{2,1}}L_{-1}\bar{L}_{-1}\left|\frac{2}{5},\frac{2}{5}\right\rangle \right)
\end{equation}
From TCSA one can see that $\left|+\right\rangle $ corresponds to
the first $AAA$ three-particle state and $\left|-\right\rangle $
to the second $A\bar{A}$ two-particle state. For the evaluation of
the counter term we therefore need to consider the following conformal
four-point functions:
\begin{itemize}
\item $\langle\Phi_{3,1}|\Phi_{2,1}\left(1,1\right)\Phi_{2,1}\left(z,\bar{z}\right)|\Phi_{3,1}\rangle$
\item $\langle\Phi_{2,1}|L_{1}\bar{L}_{1}\Phi_{2,1}\left(1,1\right)\Phi_{2,1}\left(z,\bar{z}\right)L_{-1}\bar{L}_{-1}|\Phi_{2,1}\rangle=\mathcal{D}\langle\Phi_{2,1}|\Phi_{2,1}\left(1,1\right)\Phi_{2,1}\left(z,\bar{z}\right)|\Phi_{2,1}\rangle$,
where $\mathcal{D}$ is some differential operator constructing the
descendant matrix element
\item $\langle\Phi_{2,1}|L_{1}\bar{L}_{1}\Phi_{2,1}\Phi_{2,1}|\Phi_{3,1}\rangle$ 
\item $\langle\Phi_{3,1}|\Phi_{2,1}\left(1,1\right)\Phi_{2,1}\left(z,\bar{z}\right)L_{-1}\bar{L}_{-1}|\Phi_{2,1}\rangle$
\end{itemize}
Due to the fusion rules, the last two are eventually zero. Therefore
to order $\lambda^{2}$, the counter term for both the two-particle
state and the three-particle state is
\begin{equation}
\delta e_{2,n}(r)=\delta e_{3,n}(r)=\frac{1}{2}\left(\delta e_{\left|\frac{7}{5},\frac{7}{5}\right\rangle ,n}(r)+\frac{1}{4h_{2,1}^{2}}\delta e_{L_{-1}\bar{L}_{-1}\left|\frac{2}{5},\frac{2}{5}\right\rangle ,n}(r)\right)\label{eq:aabar2_aaa_level_contrib}
\end{equation}
where the indices indicate the contributing matrix element. 

The first contribution can be calculated following the procedure in
subsection \ref{sub:The-first-neutral-two-particle-state}: it is
necessary to compute the level contributions for $\left\langle \Phi_{3,1}\left|\Phi_{2,1}\left(1,1\right)\Phi_{2,1}\left(z,\bar{z}\right)\right|\Phi_{3,1}\right\rangle $.
To obtain it one needs the following OPEs:
\begin{eqnarray}
\Phi_{2,1}\times\Phi_{2,1} & = & \mathbb{I}+\Phi_{3,1}\nonumber \\
\Phi_{2,1}\times\Phi_{3,1} & = & \Phi_{2,1}+\Phi_{4,1}\nonumber \\
\Phi_{3,1}\times\Phi_{3,1} & = & \mathbb{I}+\Phi_{3,1}
\end{eqnarray}
which lead to
\begin{eqnarray}
\langle\Phi_{3,1}|\Phi_{2,1}\left(1,1\right)\Phi_{2,1}\left(z,\bar{z}\right)|\Phi_{3,1}\rangle & = & \left(C_{\Phi_{2,1}\Phi_{3,1}}^{\Phi_{2,1}}\right)^{2}\left|\blockZ{3,1}{3,1}{2,1}{2,1}{2,1}\right|^{2}+\left(C_{\Phi_{2,1}\Phi_{3,1}}^{\Phi_{4,1}}\right)^{2}\left|\blockZ{3,1}{3,1}{2,1}{2,1}{4,1}\right|^{2}\nonumber \\
\nonumber \\
\end{eqnarray}
For the level $n$ contribution one needs the coefficient of $z^{n-h_{3,1}}$
in the first term and $z^{h_{4,1}+n-h_{2,1}-h_{3,1}}$ in the second
term. Rewriting the conformal blocks in the dual channel 
\begin{eqnarray}
\blockZ{3,1}{3,1}{2,1}{2,1}{2,1} & = & \mathcal{F}_{\phi_{2,1}\mathbb{I}}[\phi_{3,1}]\blockOnemZ{3,1}{3,1}{2,1}{2,1}{1,1}+\mathcal{F}_{\phi_{2,1}\phi_{3,1}}[\phi_{3,1}]\blockOnemZ{3,1}{3,1}{2,1}{2,1}{3,1}\nonumber \\
\nonumber \\
\blockZ{3,1}{3,1}{2,1}{2,1}{4,1} & = & \mathcal{F}_{\phi_{4,1}\mathbb{I}}[\phi_{3,1}]\blockOnemZ{3,1}{3,1}{2,1}{2,1}{1,1}+\mathcal{F}_{\phi_{4,1}\phi_{3,1}}[\phi_{3,1}]\blockOnemZ{3,1}{3,1}{2,1}{2,1}{3,1}\nonumber \\
\end{eqnarray}
and using the expansions
\begin{eqnarray}
\blockOnemZ{3,1}{3,1}{2,1}{2,1}{1,1} & = & \left(1-z\right)^{-2h_{2,1}}\left(1+\frac{2h_{2,1}h_{3,1}}{c}\left(1-z\right)^{2}+\mathcal{O}\left(\left(1-z\right)^{3}\right)\right)\nonumber \\
\nonumber \\
\blockOnemZ{3,1}{3,1}{2,1}{2,1}{3,1} & = & \left(1-z\right)^{-2h_{2,1}+h_{3,1}}\left(1+\frac{h_{3,1}}{2}\left(1-z\right)+\mathcal{O}\left(\left(1-z\right)^{2}\right)\right)\nonumber \\
\label{eq:dual_expansion2}
\end{eqnarray}
\\
one can determine the necessary coefficients. Keeping only the leading
terms in the $(1-z)$ expansion and using (\ref{eq:integral_for_extraction})
yields: 
\begin{eqnarray}
\blockZ{3,1}{3,1}{2,1}{2,1}{2,1} & = & \sum_{n=0}^{\infty}\left\{ \mathcal{F}_{\phi_{2,1}\mathbb{I}}[\phi_{3,1}]\left(\frac{\Gamma\left(n+2h_{2,1}-h_{3,1}\right)}{\Gamma\left(2h_{2,1}\right)\Gamma\left(n+1-h_{3,1}\right)}+\dots\right)\right.\nonumber \\
\\
 &  & +\left.\mathcal{F}_{\phi_{2,1}\phi_{3,1}}[\phi_{3,1}]\left(\frac{\Gamma\left(n+2h_{2,1}-2h_{3,1}\right)}{\Gamma\left(2h_{2,1}-h_{3,1}\right)\Gamma\left(n+1-h_{3,1}\right)}+\dots\right)\right\} z^{n-h_{3,1}}\nonumber 
\end{eqnarray}
\begin{eqnarray}
\blockZ{3,1}{3,1}{2,1}{2,1}{4,1} & = & \sum_{n=0}^{\infty}\left\{ \mathcal{F}_{\phi_{4,1}\mathbb{I}}[\phi_{3,1}]\left(\frac{\Gamma\left(n+h_{2,1}+h_{4,1}-h_{3,1}\right)}{\Gamma\left(2h_{2,1}\right)\Gamma\left(n+1+h_{4,1}-h_{2,1}-h_{3,1}\right)}+\dots\right)\right.\nonumber \\
\\
 &  & +\left.\mathcal{F}_{\phi_{4,1}\phi_{3,1}}[\phi_{3,1}]\left(\frac{\Gamma\left(n+h_{2,1}-h_{3,1}+h_{4,1}\right)}{\Gamma\left(2h_{2,1}-h_{3,1}\right)\Gamma\left(n+1+h_{4,1}-h_{2,1}-h_{3,1}\right)}+\dots\right)\right\} z^{h_{4,1}+n-h_{2,1}-h_{3,1}}\nonumber 
\end{eqnarray}
where the ellipsis indicate terms which are subleading for large $n$,
resulting from the subleading terms in (\ref{eq:dual_expansion2}).
Putting together the left and right moving parts, and performing the
integral (\ref{eq:excited_state_level_contrib}), the level contribution
from this channel is then 
\begin{eqnarray}
\tilde{e}_{\left|\frac{7}{5},\frac{7}{5}\right\rangle ,n} & = & -\frac{\left(C_{\Phi_{2,1}\Phi_{3,1}}^{\Phi_{2,1}}\right)^{2}}{2h_{2,1}+2\left(n-h_{3,1}\right)}\Bigg(\mathcal{F}_{\phi_{2,1}\mathbb{I}}[\phi_{3,1}]\frac{\Gamma\left(n+2h_{2,1}-h_{3,1}\right)}{\Gamma\left(2h_{2,1}\right)\Gamma\left(n+1-h_{3,1}\right)}+\dots\nonumber \\
 &  & +\mathcal{F}_{\phi_{2,1}\phi_{3,1}}[\phi_{3,1}]\frac{\Gamma\left(n+2h_{2,1}-2h_{3,1}\right)}{\Gamma\left(2h_{2,1}-h_{3,1}\right)\Gamma\left(n+1-h_{3,1}\right)}+\dots\Bigg)^{2}\nonumber \\
 &  & -\frac{\left(C_{\Phi_{2,1}\Phi_{3,1}}^{\Phi_{4,1}}\right)^{2}}{2h_{2,1}+2\left(n+h_{4,1}-h_{2,1}-h_{3,1}\right)}\Bigg(\mathcal{F}_{\phi_{4,1}\mathbb{I}}[\phi_{3,1}]\frac{\Gamma\left(n+h_{2,1}+h_{4,1}-h_{3,1}\right)}{\Gamma\left(2h_{2,1}\right)\Gamma\left(n+1+h_{4,1}-h_{2,1}-h_{3,1}\right)}+\dots\nonumber \\
 &  & +\mathcal{F}_{\phi_{4,1}\phi_{3,1}}[\phi_{3,1}]\frac{\Gamma\left(n+h_{2,1}-h_{3,1}+h_{4,1}\right)}{\Gamma\left(2h_{2,1}-h_{3,1}\right)\Gamma\left(n+1+h_{4,1}-h_{2,1}-h_{3,1}\right)}+\dots\Bigg)^{2}
\end{eqnarray}
These can be used to determine the counter term $\delta e_{\left|\frac{7}{5},\frac{7}{5}\right\rangle ,n}(r)$
following the steps in \ref{sub:Determining-the-counter-term}; we
omit the explicit form as it is quite long and not really illuminating.

For the second term one needs to repeat the computation in subsection
\ref{sub:The-first-neutral-two-particle-state}, but replacing all
objects with those pertaining to the descendant conformal block given
in (\ref{eq:desc_duality_relations},\ref{eq:desc_dualconfblockexpansion},\ref{eq:desc_confblocklevelcoeffs}).

\subsection{Power counting}

As discussed above, the leading large $n$ behaviour is the same for
all cases: $\sim n^{4h-2}$ with $h=h_{2,1}$. We remark that this
can be extracted from a simple power counting argument. The second
order cut-off dependence is determined by the short-distance contribution
to the integrated correlator 
\begin{equation}
\int d^{2}\vec{x}\langle\Psi|\Phi(\vec{x})\Phi(0,0)|\Psi\rangle_{CFT}
\end{equation}
where $\vec{x}=(\tau,x)$. In the scaling Potts model $\Phi=\Phi_{2,1}$
which has the short-distance expansion 
\begin{equation}
\Phi_{2,1}(\vec{x})\Phi_{2,1}(0,0)\sim A\left(\frac{1}{r^{4h_{2,1}}}+\mbox{descendants}\right)+B\left(\frac{\Phi_{3,1}(0,0)}{r^{4h_{2,1}-2h_{3,1}}}+\mbox{descendants}\right)
\end{equation}
where $r=\sqrt{\tau^{2}+x^{2}}$ and $A$ and $B$ are conformal OPE
coefficients. The most singular term is the one coming from the identity
(descendants always contribute terms that are less singular), and
putting a short-distance cut-off $r>1/\Lambda$ gives a leading dependence
$\Lambda^{4h_{2,1}-2}$ by simple power counting. Since the TCSA cut-off
for large $n$ is 
\begin{equation}
\Lambda=\frac{4\pi n}{L}+O(1)
\end{equation}
the expected dependence is exactly $n^{4h_{2,1}-2}$.

\section{Excited state TBA \label{sec:Excited-state-TBA}}

\subsection{The excited state TBA equations in the paramagnetic phase \label{sub:The-excited-state-TBA-paramagnetic}}

Since the Potts S-matrices in the high-temperature (paramagnetic)
phase (\ref{eq:highT_Smatrix}) are diagonal, the ground state TBA
can be written down in a straightforward manner \cite{Zamolodchikov:1989cf}:
\begin{eqnarray}
\epsilon_{1}(\theta) & = & mR\cosh\theta-\phi_{1}\star L_{1}(\theta)-\phi_{2}\star L_{2}(\theta)\nonumber \\
\epsilon_{2}(\theta) & = & mR\cosh\theta-\phi_{1}\star L_{2}(\theta)-\phi_{2}\star L_{1}(\theta)
\end{eqnarray}
where the kernels are given by the derivatives of the phase-shift
\begin{equation}
\phi_{1}(\theta)=-i\frac{d}{d\theta}\log S_{1}(\theta)=-\frac{\sqrt{3}}{1+2\cosh\theta}\qquad\phi_{2}(\theta)=-i\frac{d}{d\theta}\log S_{2}(\theta)=\frac{\sqrt{3}}{1-2\cosh\theta}
\end{equation}
and we introduced the notations
\begin{equation}
L_{i}(\theta)=\log(1+e^{-\epsilon_{i}(\theta)})\qquad A\star B(\theta)=\int\frac{d\lambda}{2\pi}A(\theta-\lambda)B(\lambda)
\end{equation}
The ground state energy can be obtained as
\begin{equation}
E_{0}(R)=-\int\frac{d\theta}{2\pi}m\cosh\theta\, L_{1}(\theta)-\int\frac{d\theta}{2\pi}m\cosh\theta\, L_{2}(\theta)
\end{equation}
The two pseudo-energy functions $\epsilon_{1,2}(\theta)$ correspond
to the two particles $A$ and $\bar{A}$. Since the ground state is
charge neutral, one has $\epsilon_{1}(\theta)=\epsilon_{2}(\theta)=\epsilon(\theta)$
and the equation for $\epsilon(\theta)$ turns out to be identical
to the TBA for the scaling Lee-Yang model, with the ground state energy
differing by a factor of $2$ \cite{Zamolodchikov:1989cf}.

Following the argument of analytic continuation as described in \cite{Dorey:1996re,Dorey:1997rb}
leads to the following general form of the excited TBA equations:
\begin{eqnarray}
\epsilon_{1}(\theta) & = & mR\cosh\theta+\sum_{k}\log\frac{S_{1}(\theta-\theta_{k}^{+})}{S_{2}(\theta-\bar{\theta}_{k}^{+})}+\sum_{l}\log\frac{S_{2}(\theta-\theta_{l}^{-})}{S_{1}(\theta-\bar{\theta}_{l}^{-})}-\phi_{1}\star L_{1}(\theta)-\phi_{2}\star L_{2}(\theta)\nonumber \\
\epsilon_{2}(\theta) & = & mR\cosh\theta+\sum_{k}\log\frac{S_{2}(\theta-\theta_{k}^{+})}{S_{1}(\theta-\bar{\theta}_{k}^{+})}+\sum_{l}\log\frac{S_{1}(\theta-\theta_{l}^{-})}{S_{2}(\theta-\bar{\theta}_{l}^{-})}-\phi_{1}\star L_{2}(\theta)-\phi_{2}\star L_{1}(\theta)\nonumber \\
 &  & e^{\epsilon_{1}(\theta_{k}^{+})}=e^{\epsilon_{1}(\bar{\theta}_{k}^{-})}=-1\nonumber \\
 &  & e^{\epsilon_{2}(\theta_{k}^{-})}=e^{\epsilon_{2}(\bar{\theta}_{k}^{+})}=-1\label{eq:highT_exc_TBA}
\end{eqnarray}
with the energy expressed as 
\begin{equation}
E(R)=-im\sum_{k}\left(\sinh\theta_{k}^{+}-\sinh\bar{\theta}_{k}^{+}\right)-im\sum_{l}\left(\sinh\theta_{l}^{-}-\sinh\bar{\theta}_{l}^{-}\right)-\int\frac{d\theta}{2\pi}m\cosh\theta\,\left(L_{1}(\theta)+L_{2}(\theta)\right)\label{eq:highT_energy_of_state}
\end{equation}
where $\theta_{k}^{\pm}$ and $\bar{\theta}_{k}^{\pm}$ are positions
of singularities picked up during the continuation. Reality of the
finite volume energy $E(R)$ requires that $\epsilon_{2}(\theta)=\epsilon_{1}(\theta)^{*}$
for real $\theta$, which in turn suggests
\begin{equation}
\bar{\theta}_{k}^{\pm}=\left(\theta_{k}^{\pm}\right)^{*}
\end{equation}
Then the independent relations for the singularity positions can be
written as
\begin{equation}
\epsilon_{1}(\theta_{k}^{+})=\pi i(2n_{k}^{+}+1)\qquad\epsilon_{2}(\theta_{k}^{-})=\pi i(2n_{k}^{-}+1)\label{eq:sing_rels}
\end{equation}
Indeed, the analysis of the infrared limit below shows that this is
the correct choice. However, when continuing to small volumes, some
branching transitions may occur for specific levels, just as observed
for the scaling Lee-Yang model in \cite{Dorey:1996re,Bazhanov:1996aq}.

\subsection{The infrared limit of the excited state TBA\label{sub:The-infrared-limit-of-TBA}}

In the infrared limit, the convolution terms can be neglected in (\ref{eq:highT_exc_TBA}).
Writing

\begin{eqnarray*}
\theta_{k}^{+}=\lambda_{k}^{+}+i\rho_{k}^{+} & \qquad & \bar{\theta}_{k}^{+}=\lambda_{k}^{+}-i\rho_{k}^{+}\\
\theta_{k}^{-}=\lambda_{k}^{-}+i\rho_{k}^{-} & \qquad & \theta_{k}^{-}=\lambda_{k}^{-}-i\rho_{k}^{-}
\end{eqnarray*}
the real part of the relations (\ref{eq:sing_rels}) read 
\begin{eqnarray}
0=mR\cosh\lambda_{r}^{+}\cos\rho_{r}^{+}+\sum_{k}\mbox{Re}\log\frac{S_{1}(\lambda_{r}^{+}-\lambda_{k}^{+}+i(\rho_{r}^{+}-\rho_{k}^{+}))}{S_{2}(\lambda_{r}^{+}-\lambda_{k}^{+}+i(\rho_{r}^{+}+\rho_{k}^{+}))}+\sum_{k}\mbox{Re}\log\frac{S_{2}(\lambda_{r}^{+}-\lambda_{k}^{-}+i(\rho_{r}^{+}-\rho_{k}^{-}))}{S_{1}(\lambda_{r}^{+}-\lambda_{k}^{-}+i(\rho_{r}^{+}+\rho_{k}^{-}))}\nonumber \\
0=mR\cosh\lambda_{r}^{-}\cos\rho_{r}^{-}+\sum_{k}\mbox{Re}\log\frac{S_{1}(\lambda_{r}^{-}-\lambda_{k}^{-}+i(\rho_{r}^{-}-\rho_{k}^{-}))}{S_{2}(\lambda_{r}^{-}-\lambda_{k}^{-}+i(\rho_{r}^{-}+\rho_{k}^{-}))}+\sum_{k}\mbox{Re}\log\frac{S_{2}(\lambda_{r}^{-}-\lambda_{k}^{+}+i(\rho_{r}^{-}-\rho_{k}^{+}))}{S_{1}(\lambda_{r}^{-}-\lambda_{k}^{+}+i(\rho_{r}^{-}+\rho_{k}^{+}))}
\end{eqnarray}
For large $R$ the first term grows arbitrarily large, therefore one
of the $S$-matrix terms must approach a pole. Now the singularity
positions with upper index $+$ (corresponding to particle species
$A$) are expected to be pairwise different, and similarly for upper
index $-$ (particle species $\bar{A}$) due to the effective exclusion
statistics of the particles resulting from $S_{1}(0)=-1$. In addition,
the singularity positions of the two species must vary independently,
as they describe the momenta of different particles. Therefore in
both equations the singularity of the $S$-matrix closest to the real
axis comes from the $S_{2}$ in the $k=r$ term of the first sums.
This forces the asymptotic behaviour 
\begin{equation}
\rho_{k}^{\pm}\rightarrow\frac{\pi}{6}\qquad\mbox{for}\qquad mR\rightarrow\infty
\end{equation}
Now we can put

\begin{eqnarray}
\rho_{k}^{+} & = & \frac{\pi}{6}+\delta_{k}^{+}\nonumber \\
\rho_{k}^{-} & = & \frac{\pi}{6}+\delta_{k}^{-}
\end{eqnarray}
and keeping only the dominant terms gives
\begin{eqnarray}
0 & = & \frac{\sqrt{3}}{2}mR\cosh\lambda_{r}^{+}+\mbox{Re}\log\left(-S_{2}\left(\frac{i\pi}{3}+2i\delta_{r}^{+}\right)\right)+\dots\nonumber \\
0 & = & \frac{\sqrt{3}}{2}mR\cosh\lambda_{r}^{-}+\mbox{Re}\log\left(-S_{2}\left(\frac{i\pi}{3}+2i\delta_{r}^{-}\right)\right)+\dots
\end{eqnarray}
Using
\begin{equation}
S_{2}\left(\frac{i\pi}{3}+2i\delta_{r}^{\pm}\right)=-\frac{\sqrt{3}}{2\delta_{r}^{\pm}}+O(1)
\end{equation}
we get the leading behavior
\begin{equation}
\left|\delta_{r}^{\pm}\right|\sim\mathcal{C}\exp\left(-\frac{\sqrt{3}}{2}mR\cosh\lambda_{r}^{\pm}\right)
\end{equation}
where the constant $\mathcal{C}$ depends on the $\lambda_{k}^{\pm}$
with $k\neq r$.

Turning now to the imaginary part of relations (\ref{eq:sing_rels}),
the $\delta_{k}^{\pm}$ can be safely put to zero:
\begin{eqnarray}
\pi(2n_{r}^{+}+1)=\frac{1}{2}mR\sinh\lambda_{r}^{+}+\sigma_{r}^{+}+\sum_{k\neq r}\mbox{Im}\log\frac{S_{1}(\lambda_{r}^{+}-\lambda_{k}^{+})}{S_{2}(\lambda_{r}^{+}-\lambda_{k}^{+}+i\frac{\pi}{3})}+\sum_{l}\mbox{Im}\log\frac{S_{2}(\lambda_{r}^{+}-\lambda_{l}^{-})}{S_{1}(\lambda_{r}^{+}-\lambda_{l}^{-}+i\frac{\pi}{3})}\nonumber \\
\pi(2n_{r}^{-}+1)=\frac{1}{2}mR\sinh\lambda_{r}^{-}+\sigma_{r}^{-}+\sum_{k}\mbox{Im}\log\frac{S_{2}(\lambda_{r}^{-}-\lambda_{k}^{+})}{S_{1}(\lambda_{r}^{-}-\lambda_{k}^{+}+i\frac{\pi}{3})}+\sum_{l\neq r}\mbox{Im}\log\frac{S_{1}(\lambda_{r}^{-}-\lambda_{l}^{-})}{S_{2}(\lambda_{r}^{-}-\lambda_{l}^{-}+i\frac{\pi}{3})}
\end{eqnarray}
where
\begin{equation}
\sigma_{r}^{\pm}=\mbox{Im}\log\left(-S_{2}\left(\frac{i\pi}{3}+2i\delta_{r}^{\pm}\right)\right)=\begin{cases}
0\quad & \delta_{r}^{\pm}>0\\
\pi\quad & \delta_{r}^{\pm}<0
\end{cases}\label{eq:sigmas}
\end{equation}
Now for real $\lambda$
\begin{eqnarray}
\mbox{Im}\log\frac{S_{1}(\lambda)}{S_{2}(\lambda+i\frac{\pi}{3})} & = & -\frac{i}{2}\log S_{1}(\lambda)-\pi\mbox{sign}(\lambda)\nonumber \\
\mbox{Im}\log\frac{S_{2}(\lambda)}{S_{1}(\lambda+i\frac{\pi}{3})} & = & -\frac{i}{2}\log S_{2}(\lambda)+\pi\mbox{sign}(\lambda)\label{eq:S_pi_contribs}
\end{eqnarray}
which leads to
\begin{eqnarray}
2\pi I_{r}^{+} & = & mR\sinh\lambda_{r}^{+}+\sum_{k\neq r}-i\log S_{1}(\lambda_{r}^{+}-\lambda_{k}^{+})+\sum_{l}-i\log S_{2}(\lambda_{r}^{+}-\lambda_{l}^{-})\nonumber \\
2\pi I_{r}^{-} & = & mR\sinh\lambda_{r}^{-}+\sum_{k}-i\log S_{2}(\lambda_{r}^{-}-\lambda_{k}^{+})+\sum_{l\neq r}-i\log S_{1}(\lambda_{r}^{-}-\lambda_{l}^{-})\label{eq:Potts_BY_paramagnetic}
\end{eqnarray}
where the quantum numbers are 
\begin{equation}
I_{r}^{\pm}=4n_{r}^{\pm}+2-\sigma_{r}^{\pm}-2(\pi\mbox{ terms from eqn. (\ref{eq:S_pi_contribs}))}
\end{equation}
Equations (\ref{eq:Potts_BY_paramagnetic}) describe the correct asymptotic
quantization conditions for particle rapidities in the paramagnetic
phase of the scaling Potts model, and the asymptotic form of the energy
of the state (\ref{eq:highT_energy_of_state}) also turns out to be
the correct one:
\begin{equation}
E(R)=m\sum_{k}\cosh\lambda_{k}^{+}+m\sum_{l}\cosh\lambda_{l}^{-}
\end{equation}

\subsection{Relation to the excited state TBA of the scaling Lee-Yang model\label{sub:Relation-to-the-LY}}

For special states where the number and rapidities of the $A$ and
$\bar{A}$ particles are identical
\begin{equation}
\{\theta_{k}^{+}\}=\{\theta_{k}^{-}\}
\end{equation}
the two pseudo-energy functions are identical $\epsilon_{1}(\theta)=\epsilon_{2}(\theta)=:\epsilon(\theta)$,
and the TBA equations (\ref{eq:highT_exc_TBA},\ref{eq:highT_energy_of_state})
reduce to
\begin{eqnarray}
\epsilon(\theta) & = & mR\cosh\theta+\sum_{k}\log\frac{S_{LY}(\theta-\theta_{k})}{S_{LY}(\theta-\bar{\theta}_{k})}-\phi_{LY}\star L(\theta)\nonumber \\
e^{\epsilon(\theta_{k})} & = & -1\label{eq:LYTBA}\\
E(R) & = & 2\left\{ -im\sum_{k}\left(\sinh\theta_{k}-\sinh\bar{\theta}_{k}\right)-\int\frac{d\theta}{2\pi}m\cosh\theta\, L(\theta)\right\} \nonumber 
\end{eqnarray}
where
\begin{eqnarray}
\theta_{k} & = & \theta_{k}^{+}=\theta_{k}^{-}\qquad\bar{\theta}_{k}=\bar{\theta}_{k}^{+}=\bar{\theta}_{k}^{-}\nonumber \\
\phi_{LY}(\theta) & = & -i\frac{d}{d\theta}\log S_{LY}(\theta)
\end{eqnarray}
and 
\begin{equation}
S_{LY}(\theta)=\frac{\sinh\theta+i\sin\frac{2\pi}{3}}{\sinh\theta-i\sin\frac{2\pi}{3}}
\end{equation}
is the well-known $S$-matrix of the scaling Lee-Yang model \cite{Cardy1989}.
The system (\ref{eq:LYTBA}) is just the excited TBA of the scaling
Lee-Yang model \cite{Dorey:1996re,Bazhanov:1996aq}, with the energy
expression multiplied by a factor of two. This correspondence is a
generalization of the relation between the ground state TBAs, which
was originally noted by Zamolodchikov \cite{Zamolodchikov:1989cf}.

\subsection{The excited state TBA equations in the ferromagnetic phase}

Due to the invariance of sector $\mathcal{H}_{0}$ under Kramers-Wannier
duality $\mu\rightarrow-\mu$, the ground state TBA in the ferromagnetic
phase is the same as in the paramagnetic one. However, there appear
two additional vacuum states in the $\mathcal{H}_{\pm}$ sectors,
which are obtained by inserting a twist operator $\mathcal{Z}^{\pm1}$
in the partition function, where $\mathcal{Z}$ is the cyclic permutation
in $\mathbb{S}_{3}$ introduced in \ref{sub:Scaling-Potts-model}.
The general vacuum TBA can be written as \cite{Martins1991c,Fendley:1991xn}
\begin{eqnarray}
\epsilon_{1}(\theta) & = & i\omega+mR\cosh\theta-\phi_{1}\star L_{1}(\theta)-\phi_{2}\star L_{2}(\theta)\nonumber \\
\epsilon_{2}(\theta) & = & -i\omega+mR\cosh\theta-\phi_{1}\star L_{2}(\theta)-\phi_{2}\star L_{1}(\theta)
\end{eqnarray}
where the vacuum states in $\mathcal{H}_{0}$ and $\mathcal{H}_{\pm}$
correspond to $\omega=0$ and $\omega=\pm2\pi/3$, respectively. The
excited state TBAs can be found by the same argument as in the other
phase, with the result
\begin{eqnarray}
\epsilon_{1}(\theta) & = & i\omega+mR\cosh\theta+\sum_{k=1}^{N^{+}}\log\frac{S_{1}(\theta-\theta_{k}^{+})}{S_{2}(\theta-\bar{\theta}_{k}^{+})}+\sum_{l=1}^{N^{-}}\log\frac{S_{2}(\theta-\theta_{l}^{-})}{S_{1}(\theta-\bar{\theta}_{l}^{-})}-\phi_{1}\star L_{1}(\theta)-\phi_{2}\star L_{2}(\theta)\nonumber \\
\epsilon_{2}(\theta) & = & -i\omega+mR\cosh\theta+\sum_{k=1}^{N^{+}}\log\frac{S_{2}(\theta-\theta_{k}^{+})}{S_{1}(\theta-\bar{\theta}_{k}^{+})}+\sum_{l=1}^{N^{-}}\log\frac{S_{1}(\theta-\theta_{l}^{-})}{S_{2}(\theta-\bar{\theta}_{l}^{-})}-\phi_{1}\star L_{2}(\theta)-\phi_{2}\star L_{1}(\theta)\nonumber \\
 &  & e^{\epsilon_{1}(\theta_{k}^{+})}=e^{\epsilon_{1}(\bar{\theta}_{k}^{-})}=-1\\
 &  & e^{\epsilon_{2}(\theta_{k}^{-})}=e^{\epsilon_{2}(\bar{\theta}_{k}^{+})}=-1\nonumber \\
E(R) & = & -im\sum_{k}\left(\sinh\theta_{k}^{+}-\sinh\bar{\theta}_{k}^{+}\right)-im\sum_{l}\left(\sinh\theta_{l}^{-}-\sinh\bar{\theta}_{l}^{-}\right)-\int\frac{d\theta}{2\pi}m\cosh\theta\,\left(L_{1}(\theta)+L_{2}(\theta)\right)\nonumber 
\end{eqnarray}
Another difference from the paramagnetic phase is that the excitations
are now kinks mediating between neighboring vacua. Due to periodic
boundary conditions the total number of kink steps must be divisible
by three, so there is the constraint
\begin{equation}
N_{+}=N_{-}\:\bmod\:3
\end{equation}
In general, sectors $\mathcal{H}_{\pm}$ contain states with twists
$\pm2\pi/3$, while sectors $\mathcal{H}_{0}/\mathcal{H}_{1}$ contain
untwisted states that are $\mathcal{C}$-even/odd. As discussed in
\ref{sub:Scaling-Potts-model}, the kink stepping in forward direction
will be identified with $A$, while the one stepping in reverse direction
with $\bar{A}$, as they can be considered to be in one-to-one correspondence
with the particle species in the paramagnetic phase.

In the ferromagnetic case, the infrared limiting quantization conditions
(\ref{eq:Potts_BY_paramagnetic}) are also modified by the presence
of the twist

\begin{eqnarray}
2\pi I_{r}^{+} & = & \omega+mR\sinh\lambda_{r}^{+}+\sum_{k\neq r}-i\log S_{1}(\lambda_{r}^{+}-\lambda_{k}^{+})+\sum_{l}-i\log S_{2}(\lambda_{r}^{+}-\lambda_{l}^{-})\nonumber \\
2\pi I_{r}^{-} & = & -\omega+mR\sinh\lambda_{r}^{-}+\sum_{k}-i\log S_{2}(\lambda_{r}^{-}-\lambda_{k}^{+})+\sum_{l\neq r}-i\log S_{1}(\lambda_{r}^{-}-\lambda_{l}^{-})\label{eq:Potts_BY_ferromagnetic}
\end{eqnarray}

\subsection{The UV limit of the TBA equations\label{sub:The-UV-limit-of-the-TBA-equations}}

The derivation of the UV limit is very technical, and is relegated
to Appendix \ref{sec:Derivation-of-the-UV-limit}. Here we summarize
the results for the states considered in the numerical analysis; all
the identifications below are indeed in accordance with the TCSA as
discussed in Section \ref{sec:Numerical-comparison}.

\subsubsection{Vacuum states}

For the vacuum state in $\mathcal{H}_{0}$ one obtains \cite{Zamolodchikov:1989cf}
\begin{equation}
c_{R}=c_{L}=\frac{2}{5}
\end{equation}
In the ferromagnetic phase, the vacuum states in $\mathcal{H}_{\pm}$,
corresponding to $\omega=\pm2\pi/3$ satisfy \cite{Martins1991c,Fendley:1991xn}
\begin{equation}
c_{R}=c_{L}=-\frac{2}{5}
\end{equation}
The corresponding conformal weights can be computed from
\begin{equation}
c_{R,L}=c-24\Delta_{R,L}
\end{equation}
and give $\Delta_{R,L}=1/15$.

\subsubsection{One-particle states}

In the paramagnetic state, the lowest energy levels in a given momentum
sector of $\mathcal{H}_{\pm}$ are expected to correspond to one-particle
states. Considering a one-particle $A$ state with a singularity located
at $\theta^{+}$such that
\begin{eqnarray}
\epsilon_{1}(\theta^{+}) & = & i\pi(2n^{+}+1)\nonumber \\
\mbox{Im}\theta^{+} & = & \frac{\pi}{6}+\delta^{+}
\end{eqnarray}
the following result is obtained for $n^{+}>0$
\begin{eqnarray}
c_{R} & = & -\frac{2}{5}-12\left(2n^{+}-\sigma^{+}\right)\qquad\sigma^{+}=\begin{cases}
0 & \quad\delta^{+}>0\\
1 & \quad\delta^{+}<0
\end{cases}\nonumber \\
c_{L} & = & -\frac{2}{5}
\end{eqnarray}
which corresponds to a right descendant of $\Phi_{\frac{1}{15},\frac{1}{15}}^{+}$
with momentum quantum number $2n^{+}-\sigma^{+}$; for $n^{+}<0$,
the result is similar, but it is a left descendant instead. 

For a stationary particle, a numerical analysis of the TBA equation
in the infrared shows that the relevant quantum numbers are $n^{+}=0$
and $\delta^{+}>0$; in such a case $\theta^{+}$ is purely imaginary.
When decreasing the volume, the position of the singularity at a critical
value $mR=r_{c}$ reaches the line
\begin{equation}
\delta^{+}=\frac{\pi}{3}
\end{equation}
Similarly to the Lee-Yang case, for $mR<r_{c}$ the equation requires
analytic continuation. We do not go into the details here; the relevant
methods can be found in \cite{Dorey:1996re,Bazhanov:1996aq}. The
UV limit can be computed simply by noticing that because the singularity
is stuck in the middle, the two kink systems become identical to the
twisted ground system (with opposite values of twists on the two sides),
therefore
\begin{equation}
c_{R}=c_{L}=-\frac{2}{5}
\end{equation}
corresponding to the primary state create by $\Phi_{\frac{1}{15},\frac{1}{15}}^{+}$.

\subsubsection{Untwisted two-particle states $A\bar{A}$}

Supposing that one of the particles is right moving ($\theta^{+}>0$),
while the other one is left-moving ($\theta^{-}<0$) with
\begin{eqnarray}
\epsilon_{1}(\theta^{+})=i\pi(2n^{+}+1) &  & \mbox{Im}\theta^{+}=\frac{\pi}{6}+\delta^{+}\nonumber \\
\epsilon_{2}(\theta^{-})=-i\pi(2n^{-}+1) &  & \mbox{Im}\theta^{-}=\frac{\pi}{6}+\delta^{-}
\end{eqnarray}
the result is
\begin{eqnarray}
c_{R} & = & \frac{2}{5}-12\left(\frac{2}{5}+2n^{+}-\sigma^{+}\right)\nonumber \\
c_{L} & = & \frac{2}{5}-12\left(\frac{2}{5}+2n^{-}-\sigma^{-}\right)\nonumber \\
 &  & \sigma^{\pm}=\begin{cases}
0 & \quad\delta^{\pm}>0\\
1 & \quad\delta^{\pm}<0
\end{cases}
\end{eqnarray}
corresponding to descendants of either $\Phi_{\frac{2}{5},\frac{2}{5}}$/$\Phi_{\frac{7}{5},\frac{7}{5}}$
(in $\mathcal{H}_{0}$), or either of $\Phi_{\frac{2}{5},\frac{7}{5}}$
/ $\Phi_{\frac{7}{5},\frac{2}{5}}$ (in $\mathcal{H}_{1}$). 

In fact there are in general two degenerate states, because charge
conjugation leaves the TBA result invariant:
\begin{eqnarray}
\frac{1}{\sqrt{2}}\left(|A(\lambda^{+})\bar{A}(\lambda^{-})\rangle+|A(\lambda^{-})\bar{A}(\lambda^{+})\rangle\right) & \in & \mathcal{H}_{0}\nonumber \\
\frac{1}{\sqrt{2}}\left(|A(\lambda^{+})\bar{A}(\lambda^{-})\rangle-|A(\lambda^{-})\bar{A}(\lambda^{+})\rangle\right) & \in & \mathcal{H}_{1}
\end{eqnarray}
with $\lambda^{\pm}=\mbox{Re}\theta^{\pm}$. These two states are
completely degenerate, which is indeed valid in TCSA up to the numerical
precision that can be attained. 

The only exception is when the state is composed of two zero momentum
particles 
\begin{equation}
|A(0)\bar{A}(0)\rangle
\end{equation}
with 
\begin{equation}
\theta^{+}=\theta^{-}=i\left(\frac{\pi}{6}+\delta\right)\qquad\delta>0
\end{equation}
This state is non-degenerate and in $\mathcal{H}_{0}$; its scaling
function is just twice the stationary one-particle scaling function
in the Lee-Yang model, as discussed in \ref{sub:Relation-to-the-LY}.

\subsubsection{Twisted $A\bar{A}$ states and $AA$/$\bar{A}\bar{A}$ states}

In the ferromagnetic phase, the lowest excited states in $\mathcal{H}_{\pm}$
are $A\bar{A}$ states with non-zero twists
\begin{equation}
\omega=\pm\frac{2\pi}{3}
\end{equation}
Using the results in Appendix \ref{sec:Derivation-of-the-UV-limit},
it turns out that these states correspond to descendants of $\Phi_{\frac{2}{3},\frac{2}{3}}$.
In the paramagnetic phase, the same levels are described in TBA as
two-particle $AA$/$\bar{A}\bar{A}$ states for $\mathcal{H}_{-}/\mathcal{H}_{+}$,
respectively.

\section{Numerical comparison\label{sec:Numerical-comparison}}

The evaluation of the TCSA spectrum consists of several steps:
\begin{enumerate}
\item First the numerical ``raw'' TCSA spectrum is determined by diagonalizing
the TCSA Hamiltonian (\ref{eq:TCSA_hamiltonian}). We used cutoffs
$n=6,7,8,9,10,11,12$ (the highest ones corresponding to several thousand
states kept in each sector), and restricted our analysis to states
with total momentum zero.
\item Next for any given energy level, the level contributions are constructed
analytically. For the vacuum we know the $n$-dependence of order
$\lambda^{2}$ contributions in a closed form. For more general excited
states, the procedure in subsection \ref{sub:Construction-of-counter-for-excited}
gives the level contributions as a series in inverse powers of $n$.
Given these level contributions, one can check whether the TCSA results
are reproduced to a sufficient precision.
\item Finally, constructing the counter terms one eliminates the cut-off
dependence of the TCSA to order $\lambda^{2}$. A useful check on
this method is to evaluate the residual order $\lambda^{2}$ cut-off
dependence of the renormalized TCSA results, which must be sufficiently
close to zero. Note that this does not eliminate all the cut-off dependence,
as it may also come from higher order in $\lambda$.
\item Finally, one can compare the renormalized TCSA data to the TBA results.
\end{enumerate}

\subsection{Level contributions and accuracy of counter terms\label{sub:Level-contributions}}

For the first and second steps listed above, we can look at the level
contributions to scaling functions $e(r)$ before and after subtraction.
We have done this analysis for all the energy levels that are considered
for the comparison to TBA in subsection \ref{sub:Comparing-the-renormalized}.
Below we show and comment on the examples of
\begin{itemize}
\item the ground state in $\mathcal{H}_{0}$, for which the exact level
contributions are known;
\item the first excited state in $\mathcal{H}_{0}$, which illustrates the
use of the expansion in the dual channel for a primary state;
\item the second and third excited states in $\mathcal{H}_{0}$, which include
two novelties: the contribution of a descendant state, and degenerate
perturbation theory.
\end{itemize}
For the other states, the picture is the same; we omit the detailed
results as they would add nothing substantial to the demonstration
of the method.

\subsubsection{Ground state}

For the ground state which is the lowest level in sector $\mathcal{H}_{0}$,
the $O(\lambda^{2})$ level contributions are known exactly for any
$n$ and are given in (\ref{eq:ground_state_level_contrib}). From
the TCSA data, the difference between two subsequent values of the
cut-off $n$ can be fitted with a function $a+b\lambda^{2}+c\lambda^{4}$
and the coefficient $b$ extracted. This was performed in the volume
range $0\leq r\leq1$ which under (\ref{eq:dimless_pcft_hamiltonian})
corresponds to $0\leq\lambda\lesssim0.113765$. To see whether the
counter term (\ref{eq:ground_state_counterterm}) really removes the
cut-of dependence, one can repeat the same procedure for the subtracted
TCSA data. The results, shown in Table \ref{tab:Level-contribution-ground-state}
demonstrate how efficient the renormalization procedure is.

\begin{table}
\centering{}%
\begin{tabular}{|c|c|c|c|}
\hline 
$n$ & Exact & TCSA & Subtracted TCSA \tabularnewline
\hline 
\hline 
$9$ & $-0.0160116158$ & $-0.0160116106$ & $7.36449\cdot10^{-8}$\tabularnewline
\hline 
$10$ & $-0.0138989447$ & $-0.0138989498$ & $2.83545\cdot10^{-8}$\tabularnewline
\hline 
$11$ & $-0.0122228492$ & $-0.0122228377$ & $2.91101\cdot10^{-8}$\tabularnewline
\hline 
$12$ & $-0.0108656858$ & $-0.0108656810$ & $1.46659\cdot10^{-8}$\tabularnewline
\hline 
\end{tabular}\caption{\label{tab:Level-contribution-ground-state} Level contribution of
the coefficient of $\lambda^{2}$ in the perturbative series for the
ground state scaling function. }
\end{table}

\subsubsection{Stationary $A\bar{A}$ pair \label{sub:Stationary--pair}}

For the first excited state in $\mathcal{H}_{0}$ which is contains
a pair of particles, both with zero momentum, one can use the counter
term constructed in subsection \ref{sub:The-first-neutral-two-particle-state}.
In contrast to the ground state, the exact $n$-dependence of the
level contributions is not available, and we use the approximation
constructed from the expansion (\ref{eq:dualconfblockexpansion})
of the conformal block in the dual channel, to obtain an approximation
in powers of $1/n$, the leading term of which is presented in (\ref{eq:aabar_level_contrib}).
Although the expansion (\ref{eq:dualconfblockexpansion}) is convergent,
the $1/n$ expansion of the level contribution resulting after the
application of the integral formula (\ref{eq:integral_for_extraction})
is only asymptotic. This means that for any $n$, including more terms
from the conformal block in the dual channel at first improves the
result, but then the error starts to grow. On the other hand, for
higher $n$ (and therefore lower $1/n$) the series starts to diverge
at higher order. This can be manifestly seen in Table \ref{tab:Level-contribution-for-AAbar},
where contributions resulting from the inclusion of the conformal
block expansion to order $n$ is labeled $\mbox{B}n$. It turns out
that for $n=12$ the $\mbox{B}5$ and $\mbox{B}6$ approximations
give essentially exact results, so they can be used to construct the
counter term. The effect of this counter term is demonstrated in Table
\ref{tab:Residual-level-contributions-for-AAbar}, which again shows
that to order $\lambda^{2}$ the truncation dependence is almost totally
eliminated. 

\begin{table}
\centering{}%
\begin{tabular}{|c|c|c|c|c|c|c|c|}
\hline 
$n$ & TCSA & B1 & B2 & B3 & B4 & B5 & B6\tabularnewline
\hline 
\hline 
$9$ & $-0.0159516$ & $-0.0159007$ & $-0.0159229$ & $-0.0159573$ & $-0.0159517$ & $-0.0159470$ & $-0.0159695$\tabularnewline
\hline 
$10$ & $-0.0138470$ & $-0.0138127$ & $-0.0138255$ & $-0.0138513$ & $-0.0138462$ & $-0.0138462$ & $-0.0138501$\tabularnewline
\hline 
$11$ & $-0.0121777$ & $-0.0121536$ & $-0.0121615$ & $-0.0121809$ & $-0.0121769$ & $-0.0121777$ & $-0.0121783$\tabularnewline
\hline 
$12$ & $-0.0108263$ & $-0.0108087$ & $-0.0108138$ & $-0.0108286$ & $-0.0108257$ & $-0.0108264$ & $-0.0108264$\tabularnewline
\hline 
\end{tabular}\caption{\label{tab:Level-contribution-for-AAbar}Level contributions for the
lowest $A\bar{A}$ level. }
\end{table}
\begin{table}
\centering{}%
\begin{tabular}{|c|c|c|}
\hline 
$n$ & B5 & B6\tabularnewline
\hline 
\hline 
$9$ & $-4.55659\cdot10^{-6}$ & $1.54134\cdot10^{-5}$\tabularnewline
\hline 
$10$ & $-7.53545\cdot10^{-7}$ & $2.90049\cdot10^{-6}$\tabularnewline
\hline 
$11$ & $-2.94012\cdot10^{-6}$ & $5.70248\cdot10^{-7}$\tabularnewline
\hline 
$12$ & $1.07722\cdot10^{-6}$ & $9.84199\cdot10^{-8}$\tabularnewline
\hline 
\end{tabular}\caption{\label{tab:Residual-level-contributions-for-AAbar}$O(\lambda^{2})$
level contributions for the lowest $A\bar{A}$ level after subtraction,
where the approximations B5 and B6 were used.}
\end{table}

\subsubsection{Second $A\bar{A}$ and first $AAA$ levels }

The second excited level is degenerate at the fixed point with the
third one. This is the pair of states described in subsection \ref{sub:The-second-AA-and-AAA}.
The level contributions for these states are shown in Table \ref{tab:Level-contributions-for-AAbar-AAA},
while Table \ref{tab:Level-contributions-for-AAbar-AAA-subtr} shows
the residuals after subtraction. The perturbative results are somewhat
less accurate for these states; however these states are higher up
on the spectrum and therefore are more affected by higher-order terms
in the cut-off and $\lambda$. Still, as we demonstrate later these
counter terms result in a spectacular improvement in the agreement
between TCSA and the TBA predictions.

\begin{table}
\centering{}%
\begin{tabular}{|c|c|c|c|c|c|c|c|c|}
\hline 
$n$ & $A\bar{A}$ & $AAA$ & B1 & B2 & B3 & B4 & B5 & B6\tabularnewline
\hline 
\hline 
$9$ & $-0.0180702$ & $-0.0180669$ & $-0.0178814$ & $-0.0179488$ & $-0.0181169$ & $-0.0181083$ & $-0.0180053$ & $-0.0186437$\tabularnewline
\hline 
$10$ & $-0.0154775$ & $-0.0154748$ & $-0.0153561$ & $-0.0153936$ & $-0.0155160$ & $-0.0154957$ & $-0.0154819$ & $-0.0155671$\tabularnewline
\hline 
$11$ & $-0.0134662$ & $-0.0134641$ & $-0.0133835$ & $-0.0134058$ & $-0.0134953$ & $-0.0134780$ & $-0.0134791$ & $-0.0134937$\tabularnewline
\hline 
$12$ & $-0.0118654$ & $-0.0118636$ & $-0.0118077$ & $-0.0118217$ & $-0.0118881$ & $-0.0118752$ & $-0.0118782$ & $-0.0118805$\tabularnewline
\hline 
\end{tabular}\caption{\label{tab:Level-contributions-for-AAbar-AAA}$O(\lambda^{2})$ level
contributions for the second $A\bar{A}$ and the first $AAA$ state}
\end{table}
\begin{table}
\centering{}%
\begin{tabular}{|c|c|c|c|c|}
\hline 
$n$ & B5 $A\bar{A}$ & B5 $AAA$ & B6 $A\bar{A}$ & B6 $AAA$\tabularnewline
\hline 
\hline 
$9$ & $-5.63362\cdot10^{-5}$ & $-5.30678\cdot10^{-5}$ & $3.32776\cdot10^{-4}$ & $3.36045\cdot10^{-4}$\tabularnewline
\hline 
$10$ & $5.13623\cdot10^{-6}$ & $7.80905\cdot10^{-6}$ & $7.29807\cdot10^{-5}$ & $7.56536\cdot10^{-5}$\tabularnewline
\hline 
$11$ & $1.30180\cdot10^{-5}$ & $1.50275\cdot10^{-5}$ & $2.58255\cdot10^{-5}$ & $2.78350\cdot10^{-5}$\tabularnewline
\hline 
$12$ & $1.29015\cdot10^{-5}$ & $1.46115\cdot10^{-5}$ & $1.48606\cdot10^{-5}$ & $1.65706\cdot10^{-5}$\tabularnewline
\hline 
\end{tabular}\caption{\label{tab:Level-contributions-for-AAbar-AAA-subtr}$O(\lambda^{2})$
level contributions for second $A\bar{A}$ and the first $AAA$ state
after subtraction}
\end{table}

\subsection{Comparing the renormalized TCSA to the TBA results \label{sub:Comparing-the-renormalized}}

The third step listed in the beginning of section \ref{sec:Numerical-comparison}
is the actual construction of counter terms. This was described in
Section \ref{sec:Renormalization-in-TCSA} and is straightforward
given the level contributions tested above. 

The last step is to compare the renormalized TCSA data to the TBA
predictions. We must take into account that the TBA and the perturbed
conformal field theory (TCSA) energy levels differ by the so-called
universal bulk energy term \cite{Zamolodchikov:1989cf}
\begin{equation}
E_{TBA}(R)=E_{TCSA}(R)-\mathcal{B}R\label{eq:TBA_TCSA_rel}
\end{equation}
where
\begin{equation}
\mathcal{B}=-\frac{1}{2\sqrt{3}}m^{2}\label{eq:univbulk}
\end{equation}
Therefore following (\ref{eq:TBA_TCSA_rel}) we compare the TBA data
to TCSA data with the predicted bulk energy contribution subtracted
(with the exception of figure \ref{fig:Comparing-ground-state}).
Some numbers are given in tables in Appendix \ref{sec:Comparison-tables};
here we only show a few plots for illustration.

\subsubsection{Energy levels \label{sub:Energy-levels}}

\begin{figure}
\begin{centering}
\includegraphics[scale=0.6]{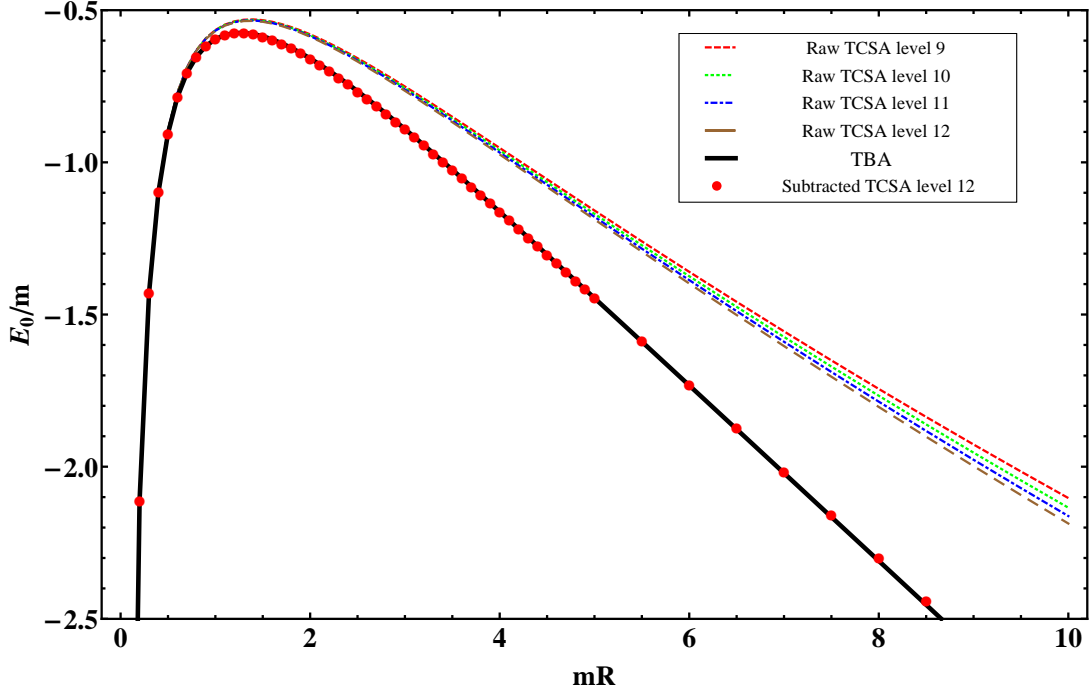}
\par\end{centering}

\caption{\label{fig:Comparing-ground-state} (color online) Comparing TCSA
and TBA for the ground state. The slow convergence of the TCSA is
apparent from the raw data; renormalized data are only presented for
level $12$, as the others would not be discernible on the plot. This
plot does not have the bulk energy subtracted to show that the renormalization
also gives back the right value for the universal bulk energy term.}

\end{figure}

\begin{figure}
\begin{centering}
\includegraphics[scale=0.6]{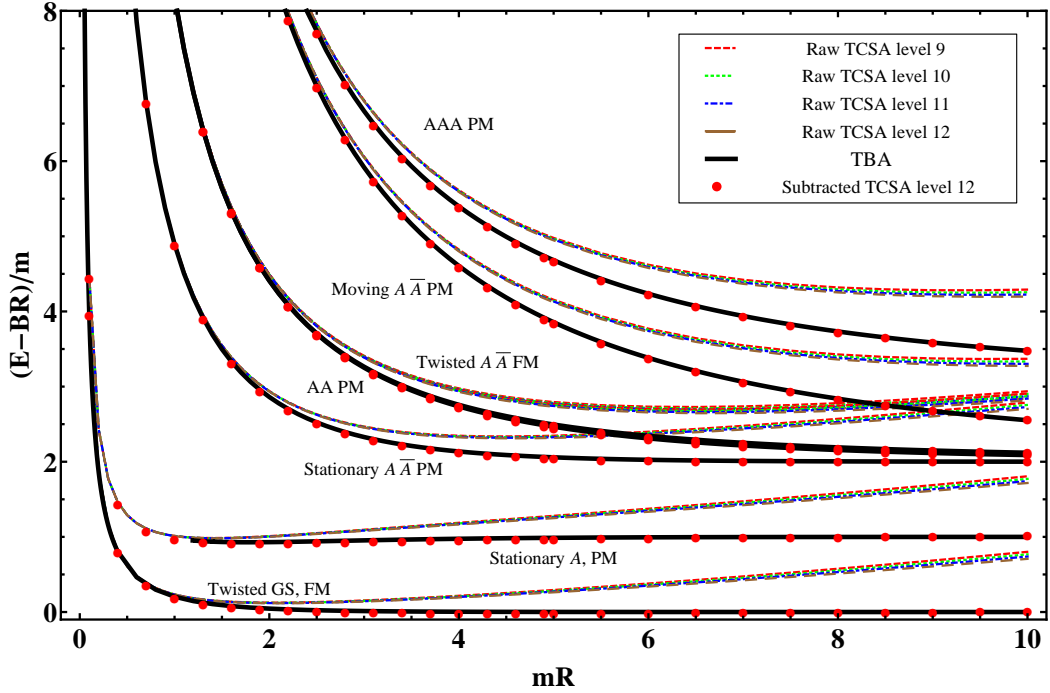}
\par\end{centering}

\caption{\label{fig:Comparing-exc-state} (color online) Comparing TCSA and
TBA for excited states. PM stands for paramagnetic, FM for ferromagnetic
phase, GS means ground state (twisted in the ferromagnetic phase).
The paramagnetic $AA$ and twisted ferromagnetic $A\bar{A}$ are so
close numerically that they eventually overlap at this resolution.}
\end{figure}

Figure \ref{fig:Comparing-ground-state} shows the comparison for
the ground state, comparing raw TCSA data for several values of the
cut-off, the renormalized TCSA and the TBA data. The renormalization
is so efficient in removing the cut-off dependence that we only show
the renormalized TCSA data for the highest cut-off, as the others
would not be discernible on the plot. The comparison for excited states
is shown in figure \ref{fig:Comparing-exc-state}; it has essentially
the same features.


\subsubsection{Two-particle phase shifts \label{sub:Two-particle-phase-shifts}}

A more sensitive test is provided by examining the phase-shift extracted
from the various two-particle states. Using the Bethe-Yang equations
(\ref{eq:Potts_BY_paramagnetic},\ref{eq:Potts_BY_ferromagnetic})
one can extract phase-shift data from the TCSA spectrum to compare
with theoretical predictions. Because the effect of the phase-shift
is subleading compared to the momentum quantum number, it is much
more sensitive to the accuracy of the numerics. From (\ref{eq:highT_Smatrix}),
we define the following phase-shift functions
\begin{eqnarray}
\delta_{AA}(\theta) & = & -i\log S_{AA}(\theta)\nonumber \\
\delta_{A\bar{A}}(\theta) & = & -i\log S_{A\bar{A}}(\theta)\label{eq:phase-shift_defs}
\end{eqnarray}
Note that the identification of the ferromagnetic phase kink states
with the paramagnetic phase particles defined in (\ref{eq:para_ferro_ident})
makes these definitions applicable in the ferromagnetic phase as well. 

The phase shifts are extracted from two-particle states with zero
total-momentum, consisting with a pair of particles with opposite
rapidities $\theta$ and $-\theta$. The ``experimental'' value
for the rapidity is determined from 
\begin{equation}
E_{\Psi}(R)-E_{0}(R)=2m\cosh\theta\label{eq:ph_shift_measure_energy}
\end{equation}
where $E_{\Psi}(R)$ and $E_{0}(R)$ are the two-particle and vacuum
levels, while the value of the phase shift at $2\theta$ is determined
from the quantization conditions (\ref{eq:Potts_BY_paramagnetic},\ref{eq:Potts_BY_ferromagnetic})
which reduce to a single equation
\begin{equation}
2\pi I=\omega+mR\sinh\theta+\delta(2\theta)\label{eq:ph_shift_measure_delta}
\end{equation}
where the twist $\omega$ is always zero in the paramagnetic phase
which contains both neutral ($A\bar{A}$) and charged ($AA$,$\bar{A}\bar{A}$)
two-particle levels. In the ferromagnetic phase it can take the values
$\omega=0,\pm2\pi/3$ ; however, in this case there are only $A\bar{A}$
levels. 

\begin{figure}
\begin{centering}
\includegraphics[width=0.8\textwidth]{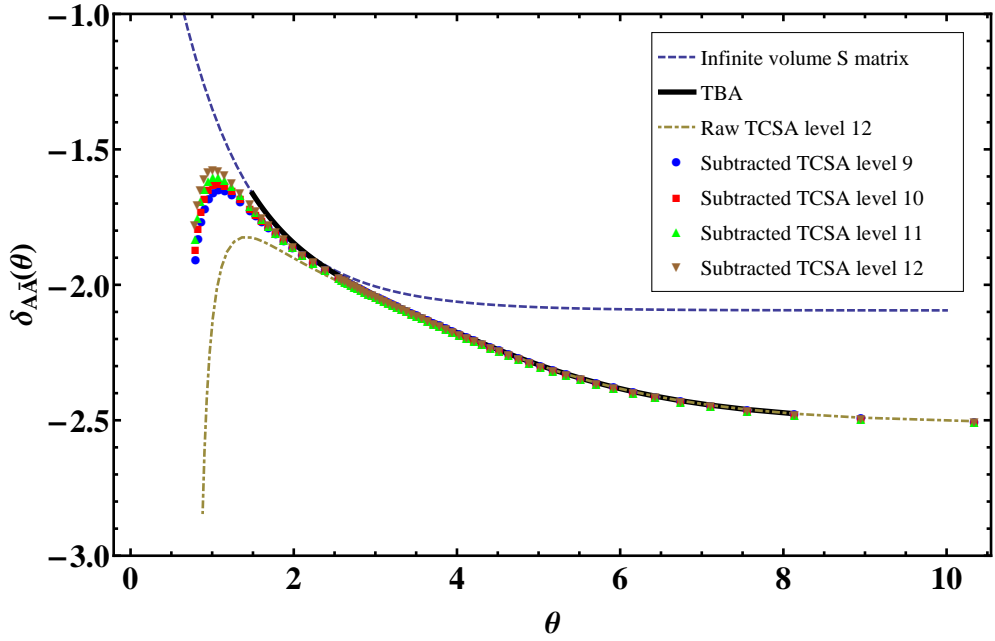}
\par\end{centering}

\caption{\label{fig:Comparing-pshiftaabar} (color online) Comparing $\delta_{A\bar{A}}(\theta)$
extracted from the third excited TCSA level in sector $\mathcal{H}_{0}$
(lowest lying moving $A\bar{A}$ state) to the scattering theory predictions
(\ref{eq:Potts_BY_paramagnetic},\ref{eq:Potts_BY_ferromagnetic})
and to TBA. }
\end{figure}

\begin{figure}
\begin{centering}
\includegraphics[width=0.8\textwidth]{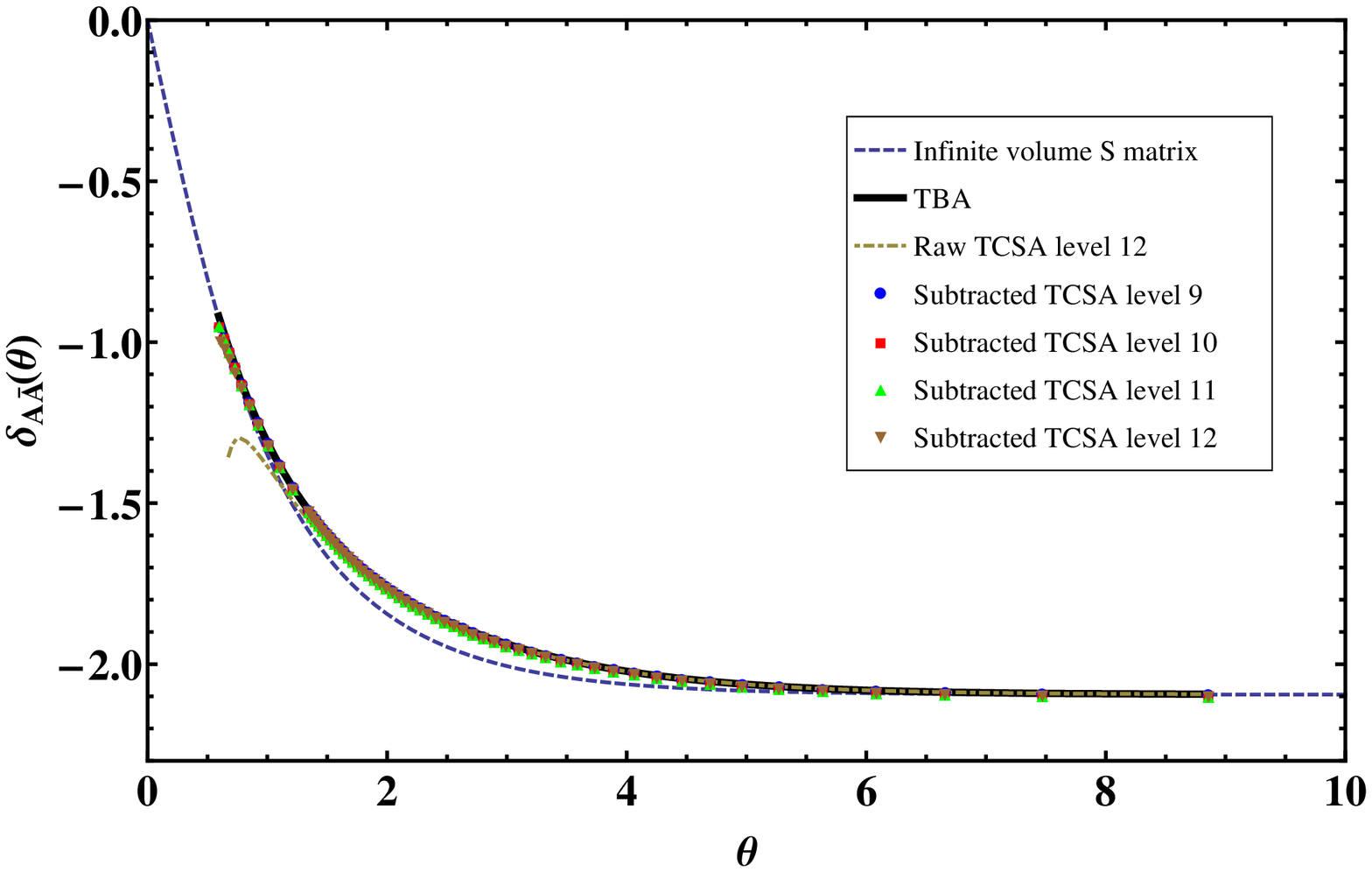}
\par\end{centering}

\caption{\label{fig:Comparing-pshifttwistedaabar} (color online) Comparing
$\delta_{A\bar{A}}(\theta)$ extracted from the first excited TCSA
level in sectors $\mathcal{H}_{\pm}$ in the ferromagnetic phase (lowest
lying twisted $A\bar{A}$ state) to the scattering theory predictions
(\ref{eq:Potts_BY_ferromagnetic}) and to TBA. }
\end{figure}

\begin{figure}
\begin{centering}
\includegraphics[width=0.8\textwidth]{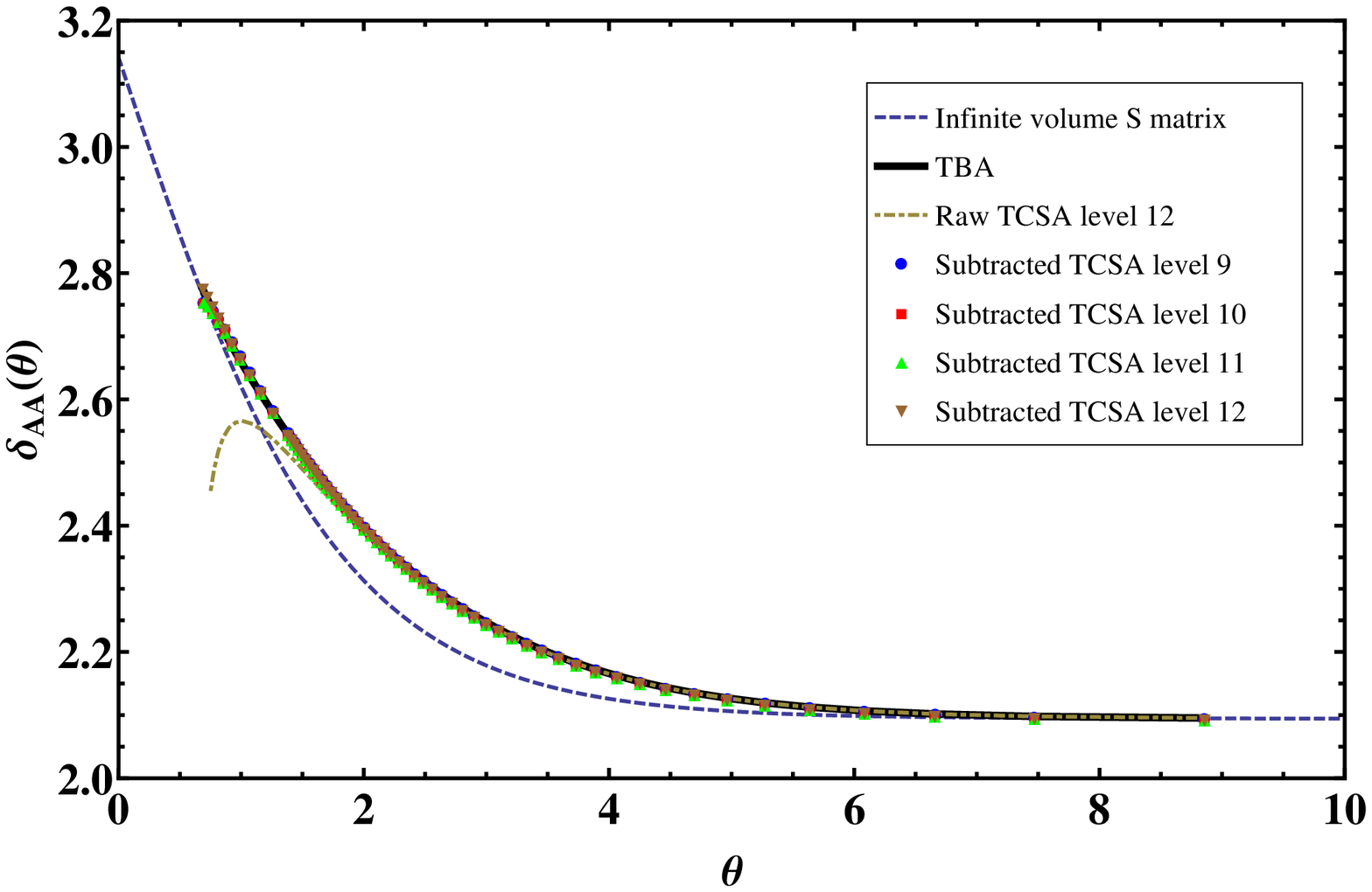}
\par\end{centering}

\caption{\label{fig:Comparing-pshiftaa} (color online) Comparing $\delta_{AA}(\theta)$
extracted from the first excited TCSA level in sectors $\mathcal{H}_{\pm}$
in the paramagnetic phase (lowest lying $AA$/$\bar{A}\bar{A}$ state)
to the scattering theory predictions (\ref{eq:Potts_BY_paramagnetic})
and to TBA. }
\end{figure}

The phase shifts extracted from the TCSA data can be compared to the
predictions of the infinite volume scattering amplitudes (\ref{eq:highT_Smatrix},\ref{eq:lowT_Smatrix}).
For large volumes, corresponding to small $\theta$ we expect truncation
effects to dominate. For small volumes the finite size corrections
decaying exponentially in the volume make up most of the deviation.
To demonstrate that, we also compare the TCSA phase shift to a ``effective
finite volume phase shift'' obtained by substituting the exact TBA
energy levels into (\ref{eq:ph_shift_measure_energy},\ref{eq:ph_shift_measure_delta}).
In contrast with the true infinite volume scattering amplitudes, the
effective finite volume phase shift is state-dependent. These comparisons
are presented for $\delta_{A\bar{A}}$ in figures \ref{fig:Comparing-pshiftaabar},
\ref{fig:Comparing-pshifttwistedaabar} and for $\delta_{AA}$ in
figure \ref{fig:Comparing-pshiftaa}. 


Note that the deviation of the TCSA phase-shift in the high energy
(small volume) regime is fully explained by TBA, which is not very
surprising in view of the excellent agreement between TCSA and TBA
demonstrated in Appendix \ref{sec:Comparison-tables}. For low energies
(large volumes) the agreement is very much improved by the renormalization
procedure. We also demonstrate that the residual cut-off dependence
is practically nonexistent except for very low energies; the remaining
deviation in that regime is expected to be due to $O(\lambda^{3})$
cut-off effects, the elimination of which would necessitate the extension
of the renormalization procedure to higher order.

\section{Discussion and outlook \label{sec:Conclusions}}

In this paper we provided a description of the finite volume spectrum
of the scaling Potts model combining two approaches: the renormalized
TCSA, the idea of which goes back to the recent papers \cite{Feverati:2006ni,Konik:2007cb,Giokas:2011ix},
and an excited TBA system which was first proposed in the present
work. We have developed the general theory of cut-off dependence and
counter terms for energy levels in TCSA, and applied it to the scaling
Potts field theory. Using comparison with the TBA predictions we have
shown that this gives a very precise tool to study the finite size
spectrum of perturbed conformal field theory. 

There are several potential applications of the results presented
here. TCSA has recent been extended to asymptotically free field theories
\cite{Beria2013}, but this line of development is still in its infancy.
In fact, a systematic understanding of the construction of counter
terms should prove very useful in this context. Another possible application
is given the application of the truncation approach to study non-equilibrium
physics in condensed matter theory \cite{Caux2012}. In addition,
integrability, finite size effects and the ideas of perturbed conformal
field theory have also appeared in the AdS/CFT correspondence (cf.
\cite{Beisert:2010jr} and references therein). We expect that the
methods developed here can be useful for these applications both by
improving numerical reliability and providing a detailed understanding
of scale dependence. 

There is also an interesting implication of the present results for
the study of the quantum Potts spin chain. In \cite{Rapp:2006} perturbative
calculations supplemented with renormalization group arguments cast
some doubt on the applicability of the factorized scattering amplitudes
in long-distance limit of the spin chain. A detailed DMRG analysis
has shown that the observed discrepancy between the factorized S matrix
and the low-energy scattering of quasi-particles in the discrete spin
chain persists even non-perturbatively \cite{Rapp2013}, and it was
speculated that this was due to the presence of an irrelevant operator
that has a large effect on the low-energy limit of the scattering
amplitudes away from the scaling limit. In this connection, first
of all we note that the raw TCSA phase-shifts in figures \ref{fig:Comparing-pshiftaabar},
\ref{fig:Comparing-pshifttwistedaabar} and \ref{fig:Comparing-pshiftaa}
show a characteristic deviation from the field theory predictions
at low energies which is very similar to that observed in the DMRG
results of \cite{Rapp2013}. In contrast to the DMRG study, in this
paper we are in a position to identify the source of this deviation:
it originates from the cut-off dependence introduced by the operators
that appear in the OPE
\[
\Phi_{2,1}(z,\bar{z})\Phi_{2,1}(0,0)=\frac{1}{(z\bar{z}){}^{4/5}}\left(\mathbb{I}+h_{2.1}^{2}\,(z\bar{z})^{2}(T\bar{T})(0,0)+\dots\right)+C_{\Phi_{2,1}\Phi_{2,1}}^{\Phi_{3,1}}\,(z\bar{z})^{3/5}\left(\Phi_{3,1}(0,0)+\dots\right)
\]
The counter term from the identity $\mathbb{I}$ is the universal
contribution shown explicitly in (\ref{eq:aabar_level_contrib}),
which only renormalizes the bulk energy density and thus makes no
contribution to the extracted phase-shifts. Therefore the dominant
part of the cut-off dependence comes from the irrelevant operators:
the leading one is $\Phi_{3,1}$ , while the first subleading one
is $T\bar{T}$. Once the counter-terms are added, all cut-off dependence
is eliminated to order $\lambda^{2}$ and the phase-shifts indeed
show a much better agreement with the field theoretical predictions.
In TCSA it is therefore clear that the cut-off dependence comes from
irrelevant operators, which makes it very plausible that the very
similar effect noticed in the DMRG data is also a result of the contribution
of the same irrelevant operators. Note also that the cut-off dependence
still remains for lower energies, which correspond to larger values
of the volume and therefore larger $\lambda$. Therefore it is clear
that these effects can only be removed by considering the counter
terms at higher order, which is out of the scope of this paper. To
sum up, our results for the cut-off dependence in the TCSA approach
strongly supports that the similar effect in the spin chain is caused
by the same irrelevant operators. A detailed matching of the phenomenology
of the spin chain with the perturbed CFT extended with irrelevant
operators, however, needs better quality data for the spin chain than
presently available, in addition to evaluating the counter terms for
the perturbed CFT Hamiltonian extended with the irrelevant fields.

Another interesting line of development is to extend the theory of
counter terms to a full renormalization group description along the
lines in \cite{Feverati:2006ni,Watts:2011cr,Giokas:2011ix}. The perturbing
operator considered in these works had an OPE of the form
\[
\Phi\Phi\sim\mathbb{I}+\Phi
\]
leading to a running coupling at order $\lambda^{2}$. We note that
at the next order the perturbing operator $\Phi_{2,1}$ appears in
the triple product of itself, which leads to a running coupling at
order $\lambda^{3}$. However, even at second order there appear running
couplings in the Potts field theory when other perturbing operators
are added, such as in the work \cite{Lepori:2009ip}. We are planning
to return to these issues in the near future.

\subsection*{Acknowledgments}

The authors acknowledge very useful discussions with G. Watts. This
work was supported by the Momentum grant LP2012-50/2013 of the Hungarian
Academy of Sciences.

\appendix

\section{CFT data \label{sec:CFT-data}}

\subsection{Conformal blocks}

The conformal blocks needed in this work are known in a closed form
for the Potts model \cite{Dotsenko:1984if}. Below we summarize the
necessary data for the renormalization computations. Considering the
following correlators
\begin{equation}
\langle\Phi_{r,s}|\Phi_{2,1}(1)\Phi_{2,1}(z)|\Phi_{r,s}\rangle
\end{equation}
the conformal blocks forming a basis for their chiral part around
$z=0$ are%
\footnote{The paper \cite{Dotsenko:1984if} gives the correlator in a different
basis, and so their conformal blocks must be transformed by an appropriate
conformal mapping to obtain the ones used here. One can also obtain
the blocks in our basis directly from the results of section 8.3.3
in the monograph \cite{difrancesco:1997nk}.%
}
\begin{eqnarray}
\blockZpos{r,s}{r,s}{2,1}{2,1}{r-1,s} & = & (1-z)^{3/5}z^{(-1-6r+5s)/10}\,_{2}F_{1}\left(\frac{6}{5}(1-r)+s,\frac{6}{5};1-\frac{6}{5}r+s\Big|z\right)\nonumber \\
 & \,\nonumber \\
 & \,\nonumber \\
\blockZpos{r,s}{r,s}{2,1}{2,1}{r+1,s} & = & (1-z)^{3/5}z^{(-1+6r-5s)/10}\,_{2}F_{1}\left(\frac{6}{5},\frac{6}{5}(1+r)-s;1+\frac{6}{5}r-s\Big|z\right)\label{eq:direct_channel_blocks}\\
\nonumber \\
\nonumber 
\end{eqnarray}
where the small $\phi$s refer to the chiral components, and 
\begin{equation}
\,_{2}F_{1}\left(\alpha,\beta;\gamma\big|z\right)
\end{equation}
denotes the standard hypergeometric function. The basis for the conformal
blocks around $z=1$ is given by
\begin{eqnarray}
\blockOnemZpos{r,s}{r,s}{2,1}{2,1}{1,1} & = & (1-z)^{-4/5}z^{(-1+6r-5s)/10}\,_{2}F_{1}\left(1+\frac{6}{5}(-1+r)+s,-\frac{1}{5};-\frac{2}{5}\Big|1-z\right)\nonumber \\
 & \,\nonumber \\
 & \,\nonumber \\
\blockOnemZpos{r,s}{r,s}{2,1}{2,1}{3,1} & = & (1-z)^{3/5}z^{(-1+6r-5s)/10}\,_{2}F_{1}\left(\frac{6}{5},\frac{6}{5}(1+r)-s;\frac{12}{5}\Big|1-z\right)\nonumber \\
\label{eq:crossed_channel_block}\\
\nonumber 
\end{eqnarray}
For clarity of conventions, the insertion points of the fields were
displayed above; in the following considerations they are suppressed.
Denoting $\Phi=\Phi_{2,1}$ and its chiral part by $\phi=\phi_{2,1}$
, the two bases are related by the following duality relations 
\begin{eqnarray}
\blockZgen{i}{i}{\phi}{\phi}{j} & = & \sum_{k}\mathcal{F}_{jk}[i]\blockOnemZgen{i}{i}{\phi}{\phi}{k}\nonumber \\
\label{eq:duality_relations}
\end{eqnarray}
where
\begin{equation}
\mathcal{F}_{jk}[i]=F_{jk}\left[\begin{array}{cc}
\phi & \phi\\
i & i
\end{array}\right]
\end{equation}
are the so-called fusion coefficients. These can be easily obtained
using the transformation formulas obeyed by the hypergeometric functions.
The ones relevant for our calculations are
\begin{eqnarray}
\mathcal{F}_{\phi_{r-1,s}\mathbb{I}}[\phi_{r,s}] & = & -\frac{2\sqrt{\frac{2}{5+\sqrt{5}}}\pi\Gamma\left(-\frac{6r}{5}+s+1\right)}{\Gamma\left(-\frac{2}{5}\right)\Gamma\left(\frac{6}{5}\right)\Gamma\left(-\frac{6r}{5}+s+\frac{6}{5}\right)}\nonumber \\
\mathcal{F}_{\phi_{r-1,s}\phi_{3,1}}[\phi_{r,s}] & = & \frac{2\sqrt{\frac{2}{5+\sqrt{5}}}\pi\Gamma\left(-\frac{6r}{5}+s+1\right)}{\Gamma\left(-\frac{1}{5}\right)\Gamma\left(\frac{12}{5}\right)\Gamma\left(-\frac{6r}{5}+s-\frac{1}{5}\right)}\nonumber \\
\mathcal{F}_{\phi_{r+1,s}\mathbb{I}}[\phi_{r,s}] & = & -\frac{2\sqrt{\frac{2}{5+\sqrt{5}}}\pi\Gamma\left(\frac{6r}{5}-s+1\right)}{\Gamma\left(-\frac{2}{5}\right)\Gamma\left(\frac{6}{5}\right)\Gamma\left(\frac{6r}{5}-s+\frac{6}{5}\right)}\nonumber \\
\mathcal{F}_{\phi_{r+1,s}\phi_{3,1}}[\phi_{r,s}] & = & \frac{2\sqrt{\frac{2}{5+\sqrt{5}}}\pi\Gamma\left(\frac{6r}{5}-s+1\right)}{\Gamma\left(-\frac{1}{5}\right)\Gamma\left(\frac{12}{5}\right)\Gamma\left(\frac{6r}{5}-s-\frac{1}{5}\right)}
\end{eqnarray}
Another necessary ingredient is the expansion of the blocks (\ref{eq:crossed_channel_block})
around $z=1$. One way this can be calculated is computing the series
expansion of the conformal blocks using the Taylor series of the hypergeometric
function and the binomial series. 

On the other hand, a model independent way to obtain the expansion
is provided by Virasoro symmetry. Recalling the notations in (\ref{eq:dual_channel_block_expansion})
\begin{eqnarray}
\blockOnemZgen{i}{i}{j}{j}{k} & = & \left(1-z\right)^{-2h_{j}+h_{k}}\sum_{r=0}^{\infty}B_{r}\left[\begin{array}{cc}
j & j\\
i & i
\end{array}k\right]\left(1-z\right)^{r}\nonumber \\
\label{eq:dualconfblockexpansion}
\end{eqnarray}
the first few coefficients can be easily obtained using the conformal
Ward identities, which give the following commutation relations between
the Virasoro generators and the primary fields:
\begin{eqnarray}
\left[L_{n}\left(z\right),\Phi\left(w,\bar{w}\right)\right] & = & h\left(n+1\right)\left(w-z\right)^{n}\Phi\left(w,\bar{w}\right)+\left(w-z\right)^{n+1}\partial_{w}\Phi\left(w,\bar{w}\right)\nonumber \\
\left[\bar{L}_{n}\left(\bar{z}\right),\Phi\left(w,\bar{w}\right)\right] & = & \bar{h}\left(n+1\right)\left(\bar{w}-\bar{z}\right)^{n}\Phi\left(w,\bar{w}\right)+\left(\bar{w}-\bar{z}\right)^{n+1}\partial_{\bar{w}}\Phi\left(w,\bar{w}\right)\label{eq:conformal_ward_identities}
\end{eqnarray}
where 
\begin{equation}
L_{n}(z)=\oint_{z}\frac{d\zeta}{2\pi i}\left(\zeta-z\right)^{n+1}T(\zeta)\qquad\bar{L}_{n}(\bar{z})=\oint_{\bar{z}}\frac{d\bar{\zeta}}{2\pi i}\left(\bar{\zeta}-\bar{z}\right)^{n+1}\bar{T}(\bar{\zeta})
\end{equation}
are the modes of the conformal energy momentum tensor located at $(z,\bar{z})$;
the modes located at $z=\infty$ are given by
\begin{equation}
L_{n}(\infty)=-\oint_{\infty}\frac{d\zeta}{2\pi i}\zeta^{-n+1}T(\zeta)=L_{-n}(0)\qquad\bar{L}_{n}(\infty)=-\oint_{\infty}\frac{d\bar{\zeta}}{2\pi i}\bar{\zeta}^{-n+1}\bar{T}(\bar{\zeta})=\bar{L}_{-n}(0)
\end{equation}
We computed the block coefficients up to $r=5$, but for the sake
of brevity we only give the first three cases:
\begin{eqnarray}
B_{0}\left[\begin{array}{cc}
j & j\\
i & i
\end{array}k\right] & = & 1\label{eq:confblocklevelcoeffs}\\
B_{1}\left[\begin{array}{cc}
j & j\\
i & i
\end{array}k\right] & = & \frac{h_{k}}{2}\nonumber \\
B_{2}\left[\begin{array}{cc}
j & j\\
i & i
\end{array}k\right] & = & \frac{h_{k}\left[(c+8)h_{k}^{2}+2(c-4)h_{k}+c+4h_{j}(h_{k}-1)+8h_{k}^{3}\right]+4h_{i}\left[h_{j}(4h_{k}+2)+\left(h_{k}-1\right)h_{k}\right]}{4\left(2(c-5)h_{k}+c+16h_{k}^{2}\right)}\nonumber 
\end{eqnarray}
where $h_{i}$,$h_{j}$ and $h_{k}$ are the conformal weights of
the respective fields.

For descendant state calculations we consider the first level only,
as this is all we need in the main text. The duality relations have
the same fusion coefficients 
\begin{eqnarray}
\blockZgen{L_{-1}i}{L_{-1}i}{\phi}{\phi}{j} & = & \sum_{k}\mathcal{F}_{jk}[i]\blockOnemZgen{L_{-1}i}{L_{-1}i}{\phi}{\phi}{k}\nonumber \\
\label{eq:desc_duality_relations}
\end{eqnarray}
and the conformal blocks in the dual channel can be expanded as
\begin{eqnarray}
\blockOnemZgen{L_{-1}i}{L_{-1}i}{j}{j}{k} & = & \left(1-z\right)^{-2h_{j}+h_{k}}\sum_{r=0}^{\infty}B_{r}\left[\begin{array}{cc}
j & j\\
L_{-1}i & L_{-1}i
\end{array}k\right]\left(1-z\right)^{r}\nonumber \\
\label{eq:desc_dualconfblockexpansion}
\end{eqnarray}
where we computed the coefficients up to $r=5$. The first three of
them are
\begin{eqnarray}
B_{0}\left[\begin{array}{cc}
j & j\\
L_{-1}i & L_{-1}i
\end{array}k\right] & = & 2h_{i}+h_{k}^{2}-h_{k}\label{eq:desc_confblocklevelcoeffs}\\
B_{1}\left[\begin{array}{cc}
j & j\\
L_{-1}i & L_{-1}i
\end{array}k\right] & = & \frac{1}{2}h_{k}\left(2h_{i}+h_{k}^{2}-h_{k}\right)\nonumber \\
B_{2}\left[\begin{array}{cc}
j & j\\
L_{-1}i & L_{-1}i
\end{array}k\right] & = & \frac{h_{k}\left(h_{k}^{2}-1\right)\left(h_{k}\left(ch_{k}+c+8h_{k}^{2}-4\right)+4h_{j}\left(h_{k}+2\right)\right)+8h_{i}^{2}\left(h_{j}\left(4h_{k}+2\right)+\left(h_{k}-1\right)h_{k}\right)}{4\left(2(c-5)h_{k}+c+16h_{k}^{2}\right)}\nonumber \\
 & + & \frac{h_{i}\left(h_{k}\left((c+12)h_{k}^{2}+2(c-5)h_{k}+c+10h_{k}^{3}-4\right)+8h_{j}\left(h_{k}^{3}+4h_{k}^{2}+3h_{k}+1\right)\right)}{2\left(2(c-5)h_{k}+c+16h_{k}^{2}\right)}\nonumber 
\end{eqnarray}

\subsection{Structure constants \label{sub:Structure-constants}}

For reference, here we list the matrix elements of the field $\Phi_{\frac{2}{5},\frac{2}{5}}$
between the primary states of the Hilbert space (\ref{eq:Dinvariantsectors}).
These can be arranged by the four sectors (\ref{eq:Hilbert_space_sectors}),
as matrix elements between different sectors vanish. The full set
of structure constants can be obtained from \cite{Runkel:1999dz};
here we present them in a basis of states which is orthonormal.

In the sector $\mathcal{H}_{0}=\mathcal{S}_{0,0}\oplus\mathcal{S}_{\frac{2}{5},\frac{2}{5}}\oplus\mathcal{S}_{\frac{7}{5},\frac{7}{5}}\oplus\mathcal{S}_{3,3}$,
the matrix of $\Phi_{\frac{2}{5},\frac{2}{5}}(1,1)$ on the basis
of primary states (ordered the same way as the modules) is
\begin{equation}
\left(\begin{array}{cccc}
0 & 1 & 0 & 0\\
1 & 0 & 0.9363044884 & 0\\
0 & 0.9363044884 & 0 & 0.8076923077\\
0 & 0 & 0.8076923077 & 0
\end{array}\right)\label{eq:opesect1}
\end{equation}
In both sectors $\mathcal{H}_{\pm}=\mathcal{S}_{\frac{1}{15},\frac{1}{15}}^{\pm}\oplus\mathcal{S}_{\frac{2}{3},\frac{2}{3}}^{\pm}$
one has
\begin{equation}
\left(\begin{array}{cc}
0.5461776182 & 2/3\\
2/3 & 0
\end{array}\right)\label{eq:opesect2}
\end{equation}
while in $\mathcal{H}_{1}=\mathcal{S}_{\frac{2}{5},\frac{7}{5}}\oplus\mathcal{S}_{\frac{7}{5},\frac{2}{5}}\oplus\mathcal{S}_{0,3}\oplus\mathcal{S}_{3,0}$
the matrix elements are
\begin{equation}
\left(\begin{array}{cccc}
0 & 0.9363044884 & 0.8987170343 & 0\\
0.9363044884 & 0 & 0 & 0.8987170343\\
0.8987170343 & 0 & 0 & 0\\
0 & 0.8987170343 & 0 & 0
\end{array}\right)\label{eq:opesect4}
\end{equation}
The above structure constants are in one-to-one correspondence with
the operator product coefficients involving the field $\Phi=\Phi_{\frac{2}{5},\frac{2}{5}}$.
Writing the operator product expansion in the form
\begin{equation}
\Phi_{\frac{2}{5},\frac{2}{5}}(z,z)A(0,0)=\sum_{B}C_{\Phi A}^{B}\frac{B(0,0)}{z^{h_{A}+2/5-h_{B}}\bar{z}^{\bar{h}_{A}+2/5-\bar{h}_{B}}}
\end{equation}
the OPE coefficients are
\begin{equation}
C_{\Phi A}^{B}=\langle B|\Phi_{\frac{2}{5},\frac{2}{5}}(1,1)|A\rangle\label{eq:OPE_coefficients}
\end{equation}
For primary states, these coefficients are given above in (\ref{eq:opesect1},\ref{eq:opesect2},\ref{eq:opesect4});
for descendant states they be constructed from the primary ones by
a recursive application of the conformal Ward identities (\ref{eq:conformal_ward_identities}).

\section{Derivation of the UV limit of the excited Potts TBA \label{sec:Derivation-of-the-UV-limit}}

Let us introduce a short-hand notation for the source terms
\begin{eqnarray*}
g(\theta|\theta^{+},\theta^{-}) & = & \sum_{k=1}^{N^{+}}\log\frac{S_{1}(\theta-\theta_{k}^{+})}{S_{2}(\theta-\bar{\theta}_{k}^{+})}+\sum_{l=1}^{N^{-}}\log\frac{S_{2}(\theta-\theta_{l}^{-})}{S_{1}(\theta-\bar{\theta}_{l}^{-})}\\
\bar{g}(\theta|\theta^{+},\theta^{-}) & = & \sum_{k=1}^{N^{+}}\log\frac{S_{2}(\theta-\theta_{k}^{+})}{S_{1}(\theta-\bar{\theta}_{k}^{+})}+\sum_{l=1}^{N^{-}}\log\frac{S_{1}(\theta-\theta_{l}^{-})}{S_{2}(\theta-\bar{\theta}_{l}^{-})}
\end{eqnarray*}
so that we can write the TBA equations in the form
\begin{eqnarray*}
\epsilon_{1}(\theta) & = & i\omega+mR\cosh\theta+g(\theta|\theta^{+},\theta^{-})-\phi_{1}\star L_{1}(\theta)-\phi_{2}\star L_{2}(\theta)\\
\epsilon_{2}(\theta) & = & -i\omega+mR\cosh\theta+\bar{g}(\theta|\theta^{+},\theta^{-})-\phi_{1}\star L_{2}(\theta)-\phi_{2}\star L_{1}(\theta)\\
 &  & e^{\epsilon_{1}(\theta_{k}^{+})}=e^{\epsilon_{1}(\bar{\theta}_{k}^{-})}=-1\\
 &  & e^{\epsilon_{2}(\theta_{k}^{-})}=e^{\epsilon_{2}(\bar{\theta}_{k}^{+})}=-1\\
E(R) & = & -im\sum_{k}\left(\sinh\theta_{k}^{+}-\sinh\bar{\theta}_{k}^{+}\right)-im\sum_{l}\left(\sinh\theta_{l}^{-}-\sinh\bar{\theta}_{l}^{-}\right)-\int\frac{d\theta}{2\pi}m\cosh\theta\,\left(L_{1}(\theta)+L_{2}(\theta)\right)
\end{eqnarray*}
where the twist parameter can take the values where
\begin{equation}
\omega=\frac{2\pi}{3}n_{\omega}\qquad n_{\omega}=-1,0,+1
\end{equation}
We only derive the right-moving conformal behaviour; the left-moving
part can be obtained in a similar way. For $mR\ll1$ the right kink
limit of the TBA is obtained by redefining
\begin{equation}
\theta\rightarrow\theta-\log\frac{1}{mR}
\end{equation}
and similarly for the positions of the sources
\begin{equation}
\theta_{k}^{\pm}\rightarrow\theta_{k}^{\pm}-\log\frac{1}{mR}\qquad\bar{\theta}_{k}^{\pm}\rightarrow\bar{\theta}_{k}^{\pm}-\log\frac{1}{mR}
\end{equation}
Those sources whose positions remain finite in the limit are called
right movers. To obtain the limit of the source terms, one can compute
\begin{eqnarray}
\lim_{\theta\rightarrow+\infty}\log\frac{S_{1}(\theta-\theta_{k}^{+})}{S_{2}(\theta-\bar{\theta}_{k}^{+})}=-\frac{2\pi}{3}i & \qquad & \lim_{\theta\rightarrow-\infty}\log\frac{S_{1}(\theta-\theta_{k}^{-})}{S_{2}(\theta-\bar{\theta}_{k}^{-})}=\frac{2\pi}{3}i\nonumber \\
\lim_{\theta\rightarrow+\infty}\log\frac{S_{2}(\theta-\theta_{k}^{+})}{S_{1}(\theta-\bar{\theta}_{k}^{+})}=\frac{2\pi}{3}i & \qquad & \lim_{\theta\rightarrow-\infty}\log\frac{S_{2}(\theta-\theta_{k}^{+})}{S_{1}(\theta-\bar{\theta}_{k}^{+})}=-\frac{2\pi}{3}i
\end{eqnarray}
Taking the limit $mR\rightarrow0$ we get that the right kink limiting
functions
\begin{equation}
\epsilon_{i}^{R}(\theta)=\lim_{R\rightarrow0}\epsilon_{i}(\theta-\log mR)
\end{equation}
satisfy the equations 
\begin{eqnarray}
\epsilon_{1}^{R}(\theta) & = & \frac{1}{2}e^{\theta}+i\omega_{R}+g_{R}(\theta|\theta^{+},\theta^{-})-\phi_{1}\star L_{1}(\theta)-\phi_{2}\star L_{2}(\theta)\nonumber \\
\epsilon_{2}^{R}(\theta) & = & \frac{1}{2}e^{\theta}-i\omega_{R}+\bar{g}_{R}(\theta|\theta^{+},\theta^{-})-\phi_{1}\star L_{2}(\theta)-\phi_{2}\star L_{1}(\theta)\nonumber \\
\epsilon_{1}^{R}(\theta_{k}^{+}) & = & i\pi\left(2n_{k}^{+}+1\right)\nonumber \\
\epsilon_{2}^{R}(\theta_{l}^{-}) & = & i\pi\left(2n_{l}^{+}+1\right)
\end{eqnarray}
where
\begin{eqnarray}
g_{R}(\theta|\theta^{+},\theta^{-}) & = & \sum_{k=1}^{N_{R}^{+}}\log\frac{S_{1}(\theta-\theta_{k}^{+})}{S_{2}(\theta-\bar{\theta}_{k}^{+})}+\sum_{l=1}^{N_{R}^{-}}\log\frac{S_{2}(\theta-\theta_{l}^{-})}{S_{1}(\theta-\bar{\theta}_{l}^{-})}\nonumber \\
\bar{g}_{R}(\theta|\theta^{+},\theta^{-}) & = & \sum_{k=1}^{N_{R}^{+}}\log\frac{S_{2}(\theta-\theta_{k}^{+})}{S_{1}(\theta-\bar{\theta}_{k}^{+})}+\sum_{l=1}^{N_{R}^{-}}\log\frac{S_{1}(\theta-\theta_{l}^{-})}{S_{2}(\theta-\bar{\theta}_{l}^{-})}
\end{eqnarray}
with sums only over the right-movers, and the effective right twist
is
\begin{equation}
\omega_{R}=\omega+\frac{2\pi}{3}\left[\left(N^{-}-N_{R}^{-}\right)-\left(N^{+}-N_{R}^{+}\right)\right]
\end{equation}
The right handed component of the effective central charge can be
written as 
\begin{equation}
c_{R}=\frac{6}{2\pi}i\sum_{k=1}^{N_{R}^{+}}\left(e^{\theta_{k}^{+}}-e^{\bar{\theta}_{k}^{+}}\right)+\frac{6}{2\pi}i\sum_{l=1}^{N_{R}^{-}}\left(e^{\theta_{l}^{-}}-e^{\bar{\theta}_{l}^{-}}\right)+\frac{3}{\pi^{2}}\int d\theta\frac{e^{\theta}}{2}\,\left(L_{1}(\theta)+L_{2}(\theta)\right)\label{eq:cR}
\end{equation}
We can rewrite these equations in the following form 
\begin{eqnarray}
\epsilon_{1}^{R}(\theta) & = & \frac{1}{2}e^{\theta}+i\omega_{R}-2i\pi m_{R}+g_{R}(\theta|\theta^{+},\theta^{-})-\phi_{1}\star L_{1}(\theta)-\phi_{2}\star L_{2}(\theta)\nonumber \\
\epsilon_{2}^{R}(\theta) & = & \frac{1}{2}e^{\theta}-i\omega_{R}+2i\pi m_{R}+\bar{g}_{R}(\theta|\theta^{+},\theta^{-})-\phi_{1}\star L_{2}(\theta)-\phi_{2}\star L_{1}(\theta)\nonumber \\
\epsilon_{1}^{R}(\theta_{k}^{+}) & = & i\pi\left(2n_{k}^{+}-2m_{R}+1\right)\nonumber \\
\epsilon_{2}^{R}(\theta_{l}^{-}) & = & i\pi\left(2n_{l}^{+}-2m_{R}+1\right)
\end{eqnarray}
where $m_{R}$ is defined from 
\begin{eqnarray}
i\omega_{R}+\lim_{\theta\rightarrow-\infty}g_{R}\left(\theta|\theta^{+},\theta^{-}\right) & = & i\frac{2\pi}{3}\left[n_{\omega}+\left(N^{-}-N_{R}^{-}\right)-\left(N^{+}-N_{R}^{+}\right)-N_{R}^{-}+N_{R}^{+}\right]\nonumber \\
 & = & i\frac{2\pi}{3}\left(3m_{R}+\tilde{n}_{\omega}\right)
\end{eqnarray}
where $m_{R}$ is an integer and $\tilde{n}_{\omega}=0$ or $\pm1$
is the remainder. One can the use the standard dilogarithm trick \cite{Zamolodchikov:1989cf,Dorey:1996re}
to write
\begin{eqnarray}
\frac{1}{2}e^{\theta} & = & \epsilon_{1}^{R}(\theta)-i\omega_{R}+2i\pi m_{R}-g_{R}(\theta|\theta^{+},\theta^{-})+\phi_{1}\star L_{1}(\theta)+\phi_{2}\star L_{2}(\theta)\nonumber \\
\frac{1}{2}e^{\theta} & = & \epsilon_{2}^{R}(\theta)+i\omega_{R}-2i\pi m_{R}-\bar{g}_{R}(\theta|\theta^{+},\theta^{-})+\phi_{1}\star L_{2}(\theta)+\phi_{2}\star L_{1}(\theta)
\end{eqnarray}
Differentiating the two sides
\begin{eqnarray}
\frac{1}{2}e^{\theta} & = & \frac{d}{d\theta}\left\{ \epsilon_{1}^{R}(\theta)-g_{R}(\theta|\theta^{+},\theta^{-})+\phi_{1}\star L_{1}(\theta)+\phi_{2}\star L_{2}(\theta)\right\} \nonumber \\
\frac{1}{2}e^{\theta} & = & \frac{d}{d\theta}\left\{ \epsilon_{2}^{R}(\theta)-\bar{g}_{R}(\theta|\theta^{+},\theta^{-})+\phi_{1}\star L_{2}(\theta)+\phi_{2}\star L_{1}(\theta)\right\} 
\end{eqnarray}
and substituting into the expression (\ref{eq:cR}) we obtain 
\begin{eqnarray}
c_{R} & = & \frac{6}{2\pi}i\sum_{k=1}^{N_{R}^{+}}\left(e^{\theta_{k}^{+}}-e^{\bar{\theta}_{k}^{+}}\right)+\frac{6}{2\pi}i\sum_{l=1}^{N_{R}^{-}}\left(e^{\theta_{l}^{-}}-e^{\bar{\theta}_{l}^{-}}\right)\nonumber \\
 &  & +\frac{3}{\pi^{2}}\int d\theta\frac{d}{d\theta}\left\{ \epsilon_{1}^{R}(\theta)-g_{R}(\theta|\theta^{+},\theta^{-})+\phi_{1}\star L_{1}(\theta)+\phi_{2}\star L_{2}(\theta)\right\} \, L_{1}(\theta)\nonumber \\
 &  & +\frac{3}{\pi^{2}}\int d\theta\frac{d}{d\theta}\left\{ \epsilon_{2}^{R}(\theta)-\bar{g}_{R}(\theta|\theta^{+},\theta^{-})+\phi_{1}\star L_{2}(\theta)+\phi_{2}\star L_{1}(\theta)\right\} \, L_{2}(\theta)\nonumber \\
 & = & \frac{6}{2\pi}i\sum_{k=1}^{N_{R}^{+}}\left(e^{\theta_{k}^{+}}-e^{\bar{\theta}_{k}^{+}}\right)+\frac{6}{2\pi}i\sum_{l=1}^{N_{R}^{-}}\left(e^{\theta_{l}^{-}}-e^{\bar{\theta}_{l}^{-}}\right)\nonumber \\
 &  & +\frac{3}{\pi^{2}}\int_{\epsilon_{1}^{R}(-\infty)}^{+\infty}d\epsilon\log(1+e^{-\epsilon})+\frac{6}{\pi^{2}}\int_{\epsilon_{2}^{R}(-\infty)}^{+\infty}d\epsilon\log(1+e^{-\epsilon})\nonumber \\
 &  & +\frac{3}{\pi^{2}}\int d\theta\frac{d}{d\theta}\left\{ -g_{R}(\theta|\theta^{+},\theta^{-})+\phi_{1}\star L_{1}(\theta)+\phi_{2}\star L_{2}(\theta)\right\} \, L_{1}(\theta)\nonumber \\
 &  & +\frac{3}{\pi^{2}}\int d\theta\frac{d}{d\theta}\left\{ -\bar{g}_{R}(\theta|\theta^{+},\theta^{-})+\phi_{1}\star L_{2}(\theta)+\phi_{2}\star L_{1}(\theta)\right\} \, L_{2}(\theta)
\end{eqnarray}
where the integrals over $\epsilon$ must be taken over an appropriate
contour in the $\epsilon$ plane which is analytically equivalent
the curves $\epsilon_{i}^{R}(\theta)$ as $\theta$ runs over the
real line. 

In the next step, we can treat the $\theta$ integrals using partial
integration:
\begin{eqnarray}
 &  & \int d\theta\frac{d}{d\theta}\left\{ -g_{R}(\theta|\theta^{+},\theta^{-})+\phi_{1}\star L_{1}(\theta)+\phi_{2}\star L_{2}(\theta)\right\} \, L_{1}(\theta)\nonumber \\
 &  & +\int d\theta\frac{d}{d\theta}\left\{ -\bar{g}_{R}(\theta|\theta^{+},\theta^{-})+\phi_{1}\star L_{2}(\theta)+\phi_{2}\star L_{1}(\theta)\right\} \, L_{2}(\theta)\nonumber \\
 & = & \int d\theta\left\{ -g_{R}'(\theta|\theta^{+},\theta^{-})L_{1}(\theta)-\bar{g}_{R}'(\theta|\theta^{+},\theta^{-})L_{2}(\theta)\right\} \nonumber \\
 &  & -\int d\theta\left\{ \phi_{1}\star L_{1}(\theta)+\phi_{2}\star L_{2}(\theta)\right\} \, L_{1}'(\theta)\nonumber \\
 &  & -\int d\theta\left\{ \phi_{1}\star L_{2}(\theta)+\phi_{2}\star L_{1}(\theta)\right\} \, L_{2}'(\theta)\nonumber \\
 &  & +\left[\left\{ \phi_{1}\star L_{1}(\theta)+\phi_{2}\star L_{2}(\theta)\right\} \, L_{1}(\theta)\right]_{-\infty}^{\infty}\nonumber \\
 &  & +\left[\left\{ \phi_{1}\star L_{2}(\theta)+\phi_{2}\star L_{1}(\theta)\right\} \, L_{2}(\theta)\right]_{-\infty}^{\infty}
\end{eqnarray}
and the fact that $L_{1,2}(\infty)=0$ to obtain
\begin{eqnarray}
c_{R} & = & \frac{6}{2\pi}i\sum_{k=1}^{N_{R}^{+}}\left(e^{\theta_{k}^{+}}-e^{\bar{\theta}_{k}^{+}}\right)+\frac{6}{2\pi}i\sum_{l=1}^{N_{R}^{-}}\left(e^{\theta_{l}^{-}}-e^{\bar{\theta}_{l}^{-}}\right)\nonumber \\
 &  & +\frac{3}{\pi^{2}}\int_{\epsilon_{1}^{R}(-\infty)}^{+\infty}d\epsilon\log(1+e^{-\epsilon})+\frac{3}{\pi^{2}}\int_{\epsilon_{2}^{R}(-\infty)}^{+\infty}d\epsilon\log(1+e^{-\epsilon})\nonumber \\
 &  & +\frac{3}{\pi^{2}}\int d\theta\left\{ -g_{R}'(\theta|\theta^{+},\theta^{-})L_{1}(\theta)-\bar{g}_{R}'(\theta|\theta^{+},\theta^{-})L_{2}(\theta)\right\} \nonumber \\
 &  & -\frac{3}{\pi^{2}}\frac{1}{2}\left[\phi_{1}\star L_{1}(-\infty)+\phi_{2}\star L_{2}(-\infty)\right]\, L_{1}(-\infty)\nonumber \\
 &  & -\frac{3}{\pi^{2}}\frac{1}{2}\left[\phi_{1}\star L_{2}(-\infty)+\phi_{2}\star L_{2}(-\infty)\right]\, L_{2}(-\infty)
\end{eqnarray}
The remaining integrals can be expressed using the kink TBA equations
for $\theta\rightarrow-\infty$:
\begin{eqnarray}
\phi_{1}\star L_{1}(-\infty)+\phi_{2}\star L_{2}(-\infty) & = & -\epsilon_{1}^{R}(-\infty)+i\omega_{R}-2i\pi m_{R}+g_{R}(-\infty|\theta^{+},\theta^{-})\nonumber \\
\phi_{1}\star L_{2}(-\infty)+\phi_{2}\star L_{1}(-\infty) & = & -\epsilon_{2}^{R}(-\infty)-i\omega_{R}+2i\pi m_{R}+\bar{g}_{R}(-\infty|\theta^{+},\theta^{-})
\end{eqnarray}
Using the definition of $m_{R}$ leads to 
\begin{eqnarray}
\phi_{1}\star L_{1}(-\infty)+\phi_{2}\star L_{2}(-\infty) & = & -\epsilon_{1}^{R}(-\infty)+i\frac{2\pi}{3}\tilde{n}_{\omega}\nonumber \\
\phi_{1}\star L_{2}(-\infty)+\phi_{2}\star L_{1}(-\infty) & = & -\epsilon_{2}^{R}(-\infty)-i\frac{2\pi}{3}\tilde{n}_{\omega}
\end{eqnarray}
For the terms involving $g_{R}'$ and $\bar{g}_{R}'$ we can write
\begin{eqnarray}
g_{R}'(\theta|\theta^{+},\theta^{-}) & = & \sum_{k=1}^{N_{R}^{+}}i\phi_{1}(\theta-\theta_{k}^{+})-i\phi_{2}(\theta-\bar{\theta}_{k}^{+})+\sum_{l=1}^{N_{R}^{-}}i\phi_{2}(\theta-\theta_{l}^{-})-i\phi_{1}(\theta-\bar{\theta}_{l}^{-})\nonumber \\
\bar{g}_{R}'(\theta|\theta^{+},\theta^{-}) & = & \sum_{k=1}^{N_{R}^{+}}i\phi_{2}(\theta-\theta_{k}^{+})-i\phi_{1}(\theta-\bar{\theta}_{k}^{+})+\sum_{l=1}^{N_{R}^{-}}i\phi_{1}(\theta-\theta_{l}^{-})-i\phi_{2}(\theta-\bar{\theta}_{l}^{-})
\end{eqnarray}
and so 
\begin{eqnarray}
 &  & \int d\theta\left\{ -g_{R}'(\theta|\theta^{+},\theta^{-})L_{1}(\theta)-\bar{g}_{R}'(\theta|\theta^{+},\theta^{-})L_{2}(\theta)\right\} \nonumber \\
 & = & -2\pi i\sum_{k=1}^{N_{R}^{+}}\left\{ \phi_{1}\star L_{1}(\theta_{k}^{+})+\phi_{2}\star L_{2}(\theta_{k}^{+})-\phi_{2}\star L_{1}(\bar{\theta}_{k}^{+})-\phi_{1}\star L_{2}(\bar{\theta}_{k}^{+})\right\} \nonumber \\
 &  & -2\pi i\sum_{l=1}^{N_{R}^{-}}\left\{ -\phi_{1}\star L_{1}(\bar{\theta}_{l}^{-})-\phi_{2}\star L_{2}(\bar{\theta}_{l}^{-})+\phi_{2}\star L_{1}(\theta_{l}^{-})+\phi_{1}\star L_{2}(\theta_{l}^{-})\right\} 
\end{eqnarray}
Now we can eliminate the convolution terms using the equations determining
the singularity positions
\begin{eqnarray}
i\pi\left(2n_{k}^{+}+1\right) & = & \frac{1}{2}e^{\theta_{k}^{+}}+i\omega_{R}+g_{R}(\theta_{k}^{+}|\theta^{+},\theta^{-})-\phi_{1}\star L_{1}(\theta_{k}^{+})-\phi_{2}\star L_{2}(\theta_{k}^{+})\nonumber \\
-i\pi\left(2n_{k}^{+}+1\right) & = & \frac{1}{2}e^{\bar{\theta}_{k}^{+}}-i\omega_{R}+\bar{g}_{R}(\bar{\theta}_{k}^{+}|\theta^{+},\theta^{-})-\phi_{1}\star L_{2}(\bar{\theta}_{k}^{+})-\phi_{2}\star L_{1}(\bar{\theta}_{k}^{+})
\end{eqnarray}
\begin{eqnarray}
-i\pi\left(2n_{l}^{-}+1\right) & = & \frac{1}{2}e^{\bar{\theta}_{l}^{-}}+i\omega_{R}+g_{R}(\bar{\theta}_{l}^{-}|\theta^{+},\theta^{-})-\phi_{1}\star L_{1}(\bar{\theta}_{l}^{-})-\phi_{2}\star L_{2}(\bar{\theta}_{l}^{-})\nonumber \\
i\pi\left(2n_{l}^{-}+1\right) & = & \frac{1}{2}e^{\theta_{l}^{-}}-i\omega_{R}+\bar{g}_{R}(\theta_{l}^{-}|\theta^{+},\theta^{-})-\phi_{1}\star L_{2}(\theta_{l}^{-})-\phi_{2}\star L_{1}(\theta_{l}^{-})
\end{eqnarray}
The end result is
\begin{eqnarray}
c_{R} & = & \frac{3}{\pi^{2}}\left\{ \int_{\epsilon_{1}^{R}(-\infty)}^{+\infty}d\epsilon\log(1+e^{-\epsilon})\right\} +\frac{3}{\pi^{2}}\left\{ \int_{\epsilon_{2}^{R}(-\infty)}^{+\infty}d\epsilon\log(1+e^{-\epsilon})\right\} \nonumber \\
 &  & -12\sum_{k=1}^{N_{R}^{+}}\left(2n_{k}^{+}+1\right)-12\sum_{l=1}^{N_{R}^{-}}\left(2n_{l}^{-}+1\right)+\frac{12}{\pi}\omega_{R}\left(N_{R}^{+}-N_{R}^{-}\right)\nonumber \\
 &  & -\frac{6}{\pi}i\sum_{k=1}^{N_{R}^{+}}\left\{ g_{R}(\theta_{k}^{+}|\theta^{+},\theta^{-})-\bar{g}_{R}(\bar{\theta}_{k}^{+}|\theta^{+},\theta^{-})\right\} -\frac{6}{\pi}i\sum_{l=1}^{N_{R}^{-}}\left\{ \bar{g}_{R}(\bar{\theta}_{l}^{-}|\theta^{+},\theta^{-})-g_{R}(\theta_{l}^{-}|\theta^{+},\theta^{-})\right\} \nonumber \\
 &  & -\frac{3}{\pi^{2}}\frac{1}{2}\left[-\log Y_{1}+i\frac{2\pi}{3}\tilde{n}_{\omega}\right]\,\log\left(1+Y_{1}^{-1}\right)-\frac{3}{\pi^{2}}\frac{1}{2}\left[-\log Y_{2}-i\frac{2\pi}{3}\tilde{n}_{\omega}\right]\,\log\left(1+Y_{2}^{-1}\right)\label{eq:cR_final}
\end{eqnarray}
where $Y_{i}=\epsilon_{i}\left(-\infty\right)$ are solutions of the
plateau equation
\begin{eqnarray}
\log Y_{1} & = & i\frac{2\pi}{3}\tilde{n}_{\omega}+\frac{1}{3}\log\left(1+Y_{1}^{-1}\right)+\frac{2}{3}\log\left(1+Y_{2}^{-1}\right)\nonumber \\
\log Y_{2} & = & -i\frac{2\pi}{3}\tilde{n}_{\omega}+\frac{2}{3}\log\left(1+Y_{1}^{-1}\right)+\frac{1}{3}\log\left(1+Y_{2}^{-1}\right)
\end{eqnarray}
From \cite{Martins1991c,Fendley:1991xn}, the solutions of these equations
are known, together with the values of the dilogarithm integrals:
\begin{eqnarray}
 &  & \frac{3}{\pi^{2}}\left\{ \int_{\epsilon_{1}^{R}(-\infty)}^{+\infty}d\epsilon\log(1+e^{-\epsilon})\right\} +\frac{3}{\pi^{2}}\left\{ \int_{\epsilon_{2}^{R}(-\infty)}^{+\infty}d\epsilon\log(1+e^{-\epsilon})\right\} \nonumber \\
 &  & -\frac{3}{\pi^{2}}\frac{1}{2}\left[-\log Y_{1}+i\frac{2\pi}{3}\tilde{n}_{\omega}\right]\,\log\left(1+Y_{1}^{-1}\right)-\frac{3}{\pi^{2}}\frac{1}{2}\left[-\log Y_{2}-i\frac{2\pi}{3}\tilde{n}_{\omega}\right]\,\log\left(1+Y_{2}^{-1}\right)\nonumber \\
 &  & =\begin{cases}
\frac{2}{5} & \quad\tilde{n}_{\omega}=0\\
-\frac{2}{5} & \quad\tilde{n}_{\omega}=\pm1
\end{cases}
\end{eqnarray}
Using standard identities for the logarithm of products, the contributions
containing the sums of $g_{R}$ and $\bar{g}_{R}$ terms in (\ref{eq:cR_final})
naively evaluate to zero. However, this result is changed by taking
care of the branch cuts of the logarithms. Using the notations of
subsection \ref{sub:The-UV-limit-of-the-TBA-equations}, the contribution
depends on the signs of $\delta_{r}^{\pm}$ of the corresponding singularities
and can quickly be evaluated individually for every state considered.

\section{Tables for the comparison between renormalized TCSA numerics and
TBA predictions\label{sec:Comparison-tables}}

\begin{samepage}

\begin{center}
\begin{tabular}{|l|c|c|c|c|c|}
\hline 
 & $r=0.1$ & $r=1$ & $r=3$ & $r=5$ & $r=7$\tabularnewline
\hline 
\hline 
TBA & $-4.1958706705$ & $-0.595088$ & $-0.8907$ & $-1.446$ & $-2.021$\tabularnewline
\hline 
raw TCSA level 12 & $-4.1947973491$ & $-0.568125$ & $-0.7649$ & $-1.188$ & $-1.603$\tabularnewline
\hline 
renormalized TCSA level 8 & $-4.1958706700$ & $-0.595083$ & $-0.8903$ & $-1.443$ & $-2.011$\tabularnewline
\hline 
renormalized TCSA level 12 & $-4.1958706700$ & $-0.595085$ & $-0.8905$ & $-1.444$ & $-2.014$\tabularnewline
\hline 
\end{tabular}\\
~\\
Ground state in $\mathcal{H}_{0}$ sector\\
~
\par\end{center}

\begin{center}
\begin{tabular}{|l|c|c|c|c|c|}
\hline 
 & $r=0.1$ (LYTCSA) & $r=1$(LYTCSA) & $r=3$(LYTCSA) & $r=5$ & $r=7$\tabularnewline
\hline 
\hline 
TBA & $46.1055595046$ & $4.8947517$ & $2.32864$ & $2.05871$ & $2.008$\tabularnewline
\hline 
raw TCSA level 12 & $46.1066313406$ & $4.9216798$ & $2.45431$ & $2.317$ & $2.429$\tabularnewline
\hline 
renormalized TCSA level 8 & $46.1055594445$ & $4.8947579$ & $2.32914$ & $2.062$ & $2.022$\tabularnewline
\hline 
renormalized TCSA level 12 & $46.1055595057$ & $4.8947566$ & $2.32897$ & $2.061$ & $2.018$\tabularnewline
\hline 
\end{tabular}\\
~\\
First excited state in $\mathcal{H}_{0}$ sector: stationary $A\bar{A}$
\par\end{center}

\noindent In the above data for small volumes, instead of analytically
continuing the TBA we simply used the correspondence with the scaling
Lee-Yang model, as the Lee-Yang TCSA is much easier to implement and
numerically precise enough for the present comparison.

~

\begin{center}
\begin{tabular}{|l|c|c|c|c|c|}
\hline 
 & $r=0.3$ & $r=1$ & $r=3$ & $r=5$ & $r=7$\tabularnewline
\hline 
\hline 
TBA & $57.0847781$ & $17.09738$ & $5.9176$ & $3.856$ & $3.064$\tabularnewline
\hline 
raw TCSA level 12 & $57.0898932$ & $17.12499$ & $6.0467$ & $4.122$ & $3.494$\tabularnewline
\hline 
renormalized TCSA level 8 & $57.0847809$ & $17.09742$ & $5.9188$ & $3.862$ & $3.079$\tabularnewline
\hline 
renormalized TCSA level 12 & $57.0847788$ & $17.09740$ & $5.9183$ & $3.859$ & $3.073$\tabularnewline
\hline 
\end{tabular}\\
~\\
Second excited state in $\mathcal{H}_{0}$ sector: moving $A\bar{A}$\\
~
\par\end{center}

\begin{center}
\begin{tabular}{|l|c|c|c|c|c|}
\hline 
 & $r=1.6$ & $r=2$ & $r=3$ & $r=5$ & $r=7$\tabularnewline
\hline 
\hline 
TBA & $11.4023947$ & $9.340653$ & $6.66642$ & $4.681$ & $3.939$\tabularnewline
\hline 
raw TCSA level 12 & $11.4556586$ & $9.413463$ & $6.79502$ & $4.946$ & $4.367$\tabularnewline
\hline 
renormalized TCSA level 8 & $11.4023746$ & $9.340648$ & $6.66667$ & $4.685$ & $3.951$\tabularnewline
\hline 
renormalized TCSA level 12 & $11.4023751$ & $9.340640$ & $6.66655$ & $4.683$ & $3.947$\tabularnewline
\hline 
\end{tabular}\\
~\\
Third excited state in $\mathcal{H}_{0}$ sector: $AAA$ three-particle
state\\
~
\par\end{center}

\begin{center}
\begin{tabular}{|l|c|c|c|c|c|}
\hline 
 & $r=1.2$ & $r=2$ & $r=3$ & $r=5$ & $r=7$\tabularnewline
\hline 
\hline 
TBA & $0.951783$ & $0.930075$ & $0.95363$ & $0.989$ & $0.997$\tabularnewline
\hline 
raw TCSA level 12 & $0.986768$ & $1.00174$ & $1.08041$ & $1.249$ & $1.422$\tabularnewline
\hline 
renormalized TCSA level 8 & $0.951786$ & $0.930273$ & $0.95457$ & $0.993$ & $1.015$\tabularnewline
\hline 
renormalized TCSA level 12 & $0.951776$ & $0.930195$ & $0.95420$ & $0.991$ & $1.009$\tabularnewline
\hline 
\end{tabular}\\
~\\
Stationary one particle state (ground state in $\mathcal{H}_{\pm}$
in the paramagnetic phase)\\
~
\par\end{center}

\begin{center}
\begin{tabular}{|l|c|c|c|c|c|}
\hline 
 & $r=0.1$ & $r=1$ & $r=3$ & $r=5$ & $r=7$\tabularnewline
\hline 
\hline 
TBA & $3.9667856906$ & $0.204269$ & $0.01284$ & $0.00128$ & $0.000145$\tabularnewline
\hline 
raw TCSA level 12 & $3.9678649202$ & $0.231387$ & $0.1395$ & $0.2621$ & $0.4221$\tabularnewline
\hline 
renormalized TCSA level 8 & $3.9667857008$ & $0.204285$ & $0.0137$ & $0.0064$ & $0.0157$\tabularnewline
\hline 
renormalized TCSA level 12 & $3.9667857002$ & $0.204278$ & $0.0133$ & $0.0042$ & $0.0088$\tabularnewline
\hline 
\end{tabular}\\
~\\
Twisted vacuum (ground state in $\mathcal{H}_{\pm}$ in the ferromagnetic
phase)\\
~
\par\end{center}

\begin{center}
\begin{tabular}{|l|c|c|c|c|c|}
\hline 
 & $r=0.1$ & $r=1$ & $r=3$ & $r=5$ & $r=7$\tabularnewline
\hline 
\hline 
TBA & $79.613697179$ & $8.1928567$ & $3.26417$ & $2.4947$ & $2.252$\tabularnewline
\hline 
raw TCSA level 12 & $79.614806090$ & $8.2207133$ & $3.39399$ & $2.7611$ & $2.682$\tabularnewline
\hline 
renormalized TCSA level 8 & $79.613697095$ & $8.1928575$ & $3.26436$ & $2.4966$ & $2.260$\tabularnewline
\hline 
renormalized TCSA level 12 & $79.613697169$ & $8.1928585$ & $3.26431$ & $2.4959$ & $2.257$\tabularnewline
\hline 
\end{tabular}\\
~\\
First $AA$ two-particle state (first excited state in $\mathcal{H}_{\pm}$
in the paramagnetic phase)\\
~
\par\end{center}

\begin{center}
\begin{tabular}{|l|c|c|c|c|c|}
\hline 
 & $r=0.1$ & $r=1$ & $r=3$ & $r=5$ & $r=7$\tabularnewline
\hline 
\hline 
TBA & $79.613694779$ & $8.191921$ & $3.25131$ & $2.4657$ & $2.216$\tabularnewline
\hline 
raw TCSA level 12 & $79.614803694$ & $8.219778$ & $3.38117$ & $2.7324$ & $2.647$\tabularnewline
\hline 
renormalized TCSA level 8 & $79.613694699$ & $8.191922$ & $3.25158$ & $2.4680$ & $2.226$\tabularnewline
\hline 
renormalized TCSA level 12 & $79.613694773$ & $8.191923$ & $3.25149$ & $2.4673$ & $2.223$\tabularnewline
\hline 
\end{tabular}\\
~\\
First twisted $A\bar{A}$ two-particle state (first excited state
in $\mathcal{H}_{\pm}$ in the ferromagnetic phase)
\par\end{center}

\end{samepage}

\bibliographystyle{../bibtex/utphys}
\bibliography{../bibtex/merge}

\end{document}